\documentclass[a4paper,12pt]{article}
\usepackage{amsmath,amsfonts,amssymb,graphicx,epsfig,pict2e}
\usepackage[usenames]{color}
\usepackage{pstricks}
\usepackage[backref=false]{hyperref}
\hypersetup{
colorlinks=true,
citecolor=red,
linkcolor=darkblue
}
\definecolor{darkblue}{rgb}{0,0,.8}
\definecolor{red}{rgb}{1,0,0}
\definecolor{lightgray}{rgb}{.8,.8,.8}

\numberwithin{equation}{section}
\newcommand{\nc}{\newcommand}
\nc{\fh}{\hat{f}}
\nc{\muh}{\hat{\mu}}
\nc{\nuh}{\hat{\nu}}
\nc{\mch}{\mathrm{ch}}
\nc{\spos}[2]{\makebox(0,0)[#1]{\small ${#2}$}}
\nc{\sm}[1]{{\scriptstyle #1}}
\nc{\ssm}[1]{{\scriptscriptstyle #1}}
\nc{\sposb}[2]{\makebox(0,0)[#1]{$ #2 $}}
\def\floor#1{\lfloor #1\rfloor}
\nc{\bib}{\bibitem}
\nc{\al}{\alpha}
\nc{\g}{\gamma}
\nc{\G}{\Gamma}
\nc{\D}{\Delta}
\nc{\eps}{\epsilon}
\nc{\la}{\lambda}
\nc{\La}{\Lambda}
\nc{\var}{\varphi}
\nc{\pa}{\partial}
\nc{\nn}{\nonumber \\ }
\nc{\hf}{\frac{1}{2}}
\nc{\dz}{\frac{dz}{2\pi i}}
\nc{\bin}[2]{\left(\!\!\!\begin{array}{c} {#1}\\ {#2} \end{array}\!\!\!\right)}
\nc{\smbin}[2]{\Big(\!\!\!\begin{array}{c} {#1}\\[-3pt] {#2} \end{array}\!\!\!\Big)}
\nc{\be}{\begin{equation}}
\nc{\ee}{\end{equation}}
\nc{\bea}{\begin{eqnarray}}
\nc{\eea}{\end{eqnarray}}
\nc{\bra}[1]{\langle {#1}|}
\nc{\ket}[1]{|{#1}\rangle}
\def\disp{\displaystyle}
\nc{\chit}{\raisebox{0.25ex}{$\chi$}}
\nc{\Db}{\mbox{\boldmath $D$}}
\nc{\Hb}{\mbox{\boldmath $H$}}
\nc{\Hc}{{\cal H}}
\nc{\Rc}{{\cal R}}
\nc{\Lc}{{\cal L}}
\nc{\Vc}{{\cal V}}
\nc{\Ib}{\mbox{\boldmath $I$}}
\nc{\qb}{\bar{q}}
\def\vvdots{\mathinner{\mkern1mu\raise1pt\vbox{\kern7pt\hbox{.}}\mkern2mu
  \raise4pt\hbox{.}\mkern2mu\raise7pt\hbox{.}\mkern1mu}}
\nc{\gauss}[2]{\left[\!\!\begin{array}{c} {#1}\\ {#2} \end{array}\!\!\right]}
\nc{\sgauss}[2]{\Big[\!\!\begin{array}{c} {#1}\\[-3pt] {#2} \end{array}\!\!\Big]}
\nc{\sbin}[2]{\left\{\!\!\!\begin{array}{c} {#1}\\ {#2} 
\end{array}\!\!\!\right\}}
\nc{\sbinlr}[2]{\Big\langle\!\!\begin{array}{c} {#1}\\ {#2} 
\end{array}\!\!\Big\rangle}
\nc{\bino}[2]{\left(\!\!\begin{array}{c} {#1}\\ {#2} \end{array}\!\!\right)}
\def\half {\mbox{$\textstyle \frac{1}{2}$}}
\def\vec#1{\mbox {\boldmath $#1$}}

\definecolor{lightblue}{rgb}{.55,.55,1}
\definecolor{midblue}{rgb}{.7,.7,1}
\definecolor{lightlightblue}{rgb}{.85,.85,1}
\definecolor{lightestblue}{rgb}{.96,.96,1}
\definecolor{lightpurple}{rgb}{1,.65,1}

\def\leftarc#1{\psarc[linecolor=blue,linewidth=1.5pt]#1{.5}{90}{270}}
\def\rightarc#1{\psarc[linecolor=blue,linewidth=1.5pt]#1{.5}{-90}{90}}
\def\loopa{
\psframe[linewidth=.25pt](0,0)(1,1)
\psarc[linewidth=1.5pt,linecolor=blue](1,0){.5}{90}{180}
\psarc[linewidth=1.5pt,linecolor=blue](0,1){.5}{-90}{0}
}
\def\loopb{
\psframe[linewidth=.25pt](0,0)(1,1)
\psarc[linewidth=1.5pt,linecolor=blue](0,0){.5}{0}{90}
\psarc[linewidth=1.5pt,linecolor=blue](1,1){.5}{180}{270}
}

\def\facegridpurple#1#2{
\psframe[fillstyle=solid,fillcolor=lightpurple,linewidth=0pt]#1#2
\psgrid[gridlabels=0pt,subgriddiv=1]#1#2}
\def\facegrid#1#2{
\psframe[fillstyle=solid,fillcolor=lightlightblue,linewidth=0pt]#1#2
\psgrid[gridlabels=0pt,subgriddiv=1]#1#2}

\nc{\fws}{\small}
\def\cat#1#2#3{C_{#1,#2}(#3)}
\def\catt#1#2#3{C_{#1,#2}'(#3)}

\def\qCat2{\mbox{qCat2}}

\nc{\sbinMm}[2]{\Big\langle\!\!\begin{array}{c} {#1}\\ {#2} 
\end{array}\!\!\Big\rangle}
\nc{\Bb}{\mbox{\boldmath$B$}}
\nc{\Kb}{\mbox{\boldmath$K$}}
\def\dddots{\mathinner{\mkern1mu\raise9pt\vbox{\kern7pt\hbox{.}}\mkern2mu
  \raise5pt\hbox{.}\mkern2mu\raise1pt\hbox{.}\mkern1mu}}
  
\def\face#1#2#3#4#5{\begin{pspicture}[shift=-.57](-.2,-.2)(1.2,1.2)
\facegrid{(0,0)}{(1,1)}
\psarc[linecolor=black](0,0){.15}{0}{90}
\rput(.5,.5){\scriptsize $#5$}
\end{pspicture}}

\def\rtri#1#2#3#4{\begin{pspicture}[shift=-1.1](-.2,-.2)(1.2,2.2)
\pspolygon[fillstyle=solid,fillcolor=lightlightblue](0,1)(1,0)(1,2)
\rput(.6,1){\scriptsize $#4$}
\end{pspicture}}
  
\def\DiagWt#1#2#3#4#5{\begin{pspicture}[shift=-1.1](-.2,-.2)(2.2,2.2)
\pspolygon[fillstyle=solid,fillcolor=lightlightblue](1,0)(2,1)(1,2)(0,1)
\rput(1,1){\scriptsize $#5$}
\psarc[linecolor=black](0,1){.15}{-45}{45}
\end{pspicture}}

\begin{document}

\topmargin -5mm
\oddsidemargin 5mm

\begin{titlepage}
\setcounter{page}{0}

\vspace{8mm}
\begin{center}
{\huge {\bf Infinitely Extended Kac Table\\[6pt] of Solvable Critical Dense Polymers}}

\vspace{10mm}
{\Large Paul A. Pearce}\\[.3cm]
{\em Department of Mathematics and Statistics, University of Melbourne}\\
{\em Parkville, Victoria 3010, Australia}\\[.4cm]
{\Large J{\o}rgen Rasmussen}\\[.3cm]
{\em School of Mathematics and Physics, University of Queensland}\\
{\em St Lucia, Brisbane, Queensland 4072, Australia}\\[.4cm]
{\Large Simon P. Villani}\\[.3cm]
{\em Department of Mathematics and Statistics, University of Melbourne}\\
{\em Parkville, Victoria 3010, Australia}\\[.4cm]
P.Pearce@ms.unimelb.edu.au, \ J.Rasmussen@uq.edu.au
\\[.05cm] 
S.Villani@ms.unimelb.edu.au
\end{center}

\vspace{10mm}
\centerline{{\bf{Abstract}}}
\vskip.4cm
\noindent
Solvable critical dense polymers is a Yang-Baxter integrable model of polymers on the square lattice.
It is the first member ${\cal LM}(1,2)$ of the family of logarithmic minimal models ${\cal LM}(p,p')$. 
The associated logarithmic conformal field theory admits an infinite family of  
Kac representations labelled by the Kac labels $r,s=1,2,\ldots$. 
In this paper, we explicitly construct the conjugate boundary conditions on the strip. The boundary operators 
are labelled by the Kac fusion \mbox{labels} $(r,s)=(r,1)\otimes (1,s)$ 
and involve a boundary field $\xi$. Tuning the field $\xi$ appropriately, we solve exactly for 
the transfer matrix eigenvalues on arbitrary finite-width strips and obtain the conformal spectra 
using the Euler-Maclaurin formula. 
The key to the solution is an inversion identity satisfied by the commuting double-row transfer matrices.
The transfer matrix eigenvalues are classified by the physical combinatorics 
of the patterns of zeros in the complex spectral-parameter plane. 
This yields selection rules for the physically relevant solutions to 
the inversion identity which takes the form of a decomposition into irreducible blocks 
corresponding combinatorially to 
finitized characters given by generalized $q$-Catalan polynomials.
This decomposition is in accord with the decomposition of the Kac characters into irreducible characters. 
In the scaling limit, we confirm the central charge $c=-2$ and the Kac formula for the conformal
weights $\Delta_{r,s}=\frac{(2r-s)^2-1}{8}$ for $r,s=1,2,3,\ldots$ in the infinitely extended Kac table.
\end{titlepage}
\newpage
\renewcommand{\thefootnote}{\arabic{footnote}}
\setcounter{footnote}{0}

\tableofcontents

\newpage
\section{Introduction}

Materials such as plastics, nylon, polyester and plexiglass 
are made from polymers~\cite{polymers,poly2,poly3} which consist of very long chain molecules 
with a large number of repeating structural units called monomers. 
Polymers exist in low- or high-temperature phases which are 
characterized as either dense or dilute. Polymers are {\em dense\/} 
if they fill a finite (non-zero) fraction of the available volume in 
the thermodynamic limit. 

The modern era of two-dimensional polymer theory began in the late eighties~\cite{Saleur87,Duplantier,SaleurSUSY} 
when Saleur and Duplantier initiated the study of polymers as a conformal field theory (CFT). 
From the viewpoint of lattice 
statistical mechanics, polymers are of interest as prototypical 
examples of systems involving (extended) non-local degrees of 
freedom. The non-local nature of these 
degrees of freedom has a profound effect on the associated conformal 
field theory (CFT) obtained in the continuum scaling limit. Indeed, the 
associated CFT is {\em logarithmic}~\cite{Gurarie} in the sense that certain 
representations of the dilatation 
Virasoro generator $L_0$ are non-diagonalizable and exhibit nontrivial Jordan 
cells.

 The first member ${\cal LM}(1,2)$ of the 
Yang-Baxter integrable family of 
logarithmic minimal models ${\cal LM}(p,p')$~\cite{PRZ} on the square lattice is solvable critical dense polymers~\cite{PR0610}. 
This model was solved exactly in \cite{PR0610} 
for certain boundary conditions of type $(1,s)$ on a strip. 
The model has also been solved exactly 
on the cylinder~\cite{PRS10} and torus~\cite{MPR2013} and its integrals of motion and Baxter $Q$-operator studied in \cite{Nigro09}. 
With appropriate ${\cal W}$-extended boundary conditions on the strip~\cite{PRR2008}, the lattice model is  
also compatible with the conformal spectra and fusion rules of {\em symplectic fermions}~\cite{Kau00}. 
We refer to the bibliographies of these papers and of~\cite{Ras1106} for related approaches to the problem and for more detailed references 
to the relevant literature. 
The logarithmic minimal model ${\cal LM}(p,p')$ is characterized by the {\it crossing parameter} $\lambda={(p'-p)\pi\over p'}$ so the crossing parameter of critical dense polymers is $\lambda={\pi\over 2}$.  An alternative lattice approach to logarithmic conformal field theory, including critical dense polymers, was introduced in~\cite{RS0701} and further developed in~\cite{GV1203}, in particular.

Results in~\cite{PRZ,PR0610} strongly suggest that
the conformal Virasoro algebra of solvable critical dense polymers admit Kac representations labelled by Kac labels $(r,s)$ in an infinitely extended Kac table 
\be
 \mbox{Kac table:}\qquad (r,s):\qquad  r,s=1,2,3,\ldots
\ee
as shown in Figure~\ref{Kac}. To be precise, by Kac representations, we here refer to the finitely-generated submodules of
Feigin-Fuchs modules~\cite{FF89} discussed in~\cite{Ras1012}. From the fusion rule $(r,s)=(r,1)\otimes(1,s)$, the finitized partition functions on a strip are
\be
 Z^{(N)}_{(1,1)|(r,s)}(q)=Z^{(N)}_{(r,s)|(1,1)}(q)=Z^{(N)}_{(r,1)|(1,s)}(q)=Z^{(N)}_{(1,s)|(r,1)}(q)=\chit^{(N)}_{r,s}(q)
\ee
where $\chit^{(N)}_{r,s}(q)$ are the finitized characters of the Kac representations. 
In the thermodynamic limit, the conformal partition functions are given by Virasoro characters
\be
 Z_{(1,1)|(r,s)}(q)=\lim_{N\to\infty} \chit^{(N)}_{r,s}(q)=\chit_{r,s}(q),\qquad r,s=1,2,3,\ldots
\ee

In this paper, we explicitly construct the conjugate boundary conditions on the strip labelled by $(r,s)$ 
and solve for the conformal spectra given by $\chit_{r,s}(q)$. 
This is achieved by studying the model on a 
finite-width strip subject to the $(r,s)=(r,1)\otimes (1,s)$ boundary conditions which involve a boundary field $\xi$.
Remarkably if, following \cite{BehrendP}, the boundary field is tuned to the special value $\xi=\lambda/2=\pi/4$, the normalized double-row transfer 
matrices satisfy a functional equation in the 
form of a single inversion identity which is {\it universal\/} in the sense that it is independent of $(r,s)$. 
This enables us to obtain the exact properties of the 
model on a finite lattice. The conformal properties are then readily 
accessible from finite-size corrections. In particular, in the 
continuum scaling limit, we confirm the central charge $c=-2$ and 
the conformal weights in the infinitely extended Kac table
\be
\Delta_{r,s}=\frac{(2r-s)^2-1}{8},\qquad r,s=1,2,3,\ldots\label{confwts}
\ee

The layout of this paper is as follows. In Section~\ref{SecLattice}, we define the solvable  
model of critical dense polymers, the planar Temperley-Lieb (TL) 
algebra~\cite{TL,Jones} on which it is built and the double-row transfer matrices. 
We also define vector spaces of link 
states on which the transfer matrices act and relate these to the $(r,s)$ 
boundary conditions. The boundary opperators are constructed in Appendix~\ref{BoundOps}. 
In Section~\ref{SecCFT}, we review the 
$c=-2$ CFT associated with critical dense polymers including a description of the 
Kac representations labelled by $(r,s)$ and the decomposition of their characters into characters of 
irreducible representations. 
In Section~\ref{SecExactSol}, we present the inversion identity. Its proof in the planar TL algebra is relegated
to Appendix~\ref{InvIdProof}.  We solve exactly the inversion identity on finite-width strips to obtain the 
general solution for the eigenvalues of the transfer matrices. The 
physically relevant solutions, in the $(r,s)$ sectors, are obtained empirically by selection rules encoded in 
irreducible blocks consisting of $q$-Catalan polynomials realized as suitable combinations of double-column diagrams.  Combinatorics of the related $q$-Narayana numbers are given in Appendix~\ref{qNarayana}. 
The finitized $(r,s)$ 
conformal characters are obtained explicitly 
from finite-width partition functions on the strip. The identities used to simplify the $q$-Catalan expression are given in 
Appendix~\ref{CatIdProofs}. 
It is also verified that the finitized Kac characters yield the quasi-rational characters of~\cite{PRZ} 
in the continuum scaling limit. 
In Section~\ref{SecFinite}, the finite-size corrections to the eigenvalues are obtained by applying 
the Euler-Maclaurin formula yielding the bulk and boundary free 
energies and confirming the central charge $c=-2$ and conformal
weights $\Delta_{r,s}=\frac{(2r-s)^2-1}{8}$. 
Lastly, we obtain the Hamiltonian limit of the 
double-row transfer matrices in the planar TL algebra and give its spectra.
We conclude with a brief discussion in Section~\ref{SecConcl}.

\section{Lattice Model}
\label{SecLattice}

\subsection{Critical dense polymers}

Solvable critical dense polymers on the square lattice describes the statistics of extended non-local degrees of freedom or connectivities 
representing the polymers. Since the polymers are very long, the degrees of freedom are not allowed to form closed local loops. 
In the language of the logarithmic minimal models~\cite{PRZ}, the loop fugacity vanishes $\beta=2\cos\lambda=0$ 
with the crossing parameter $\lambda=\pi/2$. 
In accord with being in the dense phase, the polymers or non-local degrees of freedom visit every face of the lattice 
exactly twice. More explicitly, consider a finite strip consisting of $N$ columns and $N'$ double rows of faces. An 
elementary face of the lattice can assume one of two configurations 
with different statistical weights
\be
\begin{pspicture}[shift=-.6](-.25,-.25)(1.25,1.25)
\facegrid{(0,0)}{(1,1)}
\put(0,0){\loopb}
\end{pspicture}
\quad\mbox{or}\quad
\begin{pspicture}[shift=-.6](-.25,-.25)(1.25,1.25)
\facegrid{(0,0)}{(1,1)}
\put(0,0){\loopa}
\end{pspicture}
\ee
where the arcs represent local segments of polymers. A typical configuration looks like
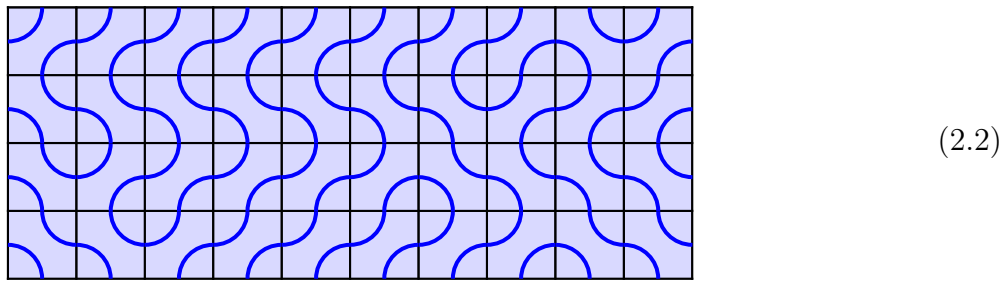
\psset{unit=.9cm}
\setlength{\unitlength}{.9cm}
\be
\begin{pspicture}[shift=-2.2](0,-.3)(10,4.4)
\facegrid{(0,0)}{(10,4)}
\rput[bl](0,0){\loopb}
\rput[bl](1,0){\loopb}
\rput[bl](2,0){\loopa}
\rput[bl](3,0){\loopa}
\rput[bl](4,0){\loopa}
\rput[bl](5,0){\loopa}
\rput[bl](6,0){\loopa}
\rput[bl](7,0){\loopa}
\rput[bl](8,0){\loopb}
\rput[bl](9,0){\loopb}
\rput[bl](0,1){\loopb}
\rput[bl](1,1){\loopa}
\rput[bl](2,1){\loopa}
\rput[bl](3,1){\loopa}
\rput[bl](4,1){\loopa}
\rput[bl](5,1){\loopa}
\rput[bl](6,1){\loopb}
\rput[bl](7,1){\loopb}
\rput[bl](8,1){\loopb}
\rput[bl](9,1){\loopb}
\rput[bl](0,2){\loopb}
\rput[bl](1,2){\loopb}
\rput[bl](2,2){\loopb}
\rput[bl](3,2){\loopb}
\rput[bl](4,2){\loopb}
\rput[bl](5,2){\loopb}
\rput[bl](6,2){\loopb}
\rput[bl](7,2){\loopa}
\rput[bl](8,2){\loopa}
\rput[bl](9,2){\loopa}
\rput[bl](0,3){\loopa}
\rput[bl](1,3){\loopa}
\rput[bl](2,3){\loopa}
\rput[bl](3,3){\loopa}
\rput[bl](4,3){\loopa}
\rput[bl](5,3){\loopa}
\rput[bl](6,3){\loopa}
\rput[bl](7,3){\loopa}
\rput[bl](8,3){\loopb}
\rput[bl](9,3){\loopa}
\end{pspicture}
\label{typconf}
\ee
Once the boundary conditions on the left and right edges have been specified, one is left with 
non-local degrees of freedom corresponding to a number of polymers on a strip. 

\subsection{Planar Temperley-Lieb algebra}

Algebraically, solvable critical dense polymers 
is described by the planar Temperley-Lieb (TL) algebra~\cite{TL,Jones}.
An elementary face of the square lattice is assigned a face weight 
according to the configuration of the face.  The two 
possible configurations with their associated weights are combined into a single face operator as
\psset{unit=.9cm}
\setlength{\unitlength}{.9cm}
\be
\begin{pspicture}[shift=-.45](-.5,-.1)(1.25,1.1)
\facegrid{(0,0)}{(1,1)}
\psarc[linewidth=.5pt](0,0){.15}{0}{90}
\rput(.5,.5){\small $u$}
\end{pspicture}
=\ \cos u\!\!
\begin{pspicture}[shift=-.45](-.5,-.1)(1.25,1.1)
\facegrid{(0,0)}{(1,1)}
\put(0,0){\loopa}
\end{pspicture}
\ +\ \sin u\!\!
\begin{pspicture}[shift=-.45](-.5,-.1)(1.25,1.1)
\facegrid{(0,0)}{(1,1)}
\put(0,0){\loopb}
\end{pspicture}\;
=\!\!\begin{pspicture}[shift=-.45](-.5,-.1)(1.25,1.1)
\facegrid{(0,0)}{(1,1)}
\psarc[linewidth=.5pt](1,0){.15}{90}{180}
\rput(.5,.5){\small $\lambda\!-\!u$}
\end{pspicture}
\label{u}
\ee
where the lower left corner has been marked to fix the orientation of 
the square. By the crossing symmetry, rotating the face by 90 degrees changes the spectral parameter from $u$ to $\lambda-u$. The polymer segments begin and end at {\em nodes} at the midpoints of the edges of the face.

The planar TL algebra is a 
diagrammatic algebra built up from elementary faces and triangles. The triangles enter through boundary conditions on the strip. 
The faces and triangles are connected such that a node of a 
face can be linked to a node of any other (or even the same)
face as long as the total set of links make up a {\em 
non-intersecting planar} web of connections.
Two relevant local properties of the planar TL algebra are the inversion relation
\psset{unit=1cm}
\setlength{\unitlength}{1cm}
\be
\begin{pspicture}[shift=-1.13](-.5,0.75)(4,3.25)
\pspolygon[fillstyle=solid,fillcolor=lightlightblue](0,2)(1,1)(2,2)(1,3)(0,2)
\pspolygon[fillstyle=solid,fillcolor=lightlightblue](2,2)(3,1)(4,2)(3,3)(2,2)
\psarc[linewidth=.5pt](0,2){.15}{-45}{45}
\psarc[linewidth=.5pt](2,2){.15}{-45}{45}
\psarc[linecolor=blue,linewidth=1.5pt](2,2){.7}{45}{135}
\psarc[linecolor=blue,linewidth=1.5pt](2,2){.7}{-135}{-45}
\rput(1,2){\small $v$}
\rput(3,2){\small $\!-v$}
\end{pspicture}
  \ \ \ =\ \ \cos^2\!v\ \
\begin{pspicture}[shift=-1.13](1,0.75)(4,3.25)
\pspolygon[fillstyle=solid,fillcolor=lightlightblue](1,2)(2,1)(3,2)(2,3)(1,2)
\psarc[linecolor=blue,linewidth=1.5pt](2,1){.7}{45}{135}
\psarc[linecolor=blue,linewidth=1.5pt](2,3){.7}{-135}{-45}
\end{pspicture}
\label{Inv}
\ee
and the Yang-Baxter equation (YBE)~\cite{BaxBook}
\psset{unit=1cm}
\setlength{\unitlength}{1cm}
\be
\begin{pspicture}[shift=-1.13](-.5,0.75)(4,3.25)
\facegrid{(2,1)}{(3,3)}
\pspolygon[fillstyle=solid,fillcolor=lightlightblue](0,2)(1,1)(2,2)(1,3)(0,2)
\psarc[linewidth=.5pt](0,2){.15}{-45}{45}
\psline[linecolor=blue,linewidth=1.5pt](1.5,1.5)(2,1.5)
\psline[linecolor=blue,linewidth=1.5pt](1.5,2.5)(2,2.5)
\psarc[linewidth=.5pt](2,1){.15}{0}{90}
\psarc[linewidth=.5pt](2,2){.15}{0}{90}
\rput(2.5,1.5){\small $u$}
\rput(2.5,2.5){\small $v$}
\rput(1,2){\small $u-v$}
\end{pspicture}
  \!\!\! =\
\begin{pspicture}[shift=-1.13](-.5,0.75)(4,3.25)
\facegrid{(0,1)}{(1,3)}
\pspolygon[fillstyle=solid,fillcolor=lightlightblue](1,2)(2,1)(3,2)(2,3)(1,2)
\psarc[linewidth=.5pt](1,2){.15}{-45}{45}
\psline[linecolor=blue,linewidth=1.5pt](1,1.5)(1.5,1.5)
\psline[linecolor=blue,linewidth=1.5pt](1,2.5)(1.5,2.5)
\psarc[linewidth=.5pt](0,1){.15}{0}{90}
\psarc[linewidth=.5pt](0,2){.15}{0}{90}
\rput(0.5,1.5){\small $v$}
\rput(0.5,2.5){\small $u$}
\rput(2,2){\small $u-v$}
\end{pspicture}
\label{YB}
\ee
These are identities for 2- and 3-tangles, respectively, where a 
$k$-tangle is an
arrangement of faces with $2k$ external nodes (at the midpoints of the external edges) where the internal polymer segments terminate. 

In the ordinary (linear) TL algebra, as defined in Section~4 of \cite{PRZ}, the generators act diagrammatically in a fixed direction, conventionally chosen to be vertical, on strings or strands via isotopy. The strands emerging from above (or below) represent an in-state, while the strands exiting below (or above) represent the corresponding out-state. In the planar setting, on the other hand, there is no fixed direction of transfer. 
In fact, any $k$ consecutive nodes of the $2k$ external nodes of a $k$-tangle can be chosen to represent the in-state. 
The remaining $k$ nodes are then also consecutive and represent the out-state. Once a direction has been fixed, expressions in the {\em planar} TL algebra reduce to equivalent expressions in the corresponding {\em linear\/} TL algebra. In the present {\em strip} context, the geometry dictates that vertical is the natural direction of transfer. It is nevertheless advantageous to work with the planar TL algebra as many proofs and diagrammatic manipulations are considerably simpler in the planar setting. In more complicated geometric set-ups, one cannot always translate expressions in the planar TL algebra into similar expressions in the linear TL algebra. This is the case when studying critical dense polymers on the cylinder~\cite{PRS10}, for example, where an extension of the periodic TL algebra is required as the `linear' replacement of the corresponding (locally) planar TL algebra.

\subsection{Double-row transfer matrix}

In the planar algebra, the double-row transfer matrix $\Db(u)$ on a strip with $(r,s)$ boundary conditions is defined as the $(N\!+\!\rho\!+\!s\!-\!2)$-tangle 
\psset{unit=1.3cm}
\setlength{\unitlength}{1.3cm}
\be
 \Db(u)=
\qquad
\begin{pspicture}[shift=-.9](0,0)(8.5,2.1)
\facegrid{(0,0)}{(8,2)}
\facegridpurple{(6,0)}{(8,2)}
\psline[linewidth=2pt,linecolor=red,linestyle=dashed](3,-.6)(3,2.3)
\psline[linewidth=2pt,linecolor=red,linestyle=dashed](6,-.6)(6,2.3)
\leftarc{(0,1)}
\rightarc{(8,1)}
\rput(0.5,.5){\fws $u$}
\rput(0.5,1.5){\fws $\lambda\!\!-\!\!u$}
\rput(1.55,.5){\fws $\cdots$}
\rput(1.55,1.5){\fws $\cdots$}
\rput(2.5,.5){\fws $u$}
\rput(2.5,1.5){\fws $\lambda\!\!-\!\!u$}
\rput(3.5,.5){\fws $u\!\!-\!\!\xi_{\rho\!-\!1}$}
\rput(3.5,1.5){\fws $-\!u\!\!-\!\!\xi_{\rho\!-\!2}$}
\rput(4.55,.5){\fws $\cdots$}
\rput(4.55,1.5){\fws $\cdots$}
\rput(5.5,.5){\fws $u\!\!-\!\!\xi_1$}
\rput(5.5,1.5){\fws $-\!u\!\!-\!\!\xi_0$}
\rput(6,0){\loopa}
\rput(6,1){\loopb}
\rput(7,0){\loopa}
\rput(7,1){\loopb}
\rput(1.5,-.3){$\underbrace{\rule{3.9cm}{0pt}}_{N}$}
\rput(4.5,-.3){$\underbrace{\rule{3.9cm}{0pt}}_{\rho-1}$}
\rput(7,-.3){$\underbrace{\rule{2.6cm}{0pt}}_{s-1}$}
\multirput(0,0)(2,0){2}{\multirput(0,0)(0,1){2}{\psarc[linewidth=.5pt](0,0){.1}{0}{90}}}
\multirput(3,0)(2,0){2}{\multirput(0,0)(0,1){2}{\psarc[linewidth=.5pt](0,0){.1}{0}{90}}}
\end{pspicture}
\label{D}
\\[.6cm]
\ee
with $N$ bulk columns and $\rho+s-2$ boundary columns where $\rho,s\ge 1$.
Two boundary seams of type $(r,1)$ and type $(1,s)$ on the right are delimited from the bulk on the left and from each other by dashed vertical lines. 
Following the fusion construction and methods of \cite{BehrendP}, it is shown in Appendix~\ref{BoundOps}, that the boundary operator \bea
{\begin{picture}(5.,3.2)
\put(1,.5){\color{lightlightblue}\rule{5\unitlength}{2\unitlength}}
\put(0.45,1.5){\makebox(0,0)[]{$=$}}
\put(.45,2.83){\spos{bc}{\color{blue}=}}
\put(-.1,2.8){\spos{bc}{\color{blue}(r,s)}}
\put(3.9,2.8){\spos{bc}{\color{blue}(r,1)}}
\put(6,2.8){\spos{bc}{\color{blue}\otimes}}
\put(-0.1,0.5){\line(0,1){2}}
\multiput(1,0.5)(1,0){4}{\line(0,1){2}}
\multiput(5,0.5)(1,0){2}{\line(0,1){2}}
\multiput(1,0.5)(0,1){3}{\line(1,0){5}}
\put(-0.6,1.5){\line(1,2){0.5}}
\put(-0.6,1.5){\line(1,-2){0.5}}
\pspolygon[linewidth=1pt,fillstyle=solid, fillcolor=lightlightblue](-0.6,1.5)(-.1,2.5)(-.1,.5)
\put(5,.5){\oval(.2,.2)[tr]}
\put(5,1.5){\oval(.2,.2)[tr]}
\multiput(1,.5)(1,0){2}{\oval(.2,.2)[tr]}
\multiput(1,1.5)(1,0){2}{\oval(.2,.2)[tr]}
\put(1.5,1){\spos{}{u\!-\!\xi_{\rho\!-\!1}}}
\put(2.5,1){\spos{}{u\!-\!\xi_{\rho\!-\!2}}}
\put(5.5,1){\spos{}{u\!-\!\xi_1}}
\put(1.5,2){\spos{}{-\!u\!\!-\!\!\xi_{\rho\!-\!2}}}
\put(2.5,2){\spos{}{-\!u\!\!-\!\!\xi_{\rho\!-\!3}}}
\multirput(0,0)(0,1){2}{\multirput(0,0)(1,0){2}{\rput(3.5,1){$\cdots$}}}
\put(5.5,2){\spos{}{-\!u\!-\!\xi_0}}
\put(-0.31,1.5){\spos{}{u}}
\multiput(0,0)(1,0){5}{\psarc[linewidth=2pt](.5,2.5){1}{0}{40}}
\multiput(0,0)(1,0){5}{\psline[linewidth=2pt](1.5,.3)(1.5,.5)}
\psline[linewidth=2pt](.8,1)(1,1)
\psline[linewidth=2pt](.8,2)(1,2)
\psline[linewidth=2pt](-.55,1)(-.35,1)
\psline[linewidth=2pt](-.55,2)(-.35,2)
\end{picture}}
{\begin{pspicture}(4.5,3.2)
\put(1,.5){\color{lightpurple}\rule{5\unitlength}{2\unitlength}}
\put(2.9,2.8){\spos{bc}{\color{blue}(1,s)}}
\put(6.5,2.8){\spos{bc}{\color{blue}(1,1)}}
\put(6,2.8){\spos{bc}{\color{blue}\otimes}}
\put(6.5,0.5){\line(0,1){2}}
\multiput(1,0.5)(1,0){4}{\line(0,1){2}}
\multiput(5,0.5)(1,0){2}{\line(0,1){2}}
\multiput(1,0.5)(0,1){3}{\line(1,0){5}}
\put(6,1.5){\line(1,2){0.5}}
\put(6,1.5){\line(1,-2){0.5}}
\pspolygon[linewidth=1pt,fillstyle=solid, fillcolor=lightlightblue](6,1.5)(6.5,2.5)(6.5,.5)
\multiput(0,0)(1,0){5}{\psarc[linewidth=2pt](.5,2.5){1}{0}{40}}
\multiput(0,0)(1,0){5}{\psline[linewidth=2pt](1.5,.3)(1.5,.5)}
\multiput(6,0.5)(0,2){2}{\makebox(0.5,0){\dotfill}}
\psarc[linewidth=2pt](5.7,1.5){.6}{-61}{61}
\rput[bl](1,.5){\loopa}
\rput[bl](2,.5){\loopa}
\rput[bl](3,.5){\loopa}
\rput[bl](4,.5){\loopa}
\rput[bl](5,.5){\loopa}
\rput[bl](1,1.5){\loopb}
\rput[bl](2,1.5){\loopb}
\rput[bl](3,1.5){\loopb}
\rput[bl](4,1.5){\loopb}
\rput[bl](5,1.5){\loopb}
\psline[linewidth=2pt,linecolor=red,linestyle=dashed](1,-.2)(1,2.7)
\end{pspicture}}
\qquad\qquad\qquad\nonumber\\[-24pt]
\mbox{}\hspace{-.2in}
\underbrace{\hspace{6.45cm}}_{\mbox{\small $\rho-1$}}\underbrace{\hspace{6.45cm}}_{\mbox{\small $s-1$}}\;\quad
\eea
constructed in the planar algebra, satisfies the Boundary Yang-Baxter Equation (BYBE)
\psset{unit=.25cm}
\setlength{\unitlength}{.25cm}
\thicklines
\def\lbybe#1#2#3#4#5#6{
\begin{pspicture}[shift=-8.5](0,-1.)(12,17)
\pspolygon[linewidth=.25pt,fillstyle=solid, fillcolor=lightlightblue](12,16)(0,4)(4,0)(12,8)(8,12)(4,8)(12,0)(12,16)
\rput(4,4){\small $#1$}
\rput(8,8){\small $#2$}
\rput(10.3,4){\small $#3$}
\rput(10.3,12){\small $#4$}
\psarc(0,4){.35}{-45}{45}
\psarc(4,8){.35}{-45}{45}
\psline[linestyle=dashed,dash=.5 .5,linewidth=.25pt](4,0)(12,0)
\rput(10.5,4){\small $#5$}
\rput(10.5,12){\small $#6$}
\end{pspicture}}
\def\rbybe#1#2#3#4#5#6{
\begin{pspicture}[shift=-8.5](0,-1)(12,17)
\pspolygon[linewidth=.25pt,fillstyle=solid, fillcolor=lightlightblue](12,0)(0,12)(4,16)(12,8)(8,4)(4,8)(12,16)(12,0)
\rput(4,12){\small $#1$}
\rput(8,8){\small $#2$}
\rput(10.3,4){\small $#4$}
\rput(10.3,12){\small $#3$}
\psarc(0,12){.35}{-45}{45}
\psarc(4,8){.35}{-45}{45}
\psline[linestyle=dashed,dash=.5 .5,linewidth=.25pt](4,16)(12,16)
\rput(10.5,4){\small $#5$}
\rput(10.5,12){\small $#6$}
\end{pspicture}}
\bea
\lbybe{u\!-\!v}{\lambda\!-\!u\!-\!v}{}{}{u}{v}\ \ =\ \ \rbybe{u\!-\!v}{\lambda\!-\!u\!-\!v}{}{}{v}{u}
\label{BYBE}
\eea
for all $\rho$, $s$ and $\xi$. This boundary operator will act on the link states introduced in the next section but the proof of the BYBE is independent of the choice of link states on which it acts. The bulk consists of $N$ columns with parity fixed by $N=\rho+s$ mod 2. The $r$-type seam consists of $\rho-1$ columns and the $s$-type seam consists of $s-1$ columns. 
All faces in the bulk and $r$-type seam have the standard orientation with the bottom-left corner marked and spectral parameters as indicated. The column inhomogeneities are 
\bea
\xi_j=\xi+j\lambda
\eea
In principle, $\xi$ is an arbitrary complex parameter. Since it lives on the boundary, we call it a {\it boundary field}. Adding a suitably scaled imaginary part to $\xi$ and varying it can, in the continuum scaling limit, induce~\cite{FPR2003} a boundary flow between different conformal boundary conditions. For the conformal boundary conditions of interest here, we eventually fix 
$\xi=\frac{\lambda}{2}=\frac{\pi}{4}$ in Section~\ref{SecExact}. 
The face configurations in the $s$-type seam are fixed as shown.  
An $s$-type seam is just an $r$-type seam in the limit $\xi\to i\infty$. 
For critical dense polymers, the crossing parameter is $\la=\tfrac{\pi}{2}$. In this case, 
the non-negative integer $\rho$ can be chosen with either even or odd parity and is related to $r$ by
\bea
r=\Big\lceil {\rho\over 2}\Big\rceil,\qquad\quad\rho=\begin{cases}
2r, &\mbox{$\rho$ even}\\
2r-1, &\mbox{$\rho$ odd}\\
\end{cases}
\label{rhor}
\eea
For the special value of the boundary field $\xi=\tfrac{\pi}{4}$, in particular, this is the appropriate choice of $\rho$ to realize the $r$-type conformal boundary conditions. 
More generally, for real $\xi$, the lattice boundary conditions can converge in the continuum scaling limit to conformal fixed points of different $r$-type boundary conditions depending on intervals of $\xi$ corresponding to respective basins of attraction.

The double-row transfer matrix $\Db(u)$
coincides with the double-row transfer matrices introduced in \cite{PRZ}
for general fugacity $\beta=2\cos\lambda$. 
In the planar algebra, multiplication of $\Db(u)$ with $\Db(v)$ means 
vertical concatenation of the two tangles. 
Using precisely the same diagrammatic arguments as in 
\cite{BPO}, it follows that $\Db(u)$ is crossing symmetric and gives rise to a commuting family of transfer matrices
\bea
   \Db\big(\lambda-u\big)\;=\;\Db(u),\qquad [\Db(u),\Db(v)]\;=\;0
\label{cross}
\eea

\subsection{Link states and matrix representations}

Since the double-row transfer matrix is constructed as a planar tangle, a matrix representation is needed to obtain spectra. 
A matrix representation of the double-row transfer tangle $\Db(u)$ is obtained by acting
with the planar tangle $\Db(u)$ from below on a suitable vector space of link states. 
We use the same notation for the tangle and the matrix representation. 
For the case of trivial boundary conditions ($\rho=s=1$) with $N=2n$ even, the $C_{n}={1\over n+1} \binom{2n}{n}$ TL link states and the diagrammatic action of the TL generators $e_j$ on them via isotopy were discussed in \cite{PRZ}. Each link state consists of $N$ nodes arranged on a line and connected in pairs from above by non-crossing (planar) connectivities. For $N=6$, the five link states are
\bea
\begin{pspicture}(6,4)
\psarc[linecolor=black,linewidth=1.5pt](-1,0){1}{0}{180}
\psarc[linecolor=black,linewidth=1.5pt](3,0){1}{0}{180}
\psarc[linecolor=black,linewidth=1.5pt](7,0){1}{0}{180}
\rput(9,-.2){,}
\end{pspicture}\qquad\qquad
\begin{pspicture}(6,4)
\psarc[linecolor=black,linewidth=1.5pt](-1,0){1}{0}{180}
\psarc[linecolor=black,linewidth=1.5pt](5,0){3}{0}{180}
\psarc[linecolor=black,linewidth=1.5pt](5,0){1}{0}{180}
\rput(9,-.2){,}
\end{pspicture}\qquad\qquad
\begin{pspicture}(6,4)
\psarc[linecolor=black,linewidth=1.5pt](1,0){3}{0}{180}
\psarc[linecolor=black,linewidth=1.5pt](1,0){1}{0}{180}
\psarc[linecolor=black,linewidth=1.5pt](7,0){1}{0}{180}
\rput(9,-.2){,}
\end{pspicture}\qquad\qquad
\begin{pspicture}(6,4)
\psarc[linecolor=black,linewidth=1.5pt](3,0){5}{0}{180}
\psarc[linecolor=black,linewidth=1.5pt](1,0){1}{0}{180}
\psarc[linecolor=black,linewidth=1.5pt](5,0){1}{0}{180}
\rput(9,-.2){,}
\end{pspicture}\qquad\qquad
\begin{pspicture}(6,4)
\psarc[linecolor=black,linewidth=1.5pt](3,0){5}{0}{180}
\psarc[linecolor=black,linewidth=1.5pt](3,0){3}{0}{180}
\psarc[linecolor=black,linewidth=1.5pt](3,0){1}{0}{180}
\end{pspicture}
\eea
The link states for the $(r,s)$ boundary conditions are a subset of the TL link states on $N+\rho+s-2$ nodes. 
To give the correct action of the boundary operator, these  
TL link states must be restricted on the $\rho+s-2$ boundary columns so that there are no half-arcs closing in the $\rho-1$ columns of the $r$-type seam and 
no half-arcs closing in the $s-1$ columns of the $s$-type seam. Half-arcs are however allowed to close {\em between} the $r$- and $s$-type seams.
Let ${\cal V}^{(N)}_{\rho,s}$ denote the vector space of these link states with $N+\rho+s-2$ nodes. Its dimension is given by
\bea
\dim{\cal V}^{(N)}_{\rho,s}\,=\,\smbin{N}{\rule{0pt}{12pt}{N-\rho+s\over 2}}-\smbin{N}{\rule{0pt}{12pt}{N-\rho-s\over 2}}
\,=\,\begin{cases}
\smbin{N}{{N-2r+s\over 2}}-\smbin{N}{{N-2r-s\over 2}},\quad &\rho=2r\\
\smbin{N}{{N-2r+s+1\over 2}}-\smbin{N}{{N-2r-s+1\over 2}},\quad &\rho=2r-1
\end{cases}
\label{countstates}
\eea
Using dashed vertical lines to separate the bulk and boundary seams, the six link states for ${\cal V}^{(N)}_{3,3}$ with $N=4$ are
\bea
\begin{array}{c}
\psset{unit=.5cm}
\begin{pspicture}(7,4)
\psarc[linecolor=black,linewidth=1.5pt](.5,0){.5}{0}{180}
\psarc[linecolor=black,linewidth=1.5pt](3.5,0){.5}{0}{180}
\psarc[linecolor=black,linewidth=1.5pt](5.5,0){.5}{0}{180}
\psarc[linecolor=black,linewidth=1.5pt](4.5,0){2.5}{0}{180}
\psline[linewidth=1.pt,linecolor=red,linestyle=dashed](3.5,-.1)(3.5,3.8)
\psline[linewidth=1.pt,linecolor=red,linestyle=dashed](5.5,-.1)(5.5,3.8)
\end{pspicture}\qquad\qquad
\begin{pspicture}(7,4)
\psarc[linecolor=black,linewidth=1.5pt](1.5,0){.5}{0}{180}
\psarc[linecolor=black,linewidth=1.5pt](3.5,0){.5}{0}{180}
\psarc[linecolor=black,linewidth=1.5pt](5.5,0){.5}{0}{180}
\psarc[linecolor=black,linewidth=1.5pt](3.5,0){3.5}{0}{180}
\psline[linewidth=1.pt,linecolor=red,linestyle=dashed](3.5,-.1)(3.5,3.8)
\psline[linewidth=1.pt,linecolor=red,linestyle=dashed](5.5,-.1)(5.5,3.8)
\end{pspicture}\qquad\qquad
\begin{pspicture}(7,4)
\psarc[linecolor=black,linewidth=1.5pt](2.5,0){.5}{0}{180}
\psarc[linecolor=black,linewidth=1.5pt](2.5,0){1.5}{0}{180}
\psarc[linecolor=black,linewidth=1.5pt](5.5,0){.5}{0}{180}
\psarc[linecolor=black,linewidth=1.5pt](3.5,0){3.5}{0}{180}
\psline[linewidth=1.pt,linecolor=red,linestyle=dashed](3.5,-.1)(3.5,3.8)
\psline[linewidth=1.pt,linecolor=red,linestyle=dashed](5.5,-.1)(5.5,3.8)
\end{pspicture}
\\[0pt]
\psset{unit=.5cm}
\begin{pspicture}(7,4)
\psarc[linecolor=black,linewidth=1.5pt](.5,0){.5}{0}{180}
\psarc[linecolor=black,linewidth=1.5pt](2.5,0){.5}{0}{180}
\psarc[linecolor=black,linewidth=1.5pt](5.5,0){.5}{0}{180}
\psarc[linecolor=black,linewidth=1.5pt](5.5,0){1.5}{0}{180}
\psline[linewidth=1.pt,linecolor=red,linestyle=dashed](3.5,-.1)(3.5,3.8)
\psline[linewidth=1.pt,linecolor=red,linestyle=dashed](5.5,-.1)(5.5,3.8)
\end{pspicture}\qquad\qquad
\begin{pspicture}(7,4)
\psarc[linecolor=black,linewidth=1.5pt](1.5,0){.5}{0}{180}
\psarc[linecolor=black,linewidth=1.5pt](1.5,0){1.5}{0}{180}
\psarc[linecolor=black,linewidth=1.5pt](5.5,0){.5}{0}{180}
\psarc[linecolor=black,linewidth=1.5pt](5.5,0){1.5}{0}{180}
\psline[linewidth=1.pt,linecolor=red,linestyle=dashed](3.5,-.1)(3.5,3.8)
\psline[linewidth=1.pt,linecolor=red,linestyle=dashed](5.5,-.1)(5.5,3.8)
\end{pspicture}\qquad\qquad
\begin{pspicture}(7,4)
\psarc[linecolor=black,linewidth=1.5pt](3.5,0){.5}{0}{180}
\psarc[linecolor=black,linewidth=1.5pt](3.5,0){1.5}{0}{180}
\psarc[linecolor=black,linewidth=1.5pt](3.5,0){2.5}{0}{180}
\psarc[linecolor=black,linewidth=1.5pt](3.5,0){3.5}{0}{180}
\psline[linewidth=1.pt,linecolor=red,linestyle=dashed](3.5,-.1)(3.5,3.8)
\psline[linewidth=1.pt,linecolor=red,linestyle=dashed](5.5,-.1)(5.5,3.8)
\end{pspicture}
\end{array}
\label{linkstates}
\eea

A link state on $2n$ nodes consists of $n$ half-arcs. We call a half-arc connecting {\em neighbouring} nodes a {\it small\/} half-arc. If there is a half-arc anywhere in a link state, then either the half-arc is itself a small half-arc or there must be a small half-arc somewhere inside the given half-arc. 
Within $r$- and $s$-type seams, small half-arcs have a push-through property. Specifically, if an $r$-type seam acts on a link state with a half-arc (and therefore a small half-arc) closing inside it, then there must also be a small half-arc closing in the out link state
\psset{unit=.8cm}
\bea
\label{pushthrough}
&&\begin{pspicture}[shift=-.67](-.2,-.3)(2.2,1.8)
\psarc[linewidth=1.5pt,linecolor=blue](1,1){.5}{0}{180}
\psline[linewidth=1.5pt,linecolor=blue](-.3,.5)(2.3,.5)
\psline[linewidth=1.5pt,linecolor=blue](.5,-.3)(.5,.3)
\psline[linewidth=1.5pt,linecolor=blue](1.5,-.3)(1.5,.3)
\facegrid{(0,0)}{(2,1)}
\put(0,0){\oval(.2,.2)[tr]}
\put(1,0){\oval(.2,.2)[tr]}
\rput[B](.5,.4){\small$v$}
\rput[B](1.5,.4){\small$v\!\!+\!\!\lambda$}
\end{pspicture}\ \;=\;
s(\lambda-v)s(\lambda+v)\ \ 
\begin{pspicture}[shift=-.67](-.2,-.3)(2.2,1.8)
\psarc[linewidth=1.5pt,linecolor=blue](1,1){.5}{0}{180}
\psline[linewidth=1.5pt,linecolor=blue](-.3,.5)(2.3,.5)
\psline[linewidth=1.5pt,linecolor=blue](.5,-.3)(.5,.3)
\psline[linewidth=1.5pt,linecolor=blue](1.5,-.3)(1.5,.3)
\facegrid{(0,0)}{(2,1)}
\psarc[linewidth=1.5pt,linecolor=blue](1,0){.5}{0}{180}
\psarc[linewidth=1.5pt,linecolor=blue](0,1){.5}{-90}{0}
\psarc[linewidth=1.5pt,linecolor=blue](2,1){.5}{180}{270}
\end{pspicture}\ +
s(\lambda-v)s(-v)\ \ 
\begin{pspicture}[shift=-.67](-.2,-.3)(2.2,1.8)
\psarc[linewidth=1.5pt,linecolor=blue](1,1){.5}{0}{180}
\psline[linewidth=1.5pt,linecolor=blue](-.3,.5)(2.3,.5)
\psline[linewidth=1.5pt,linecolor=blue](.5,-.3)(.5,.3)
\psline[linewidth=1.5pt,linecolor=blue](1.5,-.3)(1.5,.3)
\facegrid{(0,0)}{(2,1)}
\psarc[linewidth=1.5pt,linecolor=blue](1,0){.5}{90}{180}
\psarc[linewidth=1.5pt,linecolor=blue](2,0){.5}{90}{180}
\psarc[linewidth=1.5pt,linecolor=blue](0,1){.5}{270}{0}
\psarc[linewidth=1.5pt,linecolor=blue](1,1){.5}{270}{0}
\end{pspicture}\qquad\\
&\!\!+\!\!&
\beta s(v)s(-v)\ \ 
\begin{pspicture}[shift=-.67](-.2,-.3)(2.2,1.8)
\psarc[linewidth=1.5pt,linecolor=blue](1,1){.5}{0}{180}
\psline[linewidth=1.5pt,linecolor=blue](-.3,.5)(2.3,.5)
\psline[linewidth=1.5pt,linecolor=blue](.5,-.3)(.5,.3)
\psline[linewidth=1.5pt,linecolor=blue](1.5,-.3)(1.5,.3)
\facegrid{(0,0)}{(2,1)}
\psarc[linewidth=1.5pt,linecolor=blue](0,0){.5}{0}{90}
\psarc[linewidth=1.5pt,linecolor=blue](2,0){.5}{90}{180}
\psarc[linewidth=1.5pt,linecolor=blue](1,1){.5}{180}{360}
\end{pspicture}\ +
s(v)s(v+\lambda)\ \ 
\begin{pspicture}[shift=-.67](-.2,-.3)(2.2,1.8)
\psarc[linewidth=1.5pt,linecolor=blue](1,1){.5}{0}{180}
\psline[linewidth=1.5pt,linecolor=blue](-.3,.5)(2.3,.5)
\psline[linewidth=1.5pt,linecolor=blue](.5,-.3)(.5,.3)
\psline[linewidth=1.5pt,linecolor=blue](1.5,-.3)(1.5,.3)
\facegrid{(0,0)}{(2,1)}
\psarc[linewidth=1.5pt,linecolor=blue](0,0){.5}{0}{90}
\psarc[linewidth=1.5pt,linecolor=blue](1,1){.5}{180}{270}
\psarc[linewidth=1.5pt,linecolor=blue](1,0){.5}{0}{90}
\psarc[linewidth=1.5pt,linecolor=blue](2,1){.5}{180}{270}
\end{pspicture}\ =s(\lambda-v)s(\lambda+v)\ \ 
\begin{pspicture}[shift=-.67](-.2,-.3)(2.2,1.8)
\psarc[linewidth=1.5pt,linecolor=blue](1,1){.5}{0}{180}
\psline[linewidth=1.5pt,linecolor=blue](-.3,.5)(2.3,.5)
\psline[linewidth=1.5pt,linecolor=blue](.5,-.3)(.5,.3)
\psline[linewidth=1.5pt,linecolor=blue](1.5,-.3)(1.5,.3)
\facegrid{(0,0)}{(2,1)}
\psarc[linewidth=1.5pt,linecolor=blue](1,0){.5}{0}{180}
\psarc[linewidth=1.5pt,linecolor=blue](0,1){.5}{-90}{0}
\psarc[linewidth=1.5pt,linecolor=blue](2,1){.5}{180}{270}
\end{pspicture}\nonumber\smallskip
\eea
This holds for arbitrary $v, \lambda$ with $s(v)=\sin v$ and $\beta=2\cos\lambda$. The same is trivially true for $s$-type seams. We want the double-row transfer matrices to act from the space ${\cal V}^{(N)}_{\rho,s}$ back to itself. 
It follows from the push-through property that, if we restrict the out states at the bottom to ${\cal V}^{(N)}_{\rho,s}$, then the in-states at the top 
must also be restricted to ${\cal V}^{(N)}_{\rho,s}$ and likewise for all intermediate link states. This restriction preventing half-arcs closing within a seam is equivalent to the action on the seam of the Wenzl-Jones projectors~\cite{WenzlJones,Wenzl} when they exist. But the half-arc restriction makes sense even when the Wenzl-Jones projectors do not exist which is the case for critical dense polymers.

\section{Conformal Field Theory}
\label{SecCFT}

\begin{figure}[h]
{\vspace{0in}\psset{unit=1cm}
{
\small
\begin{center}
\qquad
\begin{pspicture}(0,0)(7,11)
\psframe[linewidth=0pt,fillstyle=solid,fillcolor=lightlightblue](0,0)(7,11)
\multiput(0,0)(0,2){5}{\psframe[linewidth=0pt,fillstyle=solid,fillcolor=midblue](0,1)(7,2)}
\multirput(1,1)(1,0){6}{\pswedge[fillstyle=solid,fillcolor=red,linecolor=red](0,0){.25}{180}{270}}
\multirput(1,2)(1,0){6}{\pswedge[fillstyle=solid,fillcolor=red,linecolor=red](0,0){.25}{180}{270}}
\multirput(1,2)(0,2){5}{\pswedge[fillstyle=solid,fillcolor=red,linecolor=red](0,0){.25}{180}{270}}
\psgrid[gridlabels=0pt,subgriddiv=1]
\rput(.5,10.65){$\vdots$}\rput(1.5,10.65){$\vdots$}\rput(2.5,10.65){$\vdots$}
\rput(3.5,10.65){$\vdots$}\rput(4.5,10.65){$\vdots$}\rput(5.5,10.65){$\vdots$}
\rput(6.5,10.5){$\vvdots$}\rput(.5,9.5){$\frac{63}8$}\rput(1.5,9.5){$\frac{35}8$}
\rput(2.5,9.5){$\frac{15}8$}\rput(3.5,9.5){$\frac{3}8$}\rput(4.5,9.5){$-\frac 18$}
\rput(5.5,9.5){$\frac{3}8$}\rput(6.5,9.5){$\cdots$}
\rput(.5,8.5){$6$}\rput(1.5,8.5){$3$}\rput(2.5,8.5){$1$}\rput(3.5,8.5){$0$}
\rput(4.5,8.5){$0$}\rput(5.5,8.5){$1$}\rput(6.5,8.5){$\cdots$}
\rput(.5,7.5){$\frac{35}8$}\rput(1.5,7.5){$\frac {15}8$}\rput(2.5,7.5){$\frac 38$}
\rput(3.5,7.5){$-\frac{1}8$}\rput(4.5,7.5){$\frac 38$}\rput(5.5,7.5){$\frac{15}8$}
\rput(6.5,7.5){$\cdots$}\rput(.5,6.5){$3$}\rput(1.5,6.5){$1$}\rput(2.5,6.5){$0$}\rput(3.5,6.5){$0$}
\rput(4.5,6.5){$1$}\rput(5.5,6.5){$3$}\rput(6.5,6.5){$\cdots$}
\rput(.5,5.5){$\frac{15}8$}\rput(1.5,5.5){$\frac {3}{8}$}\rput(2.5,5.5){$-\frac 18$}
\rput(3.5,5.5){$\frac{3}{8}$}\rput(4.5,5.5){$\frac {15}8$}\rput(5.5,5.5){$\frac{35}{8}$}
\rput(6.5,5.5){$\cdots$}\rput(.5,4.5){$1$}\rput(1.5,4.5){$0$}\rput(2.5,4.5){$0$}
\rput(3.5,4.5){$1$}\rput(4.5,4.5){$3$}\rput(5.5,4.5){$6$}\rput(6.5,4.5){$\cdots$}
\rput(.5,3.5){$\frac 38$}\rput(1.5,3.5){$-\frac 18$}\rput(2.5,3.5){$\frac 38$}
\rput(3.5,3.5){$\frac{15}8$}\rput(4.5,3.5){$\frac{35}8$}\rput(5.5,3.5){$\frac{63}8$}
\rput(6.5,3.5){$\cdots$}\rput(.5,2.5){$0$}\rput(1.5,2.5){$0$}\rput(2.5,2.5){$1$}\rput(3.5,2.5){$3$}
\rput(4.5,2.5){$6$}\rput(5.5,2.5){$10$}\rput(6.5,2.5){$\cdots$}
\rput(.5,1.5){$-\frac 18$}\rput(1.5,1.5){$\frac 38$}\rput(2.5,1.5){$\frac{15}8$}
\rput(3.5,1.5){$\frac{35}8$}\rput(4.5,1.5){$\frac{63}8$}\rput(5.5,1.5){$\frac{99}8$}
\rput(6.5,1.5){$\cdots$}\rput(.5,.5){$0$}\rput(1.5,.5){$1$}\rput(2.5,.5){$3$}\rput(3.5,.5){$6$}
\rput(4.5,.5){$10$}\rput(5.5,.5){$15$}\rput(6.5,.5){$\cdots$}
{\color{blue}
\rput(.5,-.5){$1$}
\rput(1.5,-.5){$2$}
\rput(2.5,-.5){$3$}
\rput(3.5,-.5){$4$}
\rput(4.5,-.5){$5$}
\rput(5.5,-.5){$6$}
\rput(6.5,-.5){$r$}
\rput(-.5,.5){$1$}
\rput(-.5,1.5){$2$}
\rput(-.5,2.5){$3$}
\rput(-.5,3.5){$4$}
\rput(-.5,4.5){$5$}
\rput(-.5,5.5){$6$}
\rput(-.5,6.5){$7$}
\rput(-.5,7.5){$8$}
\rput(-.5,8.5){$9$}
\rput(-.5,9.5){$10$}
\rput(-.5,10.5){$s$}}
\end{pspicture}
\end{center}}}
\caption{\label{Kac}Kac table of conformal weights of critical dense polymers. Each position $(r,s)$ 
of the table is associated with a Kac representation, a conformal dimension $\Delta_{r,s}$ and an 
associated Kac character $\protect\chit_{r,s}(q)$. 
The $(r,s)$ representations indicated with a red quadrant are irreducible representations
with characters $\protect\chit_{r,s}(q)=\mch_{r,s}(q)$. The irreducible representations at positions 
$(1,2k)\equiv (k,2)$, $k=2,3,\ldots$ are identified.}
\end{figure}

\subsection{Spectrum and characters of critical dense polymers}

In the continuum scaling limit, the finite-size spectra of the transfer matrices of critical dense polymers 
are described~\cite{PRZ,PR0610} by a logarithmic 
CFT with central charge $c=-2$ and conformal weights $\Delta_{r,s}$ given by (\ref{confwts}). 
This CFT is non-unitary and non-rational. 

In the Virasoro picture, the CFT data is neatly encoded in the infinitely extended Kac table shown in Figure~\ref{Kac}.
The $(r,s)$ conformal characters
\be
   \chit_{r,s}(q)\,=\,\frac{q^{-\frac{c}{24}}}{(q)_\infty}(q^{\D_{r,s}}-q^{\D_{r,-s}})
   \,=\,\frac{q^{-\frac{c}{24}+\Delta_{r,s}}}{(q)_\infty}(1-q^{rs}),\qquad  
  (q)_\infty=\,\prod_{k=1}^\infty(1-q^k)
\label{chirs}
\ee
are the generating functions for the conformal finite-size spectra of 
the double-row transfer matrices on the strip with $(r,s)$ boundary conditions. 
Associated to each such boundary condition is a Virasoro representation whose character is given in (\ref{chirs}).
However, the corresponding module structure is in general not determined by the character alone.
Following the terminology introduced in~\cite{PRZ}, we refer to these representations as Kac representations and denote them by $(r,s)$.
Their mathematical characterization~\cite{Ras1012} is recalled in Section~\ref{SecKacReps}. 
We reserve the notation $\mch_{r,s}(q)=\chit_{r,s}(q)$ with $s=1,2$ to denote the characters of the {\it irreducible} representations in the first two rows of the Kac table
\be
 \mch_{r,s}(q)=\frac{q^{-\frac{c}{24}+\Delta_{r,s}}}{(q)_\infty}(1-q^{rs}),\qquad s=1,2
\label{chrs}
\ee

The set of Kac representations is not exhaustive. Although we do not do it here, critical dense polymers can also be considered in the ${\cal W}$-extended picture with a ${\cal W}$-extended conformal symmetry~\cite{GabK,FeiginEtAl,GabRunk}. In this picture, it suffices to consider a finite number of representations and extended boundary conditions which respect the enlarged symmetry algebra. In fact, critical dense polymers with ${\cal W}$-extended boundary conditions~\cite{PRR2008} yields a lattice model whose continuum scaling limit is compatible with symplectic fermions~\cite{Kau00} at the level of conformal spectra and fusion rules. The structure of the corresponding ${\cal W}$-extended Kac representations is discussed in~\cite{Ras1106}.

\subsection{Description of Kac representations}
\label{SecKacReps}

The Virasoro character of the Kac representation $(r,s)$ is given by (\ref{chirs}).
This is recognized as the character of the quotient module
\be
 Q_{r,s}=V_{r,s}/V_{r,-s},\qquad r,s\in\mathbb{N}
\ee
obtained from the highest-weight Verma module $V_{r,s}$ of conformal weight $\D_{r,s}$
by quotienting out the submodule $V_{r,-s}$ at level $rs$. However, this does not imply that the
Kac representation $(r,s)$ can be identified with the quotient module $Q_{r,s}$.
Indeed, a characterization of the Kac representations as modules over the Virasoro algebra
was recently proposed in~\cite{Ras1012}, where it was argued that they can be understood as
finitely generated submodules of Feigin-Fuchs modules~\cite{FF89}. A necessary condition
for this conjecture to be true is that
the characters of these finitely generated submodules match the characters (\ref{chirs}) of the 
corresponding quotient modules. As seen below, this is indeed the case.

For $s$ even, the Kac representation $(r,s)=(r,2k)$ 
is fully reducible~\cite{RP0707}
\be
 (r,2k)=\bigoplus_{j=|r-k|+1,\,\mathrm{by}\,2}^{r+k-1}M_{j,2}
\label{rsM}
\ee
where $M_{m,n}$ is the irreducible highest-weight module of conformal weight $\D_{m,n}$,
and the corresponding Virasoro character is given by
\be
 \chit_{r,2k}(q)
  =\sum_{j=|r-k|+1,\,\mathrm{by}\,2}^{r+k-1}\mch_{j,2}(q)
\label{rs2k}
\ee
For $s$ odd, on the other hand, the Kac representation $(r,s)=(r,2k+1)$
is believed~\cite{Ras1012} to be a reducible but indecomposable rank 1 representation characterized by
\be
\begin{array}{rcll}
 &(r,s):&\ \
  M_{k-r+1,1}\to M_{k-r+2,1}\gets M_{k-r+3,1}\to\ldots\gets M_{k+r-1,1}\to M_{k+r,1},\ \ &2r<s
 \\[.5cm]
 &(r,s):&\ \
  M_{r-k,1}\gets M_{r-k+1,1}\to M_{r-k+2,1}\gets\ldots\gets M_{r+k-1,1}\to M_{r+k,1},\ \ &2r>s
\end{array}
\label{2k1}
\ee
The corresponding Virasoro character is given by
\be
 \chit_{r,2k+1}(q)
  =\sum_{j=|r-k-\frac{1}{2}|+\frac{1}{2}}^{r+k}\mch_{j,1}(q)
\label{rs2k1}
\ee
Using the explicit expressions (\ref{chrs}) for the irreducible characters $\mch_{r,s}(q)$, 
and observing the telescoping property of their sums in (\ref{rs2k}) and (\ref{rs2k1}), 
it is readily verified that these 
sums are equivalent to the general expression (\ref{chirs}) for the Kac characters $\chit_{r,s}(q)$.
For our purposes here, we write the decomposition of Kac characters into irreducible characters as
\bea
\chi_{r,s}(q)=\begin{cases}
\disp\sum_{k={1\over 2}(|2r-s|+1)}^{{1\over 2}(2r+s-1)}\!\!\!\!\!\!\! \mch_{k,1}(q)=
\!\!\!\sum_{k=1}^{\mbox{\scriptsize Min}[2r,s]} \!\!\!\mch_{{1\over 2}(|2r-s|-1+2k),1}(q)
=\sum_{k=1}^{s} \mch_{{1\over 2}(2r-s-1+2k),1}(q),&\mbox{$s$ odd}\\[18pt]
\disp\sum_{k={1\over 2}(|2r-s|+2)}^{{1\over 2}(2r+s-2)}\!\!\!\!\!\!\! \mch_{k,2}(q)=
\!\!\!\sum_{k=1}^{\mbox{\scriptsize Min}[r,s/2]}  \!\!\!\mch_{{1\over 2}(|2r-s|-2+4k),2}(q)
=\sum_{k=1}^{s/2} \mch_{{1\over 2}(2r-s-2+4k),2}(q), &\mbox{$s$ even}
\end{cases}
\label{irredDecomp}
\eea
The last form of these identities uses the extended definitions
\bea
\mch_{0,s}(q)=0,\qquad \mch_{r,s}(q)=-\mch_{-r,s}(q),\qquad r<0,\ \  s=1,2
\label{negr}
\eea

It is noted that the set of {\em irreducible} Kac representations is given by
\be
 (r,1),\ (r,2),\ (1,2k),\qquad r,k\in\mathbb{N}
\ee
However, since an irreducible representation is uniquely characterized by its associated
conformal weight, we identify the Kac representations
\be
 (1,2k)\equiv(k,2),\qquad k\in\mathbb{N}
\ee
The set of inequivalent irreducible Kac representations is thus associated with the first
two rows of the extended Kac table, that is, $s=1,2$ in $(r,s)$.

\section{Exact Solution on a Finite Strip}
\label{SecExactSol}

\subsection{Inversion identity}

Remarkably, the double-row transfer matrix 
satisfies an inversion identity in the planar TL algebra. This inversion identity is unique 
to critical dense polymers among the logarithmic minimal models. As we will discuss elsewhere, all other logarithmic minimal models satisfy cubic or higher-degree polynomial functional equations.
\bigskip

\noindent {\bf Inversion Identity:}\ \ \
{\em For $\lambda=\tfrac{\pi}{2}$, the planar tangle $\Db(u)$ \mbox{\rm (\ref{D})} satisfies the inversion
identity
\be
 \Db(u)\Db(u+\lambda)=-\tan^2 2u\,\Big(\!\cos^{2N}\!u\,\eta^{(\rho)}(u,\xi)-\sin^{2N}\!u\,\eta^{(\rho)}(u+\la,\xi)\!\Big)^{\!2}\Ib
\label{DDI}
\ee
where the right side is an entire function of $u$, $\Ib$ is the vertical identity tangle, and
\be
 \eta^{(\rho)}(u,\xi)=\prod_{j=1}^{\rho-1}\sin(u+\xi_{j-1})\sin(u-\xi_{j+1})
  =\prod_{j=1}^{\rho-1}\cos(u+\xi_j)\cos(u-\xi_j)
\label{etarho}
\ee
}

This inversion identity is proved in the planar algebra in Appendix~\ref{InvIdProof}. Notice that the form of the functional equation is 
{\em independent} of $s$. However, the link states, the boundary operator, the transfer matrix and its eigenvalues {\em do} 
depend on $s$. Since
\be
 \frac{\eta^{(\rho)}(u,\xi)}{\eta^{(\rho)}(u+\la,\xi)}\;=\;\begin{cases}
  1, &\rho=2r-1 \\[.1cm]
  \tan(u+\xi)\tan(u-\xi),\ &\rho=2r
 \end{cases}
\ee
the inversion identity is easily simplified for each parity of $\rho$. For $\rho$ odd, in particular, we thus find
\be
 \Db(u)\Db(u+\lambda)=-\sin^2 2u\,\eta^{(\rho)}(u,\xi)\,\eta^{(\rho)}(u+\la,\xi)
\left(\frac{\cos^{2N}\!u-\sin^{2N}\!u}{\cos^2\!u-\sin^2\!u}\right)^{\!\!2}\Ib,\qquad
  \rho=2r-1
\ee
For the special value
\be
 \xi=\frac{\la}{2}=\frac{\pi}{4}
\label{xi}
\ee
of the boundary field, we see that
\be
 \eta^{(\rho)}(u,\mbox{$\pi \over 4$})=\big(\half\cos 2u\big)^{\rho-1}
\ee
such that the inversion identity (\ref{DDI}) reduces to
\be
 \Db(u)\Db(u+\lambda)
  = -\sin^2 2u\,\big(\half\cos 2u\big)^{2\rho-2} \bigg(\frac{\cos^{2N}\!u-(-1)^{\rho-1}\sin^{2N}\!u}{\cos^2u-\sin^2u}\bigg)^{\!\!2} \Ib,\qquad
   \rho\in\mathbb{N}
\ee

\subsection{Exact solution for the eigenvalues}
\label{SecExact}

In the sequel, we will only consider the special case $\xi=\frac{\la}{2}=\frac{\pi}{4}$ (\ref{xi}). 
Let us introduce normalized transfer matrices by
\bea
\vec d(u)=\begin{cases}
\disp {2^{\rho-1}\vec D(u)\over \sin 2u \cos^{\rho-2} 2u},\quad&\rho=2r\\[18pt]
\disp {2^{\rho-1}\vec D(u)\over \sin 2u \cos^{\rho-1} 2u},&\rho=2r-1
\end{cases}
\label{bdynorm}
\eea 
The eigenvalues $d(u)$ of the normalized double-row transfer matrices $\vec d(u)$ then satisfy
\be
  d(u)d(u+\lambda)\,=\,\begin{cases}
 \big(\cos^{2N}\!u+\sin^{2N}\!u\big)^2,&\rho=2r\\[6pt]
\disp\left(\frac{\cos^{2N}\!u-\sin^{2N}\!u}{\cos^2\!u-\sin^2\!u}
   \right)^{\!2},\quad&\rho=2r-1
   \end{cases}
\label{LaLa}
\ee
subject to the initial condition and crossing symmetry
\bea
  d(0)=1,\qquad d(\lambda-u)=d(u)
\label{La0}
\eea
We observe that the form of this inversion identity for $\rho$ odd exactly agrees with the inversion identity solved in \cite{PR0610} for $r=1$. 
The $\rho$-dependent order-1 normalization terms in (\ref{bdynorm}) exactly cancel the additional boundary free energy arising from the $r$-type seam.

The finite-size analysis of (\ref{LaLa}) depends on the following factorizations
\bea
\cos^{2N}\!u+\sin^{2N}\!u&\!\!=\!\!&\begin{cases}
  \disp\prod_{j=1}^{\frac{N-1}{2}}
   \big(1-\sin^2\!\mbox{$\frac{j\pi}{N}$}\,\sin^2 2u\big),&\mbox{$\rho=2r$, $s$ odd}\\[14pt]
  \disp\prod_{j=1}^{\frac{N}{2}}
   \big(1-\sin^2\!\mbox{$\frac{(2j-1)\pi}{2N}$}\,\sin^2 2u\big),&\mbox{$\rho=2r$, $s$ even}
   \end{cases}\\[4pt]
  \frac{\cos^{2N}\!u-\sin^{2N}\!u}{\cos^2\!u-\sin^2\!u}
  &\!\!=\!\!&\begin{cases}
  \disp\prod_{j=1}^{\frac{N}{2}-1}
   \big(1-\sin^2\!\mbox{$\frac{j\pi}{N}$}\,\sin^2 2u\big),&\mbox{$\rho=2r-1$, $s$ odd}\\[14pt]
  \disp\prod_{j=1}^{\frac{N-1}{2}}
   \big(1-\sin^2\!\mbox{$\frac{(2j-1)\pi}{2N}$}\,\sin^2 2u\big),&\mbox{$\rho=2r-1$, $s$ even}
   \end{cases}
\label{cs}
\eea
Notice that $d(u)$, $d(u+\lambda)$ and the right-sides of (\ref{LaLa}) are entire functions. Observe also that the zeros of $d(u)$ and 
$d(u+\lambda)$ are the same only shifted by $\lambda={\pi\over 2}$. Since entire functions are defined up to an overall constant by their zeros, it is now straightforward~\cite{BaxBook,OPW} to solve the functional relation (\ref{LaLa}), subject to (\ref{La0}), by sharing out the known quadratic zeros of the right-sides. Recalling that $N+\rho+s$ is even, we find
\bea
  d(u)=
  \begin{cases}
  \disp\prod_{j=1}^{\floor{\frac{N-1}{2}}}\!
   \big(1+\eps_j\sin\!\mbox{$\frac{j\pi}{N}$}\,\sin 2u\big)
   \big(1+\mu_j\sin\!\mbox{$\frac{j\pi}{N}$}\,\sin 2u\big),&\mbox{$s$ odd}\\[14pt]
  \disp\prod_{j=1}^{\floor{\frac{N}{2}}}\!\!\big(1+\eps_j\sin\!\mbox{$(2j-1)\pi\over 2N$}\,\sin 2u\big)\big(1+\mu_j\sin\!\mbox{$(2j-1)\pi\over 2N$}\,\sin 2u\big),\quad
  &\mbox{$s$ even}
   \end{cases}
\label{N}
\eea
where $\eps_j^2=\mu_j^2=1$ for all $j=1,2,\ldots,M$. For later convenience, we have introduced
\bea
M=\begin{cases}
\floor{\frac{N-1}{2}},&\mbox{$s$ odd}\\[8pt]
\floor{\frac{N}{2}},\quad&\mbox{$s$ even}
\end{cases}
\label{MN}
\eea
The appearance of {\em two} sets of
parameters $\{\eps_j\}$ and $\{\mu_j\}$
stems from the overall squaring in (\ref{LaLa}) and allows for {\it double} zeros. The eigenvalue for $d(u+\lambda)$ is given by replacing $\epsilon_j\mapsto -\epsilon_j$, $\mu_j\mapsto -\mu_j$.

For either parity of $s$ in (\ref{N}),
the maximum eigenvalue is obtained for $\eps_j=\mu_j=1$ for all $j$
and corresponds to the groundstate in the $(r,s)=(1,1)$ or $(1,2)$ sector.
Excited states are generated
by switching a (finite) number of the parameters $\eps_j,\mu_j$
from $+1$ to $-1$.
The number of possible excitations following from
(\ref{N}) is $2^{2\floor{{(N-1)}/{2}}}$ and $2^{\floor{{N}/{2}}}$, respectively,
compared with the number of link states in a given $\rho, s$ sector given by (\ref{countstates}). 
The inversion identity fixes the form of the eigenvalues but not their multiplicities. 
We thus need a set of selection rules to determine the groundstate and excitations, along with their multiplicities, 
that actually appear in the spectrum in a given sector. This 
information is neatly encoded by specifying the patterns of complex 
zeros, characterized by $\{\epsilon_j,\mu_j\}$, for the allowed eigenvalues $d(u)$.

From the crossing symmetry and periodicity, the zeros $u_j$ of an eigenvalue $d(u)$ come in complex conjugate pairs 
in the complex $u$-plane
and appear with a periodicity $\pi$ in the real part of $u$. From (\ref{N}) 
\bea
  u_j=
(2+\nu_j)\frac{\pi}{4}\pm\frac{i}{2}\ln\tan\frac{t_j}{2}\ 
+\ \pi\,\mathbb{Z}
\label{uzeros}
\eea
where $\nu_j$ is $\eps_j$ or $\mu_j$, and 
\bea
  t_j=\begin{cases}
   \frac{j\pi}{N},\qquad &\mbox{$s$ odd}\\[8pt]
    \frac{(2j-1)\pi}{2N},&\mbox{$s$ even}
    \end{cases}
\label{tj}
\eea
A typical pattern for $N=12$ is
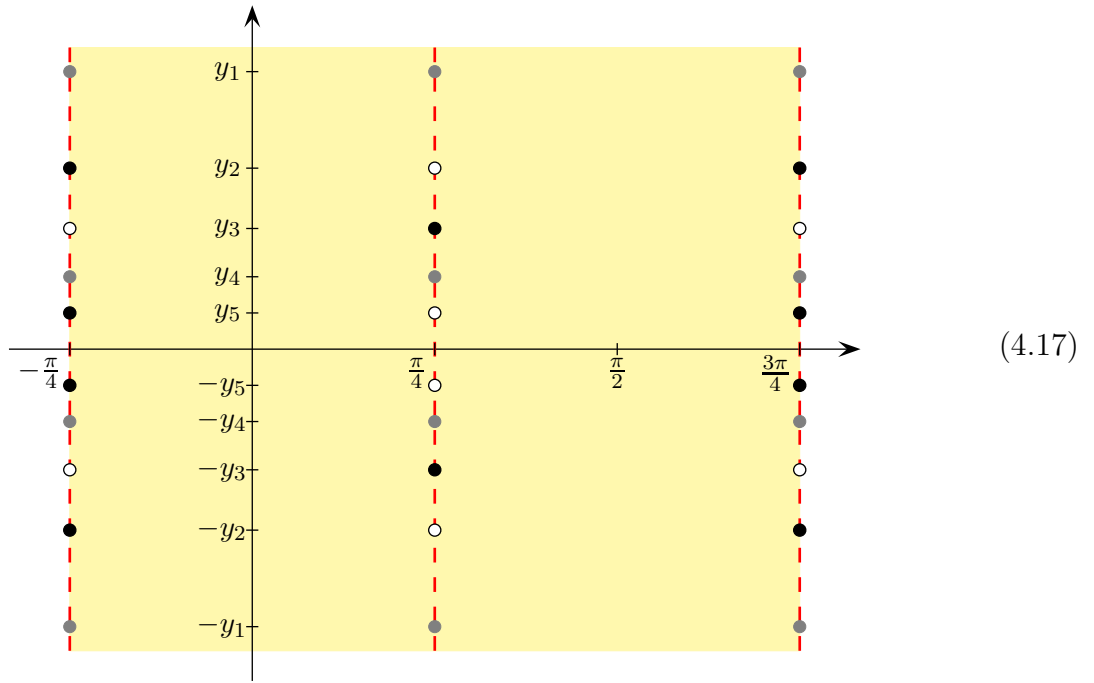
\psset{unit=.8cm}
\setlength{\unitlength}{.8cm}
\be
\begin{pspicture}[shift=-5.2](-.25,.7)(14,11.5)
\psframe[linecolor=yellow!40!white,linewidth=0pt,fillstyle=solid,fillcolor=yellow!40!white](1,1)(13,11)
\psline[linecolor=black,linewidth=.5pt,arrowsize=6pt]{->}(4,.5)(4,11.7)
\psline[linecolor=black,linewidth=.5pt,arrowsize=6pt]{->}(0,6)(14,6)
\psline[linecolor=red,linewidth=1pt,linestyle=dashed,dash=.25 .25](1,1)(1,11)
\psline[linecolor=red,linewidth=1pt,linestyle=dashed,dash=.25 .25](7,1)(7,11)
\psline[linecolor=red,linewidth=1pt,linestyle=dashed,dash=.25 .25](13,1)(13,11)
\psline[linecolor=black,linewidth=.5pt](1,5.9)(1,6.1)
\psline[linecolor=black,linewidth=.5pt](7,5.9)(7,6.1)
\psline[linecolor=black,linewidth=.5pt](10,5.9)(10,6.1)
\psline[linecolor=black,linewidth=.5pt](13,5.9)(13,6.1)
\rput(.5,5.6){\small $-\frac{\pi}{4}$}
\rput(6.7,5.6){\small $\frac{\pi}{4}$}
\rput(10,5.6){\small $\frac{\pi}{2}$}
\rput(12.6,5.6){\small $\frac{3\pi}{4}$}
\psline[linecolor=black,linewidth=.5pt](3.9,6.6)(4.1,6.6)
\psline[linecolor=black,linewidth=.5pt](3.9,7.2)(4.1,7.2)
\psline[linecolor=black,linewidth=.5pt](3.9,8.0)(4.1,8.0)
\psline[linecolor=black,linewidth=.5pt](3.9,9.0)(4.1,9.0)
\psline[linecolor=black,linewidth=.5pt](3.9,10.6)(4.1,10.6)
\psline[linecolor=black,linewidth=.5pt](3.9,5.4)(4.1,5.4)
\psline[linecolor=black,linewidth=.5pt](3.9,4.8)(4.1,4.8)
\psline[linecolor=black,linewidth=.5pt](3.9,4.0)(4.1,4.0)
\psline[linecolor=black,linewidth=.5pt](3.9,3.0)(4.1,3.0)
\psline[linecolor=black,linewidth=.5pt](3.9,1.4)(4.1,1.4)
\rput(3.6,6.6){\small $y_5$}
\rput(3.6,7.2){\small $y_4$}
\rput(3.6,8.0){\small $y_3$}
\rput(3.6,9.0){\small $y_2$}
\rput(3.6,10.6){\small $y_1$}
\rput(3.5,5.4){\small $-y_5$}
\rput(3.5,4.8){\small $-y_4$}
\rput(3.5,4.0){\small $-y_3$}
\rput(3.5,3.0){\small $-y_2$}
\rput(3.5,1.4){\small $-y_1$}
\psarc[linecolor=black,linewidth=.5pt,fillstyle=solid,fillcolor=black](1,6.6){.1}{0}{360}
\psarc[linecolor=gray,linewidth=0pt,fillstyle=solid,fillcolor=gray](1,7.2){.1}{0}{360}
\psarc[linecolor=black,linewidth=.5pt,fillstyle=solid,fillcolor=white](1,8.0){.1}{0}{360}
\psarc[linecolor=black,linewidth=.5pt,fillstyle=solid,fillcolor=black](1,9.0){.1}{0}{360}
\psarc[linecolor=gray,linewidth=0pt,fillstyle=solid,fillcolor=gray](1,10.6){.1}{0}{360}
\psarc[linecolor=black,linewidth=.5pt,fillstyle=solid,fillcolor=white](7,6.6){.1}{0}{360}
\psarc[linecolor=gray,linewidth=0pt,fillstyle=solid,fillcolor=gray](7,7.2){.1}{0}{360}
\psarc[linecolor=black,linewidth=.5pt,fillstyle=solid,fillcolor=black](7,8.0){.1}{0}{360}
\psarc[linecolor=black,linewidth=.5pt,fillstyle=solid,fillcolor=white](7,9.0){.1}{0}{360}
\psarc[linecolor=gray,linewidth=0pt,fillstyle=solid,fillcolor=gray](7,10.6){.1}{0}{360}
\psarc[linecolor=black,linewidth=.5pt,fillstyle=solid,fillcolor=black](13,6.6){.1}{0}{360}
\psarc[linecolor=gray,linewidth=0pt,fillstyle=solid,fillcolor=gray](13,7.2){.1}{0}{360}
\psarc[linecolor=black,linewidth=.5pt,fillstyle=solid,fillcolor=white](13,8.0){.1}{0}{360}
\psarc[linecolor=black,linewidth=.5pt,fillstyle=solid,fillcolor=black](13,9.0){.1}{0}{360}
\psarc[linecolor=gray,linewidth=0pt,fillstyle=solid,fillcolor=gray](13,10.6){.1}{0}{360}
\psarc[linecolor=black,linewidth=.5pt,fillstyle=solid,fillcolor=black](1,5.4){.1}{0}{360}
\psarc[linecolor=gray,linewidth=0pt,fillstyle=solid,fillcolor=gray](1,4.8){.1}{0}{360}
\psarc[linecolor=black,linewidth=.5pt,fillstyle=solid,fillcolor=white](1,4.0){.1}{0}{360}
\psarc[linecolor=black,linewidth=.5pt,fillstyle=solid,fillcolor=black](1,3.0){.1}{0}{360}
\psarc[linecolor=gray,linewidth=0pt,fillstyle=solid,fillcolor=gray](1,1.4){.1}{0}{360}
\psarc[linecolor=black,linewidth=.5pt,fillstyle=solid,fillcolor=white](7,5.4){.1}{0}{360}
\psarc[linecolor=gray,linewidth=0pt,fillstyle=solid,fillcolor=gray](7,4.8){.1}{0}{360}
\psarc[linecolor=black,linewidth=.5pt,fillstyle=solid,fillcolor=black](7,4.0){.1}{0}{360}
\psarc[linecolor=black,linewidth=.5pt,fillstyle=solid,fillcolor=white](7,3.0){.1}{0}{360}
\psarc[linecolor=gray,linewidth=0pt,fillstyle=solid,fillcolor=gray](7,1.4){.1}{0}{360}
\psarc[linecolor=black,linewidth=.5pt,fillstyle=solid,fillcolor=black](13,5.4){.1}{0}{360}
\psarc[linecolor=gray,linewidth=0pt,fillstyle=solid,fillcolor=gray](13,4.8){.1}{0}{360}
\psarc[linecolor=black,linewidth=.5pt,fillstyle=solid,fillcolor=white](13,4.0){.1}{0}{360}
\psarc[linecolor=black,linewidth=.5pt,fillstyle=solid,fillcolor=black](13,3.0){.1}{0}{360}
\psarc[linecolor=gray,linewidth=0pt,fillstyle=solid,fillcolor=gray](13,1.4){.1}{0}{360}
\end{pspicture}
\label{uplane}
\ee
where
\be
  y_j\ =\ -\frac{i}{2}\ln\tan\frac{t_j}{2}
\label{yj}
\ee
The analyticity strip $-\frac{\pi}{4}\le \mbox{Re}\,u\le\frac{3\pi}{4}$ is called the physical strip.
All zeros in the strip lie either on the boundary or on the vertical 
centre line (1-strings). If there is a zero on the left or right boundary, the same zero appears on the other boundary by periodicity. Even though these two zeros are identified, we will for convenience refer to this situation as a 2-string. In contrast to the usual situation in the 
context of Bethe ansatz, these zeros can occur either as single or 
double zeros.
A single zero is indicated by a grey dot, a double zero is indicated by
a black dot while an allowed  position with no zero is indicated by a white dot. Counting a double zero twice, the number of zeros with fixed
imaginary value $\pm y_j$ and real part either $\frac{\pi}{4}$ or 
$\frac{3\pi}{4}$
is two. This follows straightforwardly from (\ref{uzeros}) since 
$\nu_j$ can be either
$\epsilon_j$ or $\mu_j$, as depicted in (\ref{uplane}).

From complex conjugation symmetry, it follows that the full pattern of zeros is encoded in the distribution
of 1-strings in the {\em lower half-plane}.
This distribution is actually a sum of two --- one governed by
excitations administered by $\epsilon$ and one by $\mu$.
The separation of the 1-strings
into contributions coming from $\epsilon$ and $\mu$, respectively,
is superfluous but nonetheless a helpful refinement later for the 
description of the
selection rules.
It is illustrated here
\psset{unit=.7cm}
\setlength{\unitlength}{.7cm}
\be
\begin{pspicture}[shift=-2.5](-.25,-.25)(2,5)
\psframe[linewidth=0pt,fillstyle=solid,fillcolor=yellow!40!white](0,0)(2,5)
\psarc[linecolor=black,linewidth=.5pt,fillstyle=solid,fillcolor=white](1,4.5){.1}{0}{360}
\psarc[linecolor=gray,linewidth=0pt,fillstyle=solid,fillcolor=gray](1,3.5){.1}{0}{360}
\psarc[linecolor=black,linewidth=.5pt,fillstyle=solid,fillcolor=black](1,2.5){.1}{0}{360}
\psarc[linecolor=black,linewidth=.5pt,fillstyle=solid,fillcolor=white](1,1.5){.1}{0}{360}
\psarc[linecolor=gray,linewidth=0pt,fillstyle=solid,fillcolor=gray](1,0.5){.1}{0}{360}
\rput(-1,2.5){\small $\vdots$}
\rput(-1,1.5){\small $j=2$}
\rput(-1,.5){\small $j=1$}
\end{pspicture}
\hspace{.6cm}\ \ \longleftrightarrow \hspace{.6cm}
\begin{pspicture}[shift=-2.5](-.25,-.25)(2,5)
\psframe[linewidth=0pt,fillstyle=solid,fillcolor=yellow!40!white](0,0)(2,5)
\psarc[linecolor=black,linewidth=.5pt,fillstyle=solid,fillcolor=white](0.5,4.5){.1}{0}{360}
\psarc[linecolor=black,linewidth=.5pt,fillstyle=solid,fillcolor=white](0.5,3.5){.1}{0}{360}
\psarc[linecolor=gray,linewidth=0pt,fillstyle=solid,fillcolor=gray](0.5,2.5){.1}{0}{360}
\psarc[linecolor=black,linewidth=.5pt,fillstyle=solid,fillcolor=white](0.5,1.5){.1}{0}{360}
\psarc[linecolor=black,linewidth=.5pt,fillstyle=solid,fillcolor=white](0.5,0.5){.1}{0}{360}
\psarc[linecolor=black,linewidth=.5pt,fillstyle=solid,fillcolor=white](1.5,4.5){.1}{0}{360}
\psarc[linecolor=gray,linewidth=0pt,fillstyle=solid,fillcolor=gray](1.5,3.5){.1}{0}{360}
\psarc[linecolor=gray,linewidth=0pt,fillstyle=solid,fillcolor=gray](1.5,2.5){.1}{0}{360}
\psarc[linecolor=black,linewidth=.5pt,fillstyle=solid,fillcolor=white](1.5,1.5){.1}{0}{360}
\psarc[linecolor=gray,linewidth=0pt,fillstyle=solid,fillcolor=gray](1.5,0.5){.1}{0}{360}
\end{pspicture}
\label{onetwo}
\ee
where the left side corresponds to (\ref{uplane}) while the two columns
to the right encode separately the $\epsilon$ and $\mu$ excitations.
Not all such double-column configurations will appear in a given sector. A characterization of
a set of {\it admissible} two-column configurations provides a description 
of the selection rules.
This combinatorial designation of the physical states is termed 
`physical combinatorics'. As we will see,
the double-column configurations provide
a natural basis for defining the associated {\em finitized characters}. These double-column configurations 
and finitized characters also arise in the context of the dimer model~\cite{RR1207}.

\subsection{Finitized spectra, characters and partition functions}

In this section, we outline what is meant by finitized spectra, finitized characters and finitized partition functions. 
Consider a finite system of fixed size $N_0$. The eigenenergies $E_n(u)$ and eigenvalues $d_n(u)=e^{-E_n(u)}$ of this system, with $n=0,1,2,\ldots$, are transcendental numbers given by (\ref{N}). More usefully, these eigenenergies are characterized by their associated patterns of zeros. By complex conjugation, we only need to consider the patterns of zeros in the lower-half $u$-plane. The finite spectra are therefore characterized by a finite collection of these patterns of zeros in the lower-half $u$-plane. For the groundstate in the $(r,s)=(1,1)$ sector, the pattern of zeros consists of double 2-strings at all vertical positions $y_j$ given by (\ref{yj}) and (\ref{tj}). Because of the special properties of the inversion identity, these zeros lie at {\it exactly\/} these positions. In other sectors (see Figure~\ref{groundstates}), the groundstate is formed by exciting a finite number of these 2-strings to 1-strings at the same vertical heights. Within a given sector, finite excitations above the groundstate are generated by exciting an additional finite number of 2-strings to 1-strings. 

Each eigenenergy of the finite system with system size $N_0$ is {\it uniquely} associated with an eigenenergy of the infinite system through a process of {\it filling the Fermi sea}. Given the pattern of zeros associated with the given eigenenergy, we increase $N$ (respecting any parity constraints) and so increase the number of allowed vertical positions $M$ for 1-strings and 2-strings adjusting the vertical heights (\ref{yj}) accordingly. In the original $M_0$ positions, determined by the fixed size $N_0$, we preserve the given pattern of zeros $\{\epsilon_j,\mu_j\}_{j=1}^{M_0}$. In the additional positions, we just add double 2-strings so that there are no 1-strings (excitations) at these new positions and $\epsilon_j=\mu_j=1$ for $j>M_0$. In the scaling limit $N\to\infty$, this describes a finite excitation of the infinite system above the groundstate of the sector. Because the limiting system is conformal, the limiting finite-size excitation energy $E$ must be of the form (\ref{logLa})
\bea
E=\lim_{N\to\infty} {N\over 2\pi \sin 2u}\,\big(E_n(u)-E_0(u)\big)=0,1,2,3,\ldots
\eea
where $k=E$ labels the tower of conformal states above the groundstate. Starting with any eigenvalue of a finite system and filling the Fermi sea, we can therefore associate to it the unique whole number energy $E$ and the formal monomial $q^E$ where the formal parameter $q=\exp(-2\pi {N'\over N}\sin 2u)$ is identified with the modular nome (\ref{tau}) and $N'/N$ is the lattice aspect ratio. 
Doing this for the finite collection of eigenenergies of a finite system with $N=N_0$ gives the {\it finitized partition function} defined formally as the spectrum generating function
\bea
Z^{(N_0)}_{(1,1)|(r,s)}(q)=q^{-{c\over 24}+\Delta_{r,s}} {\sum_E}' q^E
\eea
where the restricted sum of formal monomials is over all the eigenvalues of the finite system and the prefactor anticipates the result (\ref{limConfPartFn}) below. Notice that this corresponds to a truncated finite set of eigenenergies $E$ of the infinite system. Setting $q=1$ gives the counting of states
\bea
Z^{(N_0)}_{(1,1)|(r,s)}(1)=\dim{\cal V}^{(N_0)}_{\rho,s}
\eea
In Section~\ref{Euler}, we use Euler-Maclaurin to show that the conformal partition functions are given by
\bea
Z_{(1,1)|(r,s)}(q)=\lim_{N_0\to\infty}Z^{(N_0)}_{(1,1)|(r,s)}(q)=q^{-{c\over 24}+\Delta_{r,s}} \sum_E q^E\label{limConfPartFn}
\eea
where the unrestricted sum is now over the complete (infinite) set of finite excitations of the infinite system.

Unlike the exact transcendental spectra of finite systems, the finitized (whole number) spectra can be described combinatorially through the associated patterns of zeros, double-column configurations and sums of $q$-Catalan polynomials of the form $\sum_E q^E$ 
as we describe in the next two subsections. The energies  $E$ are given by (\ref{wLR}) and (\ref{wLR2}) and, for small system sizes $N_0$, can be easily read off from numerical computation of the eigenvalue patterns of zeros. 
In the case $s=1$ or $s=2$, the conformal partition functions are associated with a single irreducible character (\ref{chrs}). In this case, we find that the finitized partition functions are given as in (\ref{catfin}) by single $q$-Catalan polynomials defining {\it finitized\/} irreducible characters. The eigenvalues associated with these $q$-Catalan polynomials are the irreducible building blocks for the spectra in the sense that, for $s>2$, the finitized partition functions are decomposed into sums of irreducible blocks given by the $q$-Catalan polynomials as dictated by the decomposition (\ref{irredDecomp}) of Kac characters into irreducible characters.

\subsection{Double-columns and $q$-Catalan polynomials}
\label{qCatalan}

A {\em single-column configuration} of height $M$ consists of $M$ 
sites arranged as a column. The sites are labelled from the bottom by the
integers $1,2,\ldots,M$. Each site can be occupied or unoccupied. 
The configuration is specified by listing the labels of the occupied sites in descending order. 
A {\em double-column configuration} $S$ of height $M$ consists of a pair $(L,R)$ of
single-column configurations of height $M$. 
For the double-column configuration on the right side of (\ref{onetwo}) 
\be
  S\;=\;(L,R),\ \ \ \ \ \ \ L\;=\;(3),\ \ \ \ \ \ \ R\;=\;(4,3,1)
\label{SLSR}
\ee
If $m=|L|=\half \sum_{j=1}^M(1-\epsilon_j)$ and $n=|R|=\half \sum_{j=1}^M(1-\mu_j)$ indicate the number of occupied sites
in the left and right columns, respectively,
the energy $E=E(S)$ of a double-column configuration is given by
\be
E\;=\;E(S)\;=\;E(L,R)\;=\;\sum_{j=1}^mL_j+\sum_{j=1}^nR_j
\label{wLR}
\ee
where we will ultimately take
\be
L_j=R_j=E_j=\begin{cases} j,&\mbox{$j$ is occupied with $s$ odd}\\
j-\half,&\mbox{$j$ is occupied with $s$ even}\\
0,&\mbox{$j$ is unoccupied}\end{cases}
\label{wLR2}
\ee
In finitized characters, we will also associate the monomial
\be
q^E=q^{E(S)}
\ee
to the double-column configuration $S$. The formal parameter $q$ will ultimately be the modular nome.

A double-column configuration $S=(L,R)$
is called {\em admissible} if $L\preceq R$ with respect to the partial ordering
\be
  L\ \preceq\ R\ \ \ \ \ {\rm if}\ \ \ \ \ L_j\ \leq\ R_j,\ \ j=1,\ldots,m
\label{order}
\ee
which presupposes that
\be
  0\ \leq\ m\ \leq\ n\ \leq\ M
\label{mnM}
\ee
Admissibility can be simply characterized geometrically as illustrated in the example
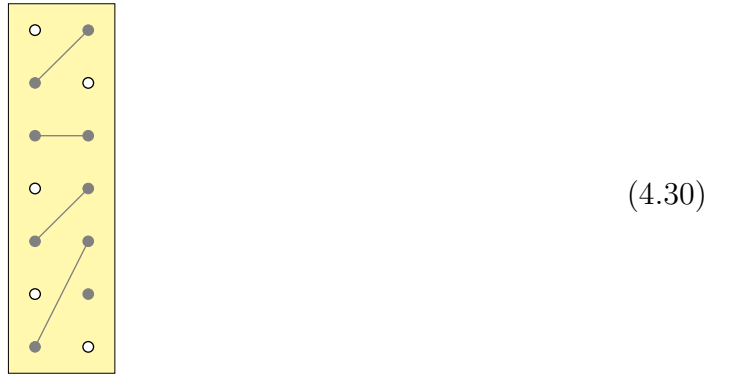
\psset{unit=.7cm}
\setlength{\unitlength}{.7cm}
\be
\begin{pspicture}[shift=-3.45](-.25,-.25)(2,7.3)
\psframe[linewidth=0pt,fillstyle=solid,fillcolor=yellow!40!white](0,0)(2,7)
\psarc[linecolor=black,linewidth=.5pt,fillstyle=solid,fillcolor=white](0.5,6.5){.1}{0}{360}
\psarc[linecolor=gray,linewidth=0pt,fillstyle=solid,fillcolor=gray](0.5,5.5){.1}{0}{360}
\psarc[linecolor=gray,linewidth=0pt,fillstyle=solid,fillcolor=gray](0.5,4.5){.1}{0}{360}
\psarc[linecolor=black,linewidth=.5pt,fillstyle=solid,fillcolor=white](0.5,3.5){.1}{0}{360}
\psarc[linecolor=gray,linewidth=0pt,fillstyle=solid,fillcolor=gray](0.5,2.5){.1}{0}{360}
\psarc[linecolor=black,linewidth=.5pt,fillstyle=solid,fillcolor=white](0.5,1.5){.1}{0}{360}
\psarc[linecolor=gray,linewidth=0pt,fillstyle=solid,fillcolor=gray](0.5,0.5){.1}{0}{360}
\psarc[linecolor=gray,linewidth=0pt,fillstyle=solid,fillcolor=gray](1.5,6.5){.1}{0}{360}
\psarc[linecolor=black,linewidth=.5pt,fillstyle=solid,fillcolor=white](1.5,5.5){.1}{0}{360}
\psarc[linecolor=gray,linewidth=0pt,fillstyle=solid,fillcolor=gray](1.5,4.5){.1}{0}{360}
\psarc[linecolor=gray,linewidth=0pt,fillstyle=solid,fillcolor=gray](1.5,3.5){.1}{0}{360}
\psarc[linecolor=gray,linewidth=0pt,fillstyle=solid,fillcolor=gray](1.5,2.5){.1}{0}{360}
\psarc[linecolor=gray,linewidth=0pt,fillstyle=solid,fillcolor=gray](1.5,1.5){.1}{0}{360}
\psarc[linecolor=black,linewidth=.5pt,fillstyle=solid,fillcolor=white](1.5,0.5){.1}{0}{360}
\psline[linecolor=gray,linewidth=.5pt](0.5,5.5)(1.5,6.5)
\psline[linecolor=gray,linewidth=.5pt](0.5,4.5)(1.5,4.5)
\psline[linecolor=gray,linewidth=.5pt](0.5,2.5)(1.5,3.5)
\psline[linecolor=gray,linewidth=.5pt](0.5,0.5)(1.5,2.5)
\end{pspicture}
\label{adm}
\ee
One draws line segments between the occupied sites of greatest height
in the two columns, then between the occupied sites of second-to-greatest
height and so on. The double-column configuration is now admissible
if it does {\em not\/} involve line segments with a {\em strictly negative slope}.
Thus, in an admissible double-column configuration, there are either no line segments ($m=0$)
or each line segment appears with a non-negative slope. 

Combinatorially, we define the (generalized) $q$-Narayana numbers $\sbinlr{M}{m,n}_{\!q}$ 
as the sum of the monomials associated to all admissible double-column configurations of height $M$ with exactly $m$ and $n$ occupied sites in 
the left and right columns respectively
\be
\sbinlr{M}{m,n}_{\!q}=\sum_{S:\,|L|=m, |R|=n} q^{E(S)}
\ee
In Appendix~\ref{qNarayana}, we show that these polynomials admit the closed-form expressions
\be
  \sbinlr{M}{m,n}_{\!q}=q^{\hf m(m+1)+\hf n(n+1)}
   \bigg(\gauss{M}{m}_{\!q}\gauss{M}{n}_{\!q}-q^{n-m+1} \gauss{M}{m-1}_{\!q}\gauss{M}{n+1}_{\!q}
   \bigg)\quad\label{sbin}
    \ee
where $\sgauss{M}{m}_{\!q}$
is a $q$-binomial (Gaussian polynomial). These coincide with
$q$-Narayana numbers~\cite{Narayana,Branden} when $m=n$.

Following~\cite{PRS10}, for $r\ge 0$, we define the following (generalized) $q$-analog of Catalan numbers
\bea
\begin{array}{ll}
\disp\cat Mrq=\sum_{m=0}^{M-r+1}\sbinlr{M}{m,m\!+\!r\!-\!1}_{\!q}=q^{\frac{r(r-1)}{2}}\frac{(1-q^r)}{(1-q^{M+1})}\gauss{2M+2}{M+1-r}_{\!q}\quad &\mbox{$s$ odd}
\label{cat1}\\[16pt]
\disp\catt Mrq=q^{-\frac{r-1}{2}}\sum_{m=0}^{M-r+1}q^{-m} \sbinlr{M}{m,m\!+\!r\!-\!1}_{\!q}
=q^{\frac{(r-1)^2}{2}}\frac{(1-q^{2r})}{(1-q^{M+r+1})}\gauss{2M+1}{M+1-r}_{\!q}\quad &\mbox{$s$ even}\label{cat2}
\end{array}\label{qCat}
\eea
It is convenient to extend these definitions using symmetry 
\bea
C_{M,0}(q)=C'_{M,0}(q)=0,\quad C_{M,r}(q)=-C_{M,-r}(q),\quad C'_{M,r}(q)=-C'_{M,-r}(q),\quad r< 0\label{negCat}
\eea
The polynomials $C_{M,1}(q)$ coincide with one common definition of $q$-Catalan numbers~\cite{Narayana}. In the sequel, we will just refer to the polynomials 
in (\ref{qCat}) as $q$-Catalan polynomials. Although we refer to these as $q$-Catalan polynomials we stress that, as defined for $r$ even, 
$\catt Mrq$ is actually a polynomial in $\sqrt{q}$ 
due to the prefactor. We assert that the $q$-Catalan polynomials (\ref{cat1}) are the spectrum generating functions of the irreducible building blocks for the finitized characters in the case that $s$ is odd or even respectively. Despite the minus signs in (\ref{sbin}) and (\ref{qCat}), the $q$-Narayana numbers and $q$-Catalan polynomials are {\em fermionic} in the sense that they only have non-negative coefficients.

Combinatorially, for $r\ge 1$, the (generalized) $q$-Catalan polynomial $\cat Mrq$ 
is the sum of the monomials associated to all admissible double-column configurations of height $M$ with $n-m=r-1$, that is, 
exactly $r-1$ more occupied sites in the right column compared to the left column. With the energy assignment $E_j=j$ appropriate for $s$ odd, we have
\be
\cat Mrq=\sum_{S:\,|R|-|L|=r-1} q^{E(S)}
\ee
Similarly, for $r\ge 1$, the (generalized) $q$-Catalan polynomial $\catt Mrq$ has precisely the same combinatorial interpretation but with the energy assignment $E_j=j-\half$ 
 appropriate for $s$ even. 
The case $M=r=2$ is illustrated in Figure~\ref{catfig}.

\begin{figure}[htbp]
\begin{center}
\psset{unit=.7cm}
\setlength{\unitlength}{.7cm}\mbox{}\quad
\begin{pspicture}[shift=-3.45](-1.25,-1.5)(2,2.3)
\psframe[linewidth=0pt,fillstyle=solid,fillcolor=yellow!40!white](0,0)(2,2)
\psarc[linecolor=black,linewidth=.5pt,fillstyle=solid,fillcolor=white](0.5,1.5){.1}{0}{360}
\psarc[linecolor=black,linewidth=.5pt,fillstyle=solid,fillcolor=white](0.5,.5){.1}{0}{360}
\psarc[linecolor=black,linewidth=.5pt,fillstyle=solid,fillcolor=white](1.5,1.5){.1}{0}{360}
\psarc[linecolor=gray,linewidth=0pt,fillstyle=solid,fillcolor=gray](1.5,0.5){.1}{0}{360}
\rput[b](1,-1){$q$}
\rput[b](1,-2){$q^{1/2}$}
\rput[b](-1.7,-1){$\cat 22q\ \ =$}
\rput[b](-1.7,-2){$\catt 22q\ \ =$}
\multirput[b](3.2,-1)(4.5,0){3}{$+$}
\multirput[b](3.2,-2)(4.5,0){3}{$+$}
\end{pspicture}\qquad
\begin{pspicture}[shift=-3.45](-1.25,-1.5)(2,2.3)
\psframe[linewidth=0pt,fillstyle=solid,fillcolor=yellow!40!white](0,0)(2,2)
\psarc[linecolor=black,linewidth=.5pt,fillstyle=solid,fillcolor=white](0.5,1.5){.1}{0}{360}
\psarc[linecolor=black,linewidth=.5pt,fillstyle=solid,fillcolor=white](0.5,.5){.1}{0}{360}
\psarc[linecolor=gray,linewidth=0pt,fillstyle=solid,fillcolor=gray](1.5,1.5){.1}{0}{360}
\psarc[linecolor=black,linewidth=.5pt,fillstyle=solid,fillcolor=white](1.5,.5){.1}{0}{360}
\rput[b](1,-1){$q^2$}
\rput[b](1,-2){$q^{3/2}$}
\end{pspicture}\qquad
\begin{pspicture}[shift=-3.45](-1.25,-1.5)(2,2.3)
\psframe[linewidth=0pt,fillstyle=solid,fillcolor=yellow!40!white](0,0)(2,2)
\psarc[linecolor=black,linewidth=.5pt,fillstyle=solid,fillcolor=white](0.5,1.5){.1}{0}{360}
\psarc[linecolor=gray,linewidth=0pt,fillstyle=solid,fillcolor=gray](0.5,0.5){.1}{0}{360}
\psarc[linecolor=gray,linewidth=0pt,fillstyle=solid,fillcolor=gray](1.5,1.5){.1}{0}{360}
\psarc[linecolor=gray,linewidth=0pt,fillstyle=solid,fillcolor=gray](1.5,0.5){.1}{0}{360}
\psline[linecolor=gray,linewidth=.5pt](0.5,0.5)(1.5,1.5)
\rput[b](1,-1){$q^4$}
\rput[b](1,-2){$q^{5/2}$}
\end{pspicture}\qquad
\begin{pspicture}[shift=-3.45](-1.25,-1.5)(2,2.3)
\psframe[linewidth=0pt,fillstyle=solid,fillcolor=yellow!40!white](0,0)(2,2)
\psarc[linecolor=gray,linewidth=0pt,fillstyle=solid,fillcolor=gray](.5,1.5){.1}{0}{360}
\psarc[linecolor=black,linewidth=.5pt,fillstyle=solid,fillcolor=white](0.5,.5){.1}{0}{360}
\psarc[linecolor=gray,linewidth=0pt,fillstyle=solid,fillcolor=gray](1.5,1.5){.1}{0}{360}
\psarc[linecolor=gray,linewidth=0pt,fillstyle=solid,fillcolor=gray](1.5,0.5){.1}{0}{360}
\psline[linecolor=gray,linewidth=.5pt](0.5,1.5)(1.5,1.5)
\rput[b](1,-1){$q^5$}
\rput[b](1,-2){$q^{7/2}$}
\end{pspicture}
\end{center}
\caption{Illustration of the combinatorial interpretation of the $q$-Catalan polynomials $\cat 22q=q+q^2+q^4+q^5$ 
and $\catt 22q=\sqrt{q}(1+q+q^2+q^3)$ for $M=r=2$. The energy assignment is $E_j=j, j-\half$ as appropriate for $s$ odd, $s$ even respectively. 
The sum is over all admissible double-column configurations of height $M$ with $r-1$ more occupied sites in the right column.\label{catfig}}
\end{figure}

The $q$-Catalan polynomials are simply related to finitized irreducible characters
\bea
\mch_{r,1}^{(N)}(q)=q^{-{c\over 24}} \cat Mrq,\qquad \mch_{r,2}^{(N)}(q)=q^{-{c\over 24}-{1\over 8}} \catt Mrq
\label{catfin}
\eea
where $M$ and $N$ are related by (\ref{MN}). Using the result
\bea
\lim_{M\to\infty} \gauss{M}{m}_{\!q}={1\over (q)_m},\qquad (q)_m=\prod_{k=1}^m (1-q^k)\label{qbinlimit}
\eea
it follows from (\ref{qCat}) that in the thermodynamic limit
\bea
\lim_{N\to\infty} \mch_{r,1}^{(N)}(q)=\mch_{r,1}(q),\qquad \lim_{N\to\infty} \mch_{r,2}^{(N)}(q)=\mch_{r,2}(q)
\label{catfinlimit}
\eea

\subsection{Selection rules}
\nc{\qqcat}[3]{C_{#1,#2}(#3)}
\nc{\qqcatt}[3]{C_{#1,#2}'(#3)}

We now turn to the description of the selection rules.
They depend on the parities of $\rho$ and $s$ with $N+\rho+s$ even. The relationship between $\rho$ and $r$ is as given in (\ref{rhor}).
Based on a body of empirical data obtained
by examining patterns of zeros by numerically diagonalizing the double-row
transfer matrices out to sizes $N=14$ with $1\le\rho, s\le 8$,
we make the following conjecture for the selection rules yielding the finitized partition functions $Z_{(1,1)|(r,s)}^{(N)} (q)$.
\\[.3cm]
\noindent {\bf Selection rules:}\ \
{\em For a system with $N$ bulk columns, $\rho\!-\!1$ $r$-type and $s\!-\!1$ $s$-type boundary columns, 
 the selection rules in the $(r,s)$ sectors are equivalent to singling out the following sets
of admissible double-column configurations in terms of the irreducible building blocks given by $q$-Catalan polynomials}
\bea
s\ {\rm odd:} \ Z_{(1,1)|(r,s)}^{(N)} (q) \!\!\!&=&\!\!\!
\begin{cases}
\disp q^{-\frac{c}{24}}\sum_{k=1}^{s} \qqcat{\frac{N-1}{2}}{\frac{\rho-s-1+2k}{2}}{q}, & \quad\rho=2r \\[14pt]
\disp q^{-\frac{c}{24}}\sum_{k=1}^{s} \!\big[\qqcat{\frac{N-2}{2}}{\frac{\rho-s+2k}{2}}{q}+
q^{\frac{N}{2}} \qqcat{\frac{N-2}{2}}{\frac{\rho-s-2+2k}{2}}{q} \big], & \quad\rho=2r\!-\!1
\end{cases}
\label{catId1}\\[8pt]
s\ {\rm even:} \ Z_{(1,1)|(r,s)}^{(N)} (q) \!\!\!&=&\!\!\!
\begin{cases}
\disp q^{-\frac{c}{24}-\frac{1}{8}}\sum_{k=1}^{s/2} \qqcatt{\frac{N}{2}}{\frac{\rho-s-2+4k}{2}}{q}, & \rho=2r \\[14pt]
\disp q^{-\frac{c}{24}-\frac{1}{8}}\sum_{k=1}^{s/2} \big[\qqcatt{\frac{N-1}{2}}{\frac{\rho-s-1+4k}{2}}{q}+q^{\frac{N}{2}} 
\qqcatt{\frac{N-1}{2}}{\frac{\rho-s-3+4k}{2}}{q}\big], & \rho=2r\!-\!1\qquad
\end{cases}\label{catId2}
\eea
If $s>2r$, each $q$-Catalan polynomial $C_{M,r}(q)$ or $C'_{M,r}(q)$ with $r< 0$ explicitly cancels out  in these sums against the same polynomial with $r>0$. 
Although we do not do so, the sums in these cases can be restricted to the values of $k$ that do not cancel out. In the first sum with $s>2r$ odd and $\rho$ even, for example, 
the lower terminal could be replaced with $k=s-2r+1$.

Notice that the excitations form conformal towers above the groundstate with integer spacing. To see this, let us define the excitation energies
\bea
E_j=\begin{cases}
j,&\mbox{$s$ odd}\\
j-\half,&\mbox{$s$ even}
\end{cases}
\eea
and the modular nome $q$
\be
  q\ =\ e^{-2\pi\tau},\ \ \ \ \ \ \ \tau\ =\ \frac{N'}{N}\sin 2u
\label{tau}
\ee
where $N'/N$ is the aspect ratio and $\vartheta=2u$ is the 
anisotropy angle~\cite{KimP} related to the geometry of the lattice.
Then, for $s$ odd, taking the ratio of (\ref{N}) with precisely one $\epsilon_j$ 
or $\mu_j=-1$ to (\ref{N}) with all 
$\epsilon_j=\mu_j=+1$, and taking the limit $N,N'\to\infty$ 
with a fixed aspect ratio $\delta=N'/N$ gives
\bea
 \lim_{N,N'\to\infty}\bigg(\frac{1-\sin\frac{j\pi}{N}\,\sin 2u}{{1+\sin\frac{j\pi}{N}\,\sin 2u}}\bigg)^{N'}\!\!
 =\exp[-2j \pi\,\delta\,\sin 2u]\;=\;q^{E_j},\qquad E_j=j
\eea
Similarly, for $s$ even, taking the ratio of (\ref{N}) with precisely 
one $\epsilon_j$ or $\mu_j=-1$ to (\ref{N}) with all 
$\epsilon_j=\mu_j=+1$, and taking the limit $N,N'\to\infty$ with a 
fixed aspect ratio $\delta=N'/N$ gives
\bea
\lim_{N,N'\to\infty}\bigg(\frac{1-\sin\frac{(2j-1)\pi}{2N}}{1+\sin\frac{(2j-1)\pi}{2N}}\bigg)^{N'}\!\!
=\exp[-(2j-1)\pi\,\delta\,\sin 2u]\;=\;q^{E_j},\qquad E_j=j-\half
\eea

The groundstates in the various sectors are given by the zero patterns associated with the minimum surviving energy configurations in (\ref{catId1}) and (\ref{catId2}) as shown in Figure~\ref{groundstates}. For $2r\ge s$, this occurs in the $q$-Catalan polynomial with $k=1$. In all cases, for $\rho$ even or odd, it occurs in the leading $q$-Catalan polynomials $C_{M,n-m+1}(q)$ or $C'_{M,n-m+1}(q)$ where the excess occupation in the 
right column over the left column is given by
\bea
n-m=|R|-|L|=\begin{cases}
\half(|2r-s|-1),&\mbox{$s$ odd}\\[4pt]
\half(|2r-s|),&\mbox{$s$ even}
\end{cases}
\label{excess}
\eea
The conformal weights associated to the minimal configurations are
\bea
\Delta_{r,s}=\begin{cases}
\Delta_{n-m+1,1},\quad&\mbox{$s$ odd}\\
\Delta_{n-m+1,2},&\mbox{$s$ even}\\
\end{cases}
\label{excessConfWt}
\eea

\psset{unit=.5cm}
\setlength{\unitlength}{.6cm}
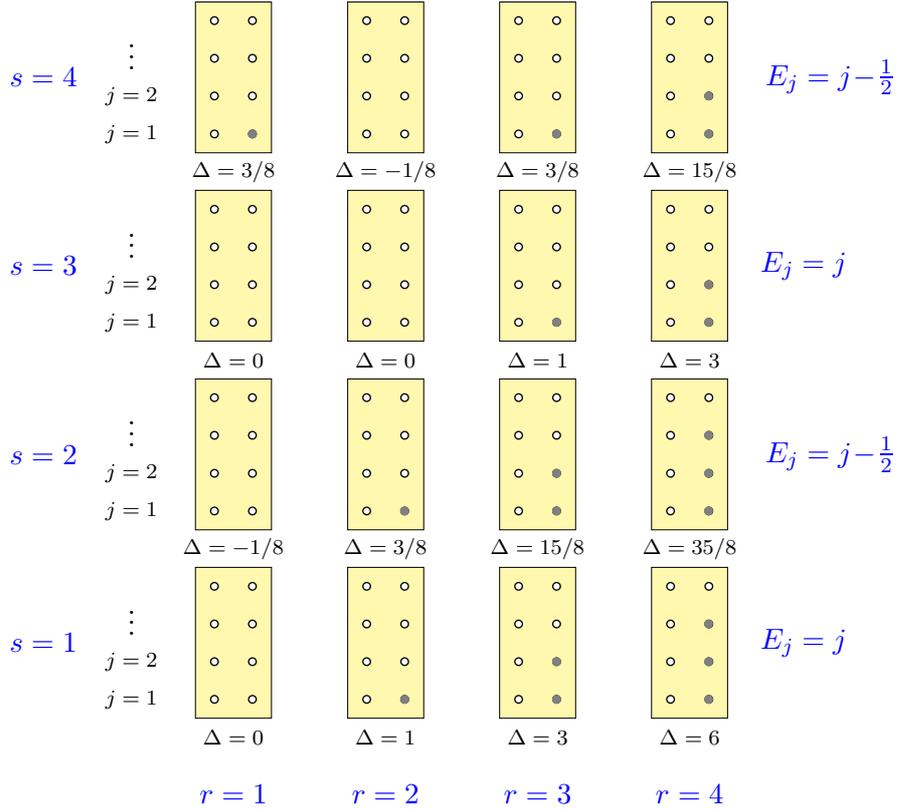
\begin{figure}[tb]
\begin{center}
\begin{pspicture}[shift=-5](-.25,-2.25)(.6,5)
\multirput(0,0)(0,5){4}{\rput(0,.5){\scriptsize $j=1$}
\rput(0,1.5){\scriptsize $j=2$}
\rput(0,2.75){$\vdots$}}
\end{pspicture}
\hspace{4pt}
\begin{pspicture}[shift=-5](-.25,-2.25)(15,19)
\multirput(0,0)(0,5){4}{\multirput(0,0)(4,0){4}{\psframe[linewidth=0pt,fillstyle=solid,fillcolor=yellow!40!white](0,0)(2,4)}}
\multirput(0,0)(4,0){4}{
\multirput(0,0)(0,5){4}{\multirput(0,0)(0,1){4}{\multirput(0,0)(1,0){2}{\psarc[linecolor=black,linewidth=.5pt,fillstyle=solid,fillcolor=white](0.5,0.5){.1}{0}{360}}}}}
\multirput(0,0)(0,5){2}{
\multirput(4,0)(4,0){3}{\psarc[linecolor= gray,linewidth=.5pt,fillstyle=solid,fillcolor=gray](1.5,.5){.1}{0}{360}}
\multirput(8,1)(4,0){2}{\psarc[linecolor= gray,linewidth=.5pt,fillstyle=solid,fillcolor=gray](1.5,.5){.1}{0}{360}}
\multirput(12,2)(4,0){1}{\psarc[linecolor= gray,linewidth=.5pt,fillstyle=solid,fillcolor=gray](1.5,.5){.1}{0}{360}}
}
\multirput(0,10)(0,5){2}{
\multirput(8,0)(4,0){2}{\psarc[linecolor= gray,linewidth=.5pt,fillstyle=solid,fillcolor=gray](1.5,.5){.1}{0}{360}}
\multirput(12,1)(4,0){1}{\psarc[linecolor= gray,linewidth=.5pt,fillstyle=solid,fillcolor=gray](1.5,.5){.1}{0}{360}}
}
\rput(0,15){\psarc[linecolor= gray,linewidth=.5pt,fillstyle=solid,fillcolor=gray](1.5,.5){.1}{0}{360}}
\rput(1,-.5){\scriptsize $\Delta=0$}
\rput(5,-.5){\scriptsize $\Delta=1$}
\rput(9,-.5){\scriptsize $\Delta=3$}
\rput(13,-.5){\scriptsize $\Delta=6$}
\rput(1,4.5){\scriptsize $\Delta=-1/8$}
\rput(5,4.5){\scriptsize $\Delta=3/8$}
\rput(9,4.5){\scriptsize $\Delta=15/8$}
\rput(13,4.5){\scriptsize $\Delta=35/8$}
\rput(1,9.5){\scriptsize $\Delta=0$}
\rput(5,9.5){\scriptsize $\Delta=0$}
\rput(9,9.5){\scriptsize $\Delta=1$}
\rput(13,9.5){\scriptsize $\Delta=3$}
\rput(1,14.5){\scriptsize $\Delta=3/8$}
\rput(5,14.5){\scriptsize $\Delta=-1/8$}
\rput(9,14.5){\scriptsize $\Delta=3/8$}
\rput(13,14.5){\scriptsize $\Delta=15/8$}
\rput(-4,2){\small \color{blue} $s=1$}
\rput(-4,7){\small \color{blue} $s=2$}
\rput(-4,12){\small \color{blue} $s=3$}
\rput(-4,17){\small \color{blue} $s=4$}
\rput(1,-2){\small \color{blue} $r=1$}
\rput(5,-2){\small \color{blue} $r=2$}
\rput(9,-2){\small \color{blue} $r=3$}
\rput(13,-2){\small \color{blue} $r=4$}
\rput(16,2){\small \color{blue} $E_j=j$}
\rput(16.7,7){\small \color{blue} $E_j=j\!-\!\half$}
\rput(16,12){\small \color{blue} $E_j=j$}
\rput(16.7,17){\small \color{blue} $E_j=j\!-\!\half$}
\end{pspicture}
\vspace{-.1in}
\end{center}
\caption{Groundstate double-column configurations arranged in a Kac table for $r,s=1,2,3,4$. 
The continuation of the pattern for larger values of $r$ and $s$ is clear. 
The solid grey dots represent single 1-strings in the center of the analyticity strip. The excess occupation in the right column over the left column is given by (\ref{excess}) and the conformal weights by (\ref{excessConfWt}). There are no double zeros in the center of the analyticity strip for these groundstates.
}
\label{groundstates}
\end{figure}

Applying the identities proved in Appendix~\ref{CatIdProofs}, the $q$-Catalan decompositions (\ref{catId1}) and (\ref{catId2}) reduce to simple finitized characters
\bea
Z_{(1,1)|(r,s)}^{(N)} (q) = \chit_{r,s}^{(N)}(q)=
\begin{cases}
q^{-\frac{c}{24}+\Delta_{r,s}} \Big(\sgauss{N}{{N-2r+s\over 2}}_{\!q}-q^{rs} \sgauss{N}{{N-2r-s\over 2}}_{\!q}\Big), & \rho=2r\\[12pt]
q^{-\frac{c}{24}+\Delta_{r,s}} \Big(\sgauss{N}{{N-2r+s+1\over 2}}_{\!q}-q^{rs} \sgauss{N}{{N-2r-s+1\over 2}}_{\!q}\Big), & \rho=2r-1\qquad
\end{cases}
\label{finitizedZ}
\eea
The results with $\rho=r=1$ and $s$ arbitrary exactly agree with the results of~\cite{PR0610}. 
We also observe that the limit $q\to 1$ gives the correct counting of states (\ref{countstates}).
Taking the thermodynamic limit and again using (\ref{qbinlimit}), we obtain the desired result
\bea
Z_{(1,1)|(r,s)}(q) = \lim_{N\to\infty}\chit_{r,s}^{(N)}(q)=\chit_{r,s}(q),\qquad r,s=1,2,3,\ldots
\eea
Similarly, taking the thermodynamic limit directly on (\ref{catId1}) and (\ref{catId2}) using (\ref{catfin}) and (\ref{catfinlimit}) gives the irreducible decomposition (\ref{irredDecomp}) of the Kac character $\chit_{r,s}(q)$. 
While these results depend on the conjectured selection rules, we point out that these selection rules were previously conjectured~\cite{PR0610} for the case 
 $\rho=r=1$ with $s$ arbitrary and subsequently proved by Morin-Duchesne~\cite{Morin}.

\section{Finite-Size Corrections}
\label{SecFinite}

\subsection{Euler-Maclaurin}
\label{Euler}

The partition function of critical dense polymers on a lattice of $N$ 
columns and $N'$ double
rows is defined by
\be
  Z_{N,N'}\ =\ \mathop{\rm Tr}\vec d(u)^{N'}\ =\ \sum_n d_{n}(u)^{N'}
   \ =\ \sum_ne^{-N'E_{n}(u)}
\label{ZNM}
\ee
where the sum is over all eigenvalues of the normalized transfer matrix $\vec d(u)$ including possible
multiplicities and $E_{n}(u)$ is the energy associated to the eigenvalue
$d_{n}(u)$. Although we suppress the dependence on $(r,s)$, the form of the partition function (\ref{ZNM}) holds in all sectors.
Conformal invariance of the model in the continuum scaling limit dictates 
\cite{BCN,Aff} that the leading finite-size corrections for large $N$ in the $(r,s)$ sector
are of the form
\be
  E_{n}(u)\ =\ -\ln d_{n}(u)\ \simeq\ 2Nf_{bulk}+f_{bdy}
   +\frac{2\pi\sin 2u}{N}\Big(\!\!-\!\frac{c}{24}+\Delta_{r,s}+k\Big)
\label{logLa}
\ee
Here $f_{bulk}$ is the bulk free energy per face and $f_{bdy}$ is the 
boundary free energy (excluding the additional boundary free energy of the $r$-type seam which was explicitly removed by normalization of $\vec d(u)$).  
The central charge of the CFT is $c$ while the 
spectrum of conformal weights is given by the possible values of $\Delta_{r,s}$
with excitations or descendants labelled by the non-negative integers $k$.
As the analysis below will confirm, the asymptotic behaviour of the
eigenvalues (\ref{N}) are in accordance with (\ref{logLa}).

Let us introduce the function
\bea
  F(t)=\ln(1+\sin t \sin 2u)
\label{F}
\eea
where for simplicity the $u$ dependence has been suppressed.
As a partial evaluation of the energies (\ref{logLa}), we find
\bea
  \ln\prod_j\big(1+\eps_j\sin t_j\,\sin 2u\big)
  \ \simeq\ \sum_jF(t_j)-2\sin 2u\sum_{j\in{\cal E}_n}t_j
\label{sumE}
\eea
where ${\cal E}_n$ is the subset of $j$ indices for which
$\eps_j=-1$ and ${\cal E}_n$ remains finite as $N\rightarrow\infty$.
Likewise, ${\cal M}_n$ denotes the subset of $j$ indices for which
$\mu_j=-1$ and it also remains finite as $N\rightarrow\infty$.
The argument $t_j$ is defined in (\ref{tj}). 
The endpoint and midpoint Euler-Maclaurin formulas are
\bea
\begin{array}{rcl}
&&\mbox{}\hspace{-.25in}\disp\sum_{k=0}^{m}F(a+kh)=\frac{1}{h}\int_a^b
F(t)\,dt+{1\over 2}[F(b)+F(a)]+\frac{h}{12}
[F'(b)-F'(a)]+\mbox{O}\big(h^2\big)\\
&&\mbox{}\hspace{-.0in}\disp\sum_{k=1}^{m}F\big(a+(k-\half)h\big)=\frac{1}{h}\int_a^b
F(t)\,dt-\frac{h}{24}
[F'(b)-F'(a)]+\mbox{O}\big(h^2\big)
\end{array}
\label{EM}
\eea
where $h=(b-a)/m$. These formulas are valid since $F(t)$ and its first two derivatives are continuous on the
closed interval $[a,b]$. 
Applying these formulas, we find
\bea
  E_{n}(u)&\simeq&2Nf_{bulk}+f_{bdy}\nn
  &+&\frac{2\pi\sin 2u}{N}
\begin{cases}
\displaystyle\
   \Big(\frac{2}{24}+\sum_{j\in{\cal E}_{n}}j
    +\sum_{j\in{\cal M}_{n}}j\Big),&\mbox{$s$ odd}\\[18pt]
\displaystyle\
   \Big(\frac{2}{24}
    -\frac{1}{8}+\sum_{j\in{\cal E}_{n}}(j-\half)
    +\sum_{j\in{\cal M}_{n}}(j-\half)\Big),&\mbox{$s$ even}
    \end{cases}
\label{logLafin}
\eea
In these expressions, the bulk free energy per face is given by
\be
  f_{bulk}=-\frac{1}{\pi}\int_{0}^{\pi/2}\ln(1+\sin t\,\sin 2u)dt
   =\half\ln2-\frac{1}{\pi}\int_{0}^{\pi/2}
   \ln\Big(\frac{1}{\sin t}+\sin 2u\Big)dt
\label{fb}
\ee
whereas the boundary free energy is given by
\bea
f_{bdy}\,=\,\begin{cases}
-2  f_{bulk}-F({\pi\over 2}),&\mbox{$\rho=2r$,\quad\ \ $s$ odd}\\
0,&\mbox{$\rho=2r$,\quad\ \ $s$ even}\\
-4  f_{bulk}-F({\pi\over 2}),&\mbox{$\rho=2r\!-\!1$, $s$ odd}\\
-2  f_{bulk},&\mbox{$\rho=2r\!-\!1$, $s$ even}
\end{cases}
\label{fs}
\eea
with $F({\pi\over 2})=\ln(1+\sin 2u)$.

The link between the lattice model and the characters of the
CFT is governed by the modular nome $q$ defined by (\ref{tau}).

\subsection{Hamiltonian limit}

Since the normalized transfer tangle $\vec d(u)$ reduces to the identity tangle for $u=0$,
we define the Hamiltonian limit of $\vec d(u)$ as the tangle appearing as the 
first-order term in the Taylor series of $\vec d(u)$ centred at $u=0$. We thus consider
\be
  \vec d(u)\, =\, \Ib\, -\, 2u\Hb\, +\, {\cal O}(u^2),\qquad \Hb=-{1\over 2}\,{d\over du}\ln \vec d(u)\Big|_{u=0}
\label{DIH}
\ee
where the factor of $2$ reflects that $\vec d(u)$ is a {\em double}-row transfer tangle.
Employing techniques used in~\cite{PRZ, PR0610}, we find the Hamiltonian
\be
 \vec H\,=\,-\sum_{j=1}^{N-1}e_j+\frac{1}{s(\xi)s(\xi_\rho)}P_N^{(\rho)}
\ee
here written in terms of the linear TL algebra, with loop fugacity $\beta=0$, acting vertically 
on the vector space ${\cal V}^{(N)}_{\rho,s}$ of link states with $N+\rho+s-2$ nodes. The operator
\be
P_N^{(\rho)}=
\begin{cases}
\disp\sum_{k=0}^{(\rho-2)/2}\!(-1)^{(\rho-2k-2)/2}\, e_N^{(2k)},&\rho=2r\\[14pt]
\disp\sum_{k=1}^{(\rho-1)/2}\!(-1)^{(\rho-2k+1)/2}\, e_N^{(2k-1)},&\rho=2r\!-\!1
\end{cases}
\ee
where
\be
e_N^{(k)}= e_Ne_{N+1}\ldots e_{N+k}=\prod_{j=0}^k e_{N+j}
\ee
is the generalized Temperley-Lieb projector discussed in Appendix~\ref{BoundOps}.
The set of normalized eigenvectors of $\Hb$ is the same
as that of $\vec d(u)$. While the eigenvalues of $\Hb$ and $\vec d(u)$ are different,  
the finite-size corrections are still given by (\ref{logLa}) but with the bulk and boundary free energies replaced with their derivatives evaluated at $u=0$, as in (\ref{DIH}), and without the factor $\sin 2u$ in the conformal term. 
The {\em conformal} spectra are thus precisely the same.

\section{Conclusion}
\label{SecConcl}

In this paper, we have solved exactly a model of critical dense 
polymers on strips of arbitrary {\it finite} width. The spectra of this model have been obtained within the framework of Yang-Baxter integrabilty 
by solving exactly a functional equation in the form of an inversion identity satisfied by the double-row transfer matrices. 
The calculations have been carried out with the vacuum boundary condition $(1,1)$ on the left edge of the strip and 
for an infinite family of integrable boundary conditions, labelled by the Kac labels $(r,s)$, on the right edge.
Our study of the physical combinatorics, which is the classification 
of the physical states using the combinatorics of patterns of zeros in the complex $u$-plane, has revealed some 
interesting connections to \mbox{$q$-Narayana} numbers and $q$-Catalan polynomials.
The conformal spectra have been obtained from finite-size corrections using the Euler-Maclaurin formula. 
The results confirm the central charge $c=-2$, the conformal weights 
$\Delta_{r,s}=\frac{(2r-s)^2-1}{8}$ and the form of the conformal partition functions and Kac characters
\bea
Z_{(1,1)|(r,s)}(q)=\chit_{r,s}(q)=\frac{q^{-\frac{c}{24}+\Delta_{r,s}}}{(q)_\infty}(1-q^{rs}),\qquad r,s=1,2,3,\ldots
\eea
Importantly, this analysis fully establishes the infinitely extended Kac table of Figure~1 for ${\cal LM}(1,2)$~\cite{PRZ}. 

We observe that, for $(r,s)$ satisfying $(2r-s)^2<8$, the conformal dimensions also agree with the critical exponents $\beta_{r,s}=(2-\alpha)\Delta_{r,s}=\Delta_{r,s}$ associated with generalized order parameters of the exactly solvable $\varphi_{1,3}$ off-critical perturbation of dense polymers~\cite{OffCrit}. In closing, we further conjecture that the $r$-type boundary conditions for the general ${\cal LM}(p,p')$ models are realized for
\bea
\rho=\rho(r)=\Big\lfloor {r p'\over p}\Big\rfloor
\eea
This coincides with the heights of the groundstates, labelled by $(\rho,\rho+1)$, in the off-critical Corner Transfer Matrix (CTM) calculations of \cite{OffCrit}. Evidence to support this conjecture will be presented elsewhere.
\vskip.5cm
\noindent{\em Acknowledgments}
\vskip.1cm
\noindent
This work is supported by the Australian Research Council (ARC). JR is supported under the ARC Future Fellowship scheme, project number FT100100774. The authors thank Alexi Morin-Duchesne for comments.

\appendix

\section{Construction and Expansion of Boundary Operators}
\label{BoundOps}

In this Appendix, we construct the $r$-type boundary operators $K^{(\rho)}(u,\xi)$ for $\rho-1$ columns. The construction is general giving the $r$-type boundary operators for all of the logarithmic minimal models ${\cal LM}(p,p')$. So, in particular, in this appendix the crossing parameter 
$\lambda$ is not restricted to $\pi/2$ and we use the notation
\bea
s(u)={\sin u\over \sin\lambda},\qquad\quad \lambda={p'-p\over p'}\,\pi
\eea

The boundary operator consists of a planar diagram along with 
a restriction on the link states. The planar boundary diagram is
\psset{unit=1.3cm}
\setlength{\unitlength}{1.3cm}
\be
 K^{(\rho)}(u,\xi)={1\over \eta^{(\rho)}(u,\xi)}
\quad\
\begin{pspicture}[shift=-1.1](0,-.2)(4.5,2.1)
\multirput(0,0)(1,0){4}{\psline[linewidth=1.5pt,linecolor=blue](.5,-.2)(.5,2.2)}
\facegrid{(0,0)}{(4,2)}
\rput(.5,.5){\fws $u\!\!-\!\!\xi_{\rho\!-\!1}$}
\rput(.5,1.5){\fws $-\!u\!\!-\!\!\xi_{\rho\!-\!2}$}
\rput(1.55,.5){\fws $\cdots$}
\rput(1.55,1.5){\fws $\cdots$}
\rput(2.5,.5){\fws $u\!\!-\!\!\xi_2$}
\rput(2.5,1.5){\fws $-\!u\!\!-\!\!\xi_1$}
\rput(3.5,.5){\fws $u\!\!-\!\!\xi_1$}
\rput(3.5,1.5){\fws $-\!u\!\!-\!\!\xi_0$}
\psarc[linewidth=.25pt](0,0){.15}{0}{90}
\psarc[linewidth=.25pt](0,1){.15}{0}{90}
\psarc[linewidth=.25pt](2,0){.15}{0}{90}
\psarc[linewidth=.25pt](2,1){.15}{0}{90}
\psarc[linewidth=.25pt](3,0){.15}{0}{90}
\psarc[linewidth=.25pt](3,1){.15}{0}{90}
\psline[linewidth=1.5pt,linecolor=blue](-0.3,0.5)(0,0.5)
\psline[linewidth=1.5pt,linecolor=blue](-0.3,1.5)(0,1.5)
\rightarc{(4,1)}
\end{pspicture}
\label{Bxi}
\ee
where there are $\rho-1$ columns, the column inhomogeneities are $\xi_j=\xi+j\la$ and $\xi$ is a boundary field. By construction, following \cite{BehrendP} and using the push-through property (\ref{pushthrough}), the boundary operator $K^{(\rho)}(u,\xi)$ is a solution to the boundary Yang-Baxter equation (\ref{BYBE}) for all $\rho$, $u$ and $\xi$
\smallskip
\psset{unit=.85cm}
\begin{align}
\mbox{LHS}\;&=\quad\raisebox{-2.6cm}{\begin{pspicture}(-1.5,-0.5)(12.5,5.5)
\psset{unit=1.35}
\rput(0,1){\DiagWt {a_{\rho}}~~{b_{\rho+1}}{u-v}}
\rput(1,2){\DiagWt {b_{\rho}}~{d_{\rho}}{c_{\rho+1}}{\lambda-u-v}}
\pspolygon[fillstyle=solid,fillcolor=lightlightblue](4,1)(4,2)(6,2)(6,1)
\pspolygon[fillstyle=solid,fillcolor=lightlightblue](4,0)(4,1)(6,1)(6,0)
\rput(2.5,1.5){\face {b_{\rho}}~~{c_{\rho}}{-u\!-\!\xi_{\rho-2}}}
\rput(2.5,0.5){\face {a_{\rho}}~~~{u\!-\!\xi_{\rho-1}}}
\rput(3.5,1.5){\face {b_{\rho-1}}{b_{\rho-2}}~~{-u\!-\!\xi_{\rho-3}}}
\rput(3.5,0.5){\face {a_{\rho-1}}{a_{\rho-2}}~~{u\!-\!\xi_{\rho-2}}}
\rput(6.5,1.5){\face {b_{3}}~~!{-u\!-\!\xi_1}}
\rput(6.5,0.5){\face {a_{3}}~~~{u\!-\!\xi_2}}
\rput(7.5,1.5){\face {b_2}{b_1}{c_1}{c_2}{-u\!-\!\xi_0}}
\rput(7.5,0.5){\face {a_2}{a_1}~~{u\!-\!\xi_{1}}}
\rput(8.5,1){\rtri {a_1}~{c_1}~}
\pspolygon[fillstyle=solid,fillcolor=lightlightblue](4,3)(4,4)(6,4)(6,3)
\pspolygon[fillstyle=solid,fillcolor=lightlightblue](4,2)(4,3)(6,3)(6,2)
\rput(2.5,3.5){\face {d_{\rho}}~~{e_{\rho}}{-v\!-\!\xi_{\rho-2}}}
\rput(2.5,2.5){\face ~~~~{v\!-\!\xi_{\rho-1}}}
\rput(3.5,3.5){\face {d_{\rho-1}}{d_{\rho-2}}{e_{\rho-2}}{e_{\rho-1}}{-v\!-\!\xi_{\rho-3}}}
\rput(3.5,2.5){\face {c_{\rho-1}}{c_{\rho-2}}~~{v\!-\!\xi_{\rho-2}}}
\rput(6.5,3.5){\face {d_{3}}~~{e_{3}}{-v\!-\!\xi_{1}}}
\rput(6.5,2.5){\face {c_{1}}~~~{v\!-\!\xi_2}}
\rput(7.5,3.5){\face {d_2}{d_1}{e_1}{e_2}{-v\!-\!\xi_0}}
\rput(7.5,2.5){\face {c_2}~~~{v\!-\!\xi_1}}
\rput(8.5,3){\rtri ~~{e_1}~}
\psline[linestyle=dashed](8,4)(9,4)
\psline[linestyle=dashed](8,2)(9,2)
\psline[linestyle=dashed](8,0)(9,0)
\psline[linestyle=dashed](0,0)(2,0)
\psline[linestyle=dashed](1,1)(2,1)
\psline[linestyle=dashed](1,3)(2,3)
\end{pspicture}}\\
&=\quad\ \ \raisebox{-2.6cm}{\begin{pspicture}(-2.5,-0.5)(11.5,5.5)
\psset{unit=1.35}
\psline[linestyle=dashed](0,0)(8,0)
\psline[linestyle=dashed](0,1)(8,1)
\psline[linestyle=dashed](0,3)(8,3)
\psline[linestyle=dashed](0,2)(8,2)
\psline[linestyle=dashed](0,4)(8,4)
\rput(6,1){\DiagWt {a_1}~{c_2}~{u-v}}
\rput(7,2){\DiagWt {b_1}~{d_1}{}{\lambda-u-v}}
\pspolygon[fillstyle=solid,fillcolor=lightlightblue](1,1)(1,2)(3,2)(3,1)
\pspolygon[fillstyle=solid,fillcolor=lightlightblue](1,0)(1,1)(3,1)(3,0)
\rput(-0.5,1.5){\face {b_{\rho+1}}{}{}~{u\!-\!\xi_{\rho-1}}}
\rput(-0.5,0.5){\face {a_{\rho}}~~~{v\!-\!\xi_{\rho-1}}}
\rput(0.5,1.5){\face {b_{\rho}}{b_{\rho-1}}~~{u\!-\!\xi_{\rho-2}}}
\rput(0.5,0.5){\face {a_{\rho-1}}{a_{\rho-2}}~~{v\!-\!\xi_{\rho-2}}}
\rput(3.5,1.5){\face {b_{4}}~~!{u\!-\!\xi_2}}
\rput(3.5,0.5){\face {a_{3}}~~~{v\!-\!\xi_2}}
\rput(4.5,1.5){\face {b_{3}}{b_2}{c_2}{}{u\!-\!\xi_1}}
\rput(4.5,0.5){\face {a_2}{a_1}~~{v\!-\!\xi_1}}
\rput(7.5,1){\rtri {a_1}~{c_1}~}
\pspolygon[fillstyle=solid,fillcolor=lightlightblue](1,3)(1,4)(3,4)(3,3)
\pspolygon[fillstyle=solid,fillcolor=lightlightblue](1,2)(1,3)(3,3)(3,2)
\rput(-0.5,3.5){\face {d_{\rho}}{d_{\rho-1}}~{e_{\rho}}{-v\!-\!\xi_{\rho-2}}}
\rput(-0.5,2.5){\face {c_{\rho+1}}~~~{-u\!-\!\xi_{\rho-2}}}
\rput(0.5,3.5){\face {d_{\rho-1}}{d_{\rho-2}}{e_{\rho-2}}{e_{\rho-1}}{-v\!-\!\xi_{\rho-3}}}
\rput(0.5,2.5){\face {c_{\rho}}{c_{\rho-1}}~~{-u\!-\!\xi_{\rho-3}}}
\rput(3.5,3.5){\face {d_{3}}~~{e_{3}}{-v\!-\!\xi_1}}
\rput(3.5,2.5){\face {c_{4}}~~~{-u\!-\!\xi_1}}
\rput(4.5,3.5){\face {d_2}{d_1}{e_1}{e_2}{-v\!-\!\xi_0}}
\rput(4.5,2.5){\face {c_{2}}~~~{-u\!-\!\xi_0}}
\rput(7.5,3){\rtri ~~{e_1}~}
\end{pspicture}}\\
&=\quad\ \ \raisebox{-2.6cm}{\begin{pspicture}(-2.5,-0.5)(11.5,5.5)
\psset{unit=1.35}
\psline[linestyle=dashed](0,0)(8,0)
\psline[linestyle=dashed](0,1)(8,1)
\psline[linestyle=dashed](0,3)(8,3)
\psline[linestyle=dashed](0,2)(8,2)
\psline[linestyle=dashed](0,4)(8,4)
\rput(6,3){\DiagWt {c_2}~{e_1}~{u-v}}
\rput(7,2){\DiagWt {b_1}~{d_1}{}{\lambda-u-v}}
\pspolygon[fillstyle=solid,fillcolor=lightlightblue](1,1)(1,2)(3,2)(3,1)
\pspolygon[fillstyle=solid,fillcolor=lightlightblue](1,0)(1,1)(3,1)(3,0)
\rput(-0.5,1.5){\face {b_{\rho}}{b_{\rho-1}}{c_{\rho-1}}~{u\!-\!\xi_{\rho-1}}}
\rput(-0.5,0.5){\face {a_{\rho}}~~~{v\!-\!\xi_{\rho-1}}}
\rput(0.5,1.5){\face {b_{\rho-1}}{b_{\rho-2}}~~{u\!-\!\xi_{\rho-2}}}
\rput(0.5,0.5){\face {a_{\rho-1}}{a_{\rho-2}}~~{v\!-\!\xi_{\rho-2}}}
\rput(3.5,1.5){\face {b_{3}}~~~{u\!-\!\xi_2}}
\rput(3.5,0.5){\face {a_{3}}~~~{v\!-\!\xi_2}}
\rput(4.5,1.5){\face {b_{2}}{b_1}{c_2}2{u\!-\!\xi_1}}
\rput(4.5,0.5){\face {a_2}{a_1}~~{v\!-\!\xi_1}}
\rput(7.5,1){\rtri {a_1}~{c_1}~}
\pspolygon[fillstyle=solid,fillcolor=lightlightblue](1,3)(1,4)(3,4)(3,3)
\pspolygon[fillstyle=solid,fillcolor=lightlightblue](1,2)(1,3)(3,3)(3,2)
\rput(-0.5,3.5){\face {d_{\rho+1}}{d_{\rho}}~{e_{\rho}}{-v\!-\!\xi_{\rho-2}}}
\rput(-0.5,2.5){\face {c_{\rho+1}}~~~{-u\!-\!\xi_{\rho-2}}}
\rput(0.5,3.5){\face {d_{\rho-1}}{d_{\rho-2}}{e_{\rho-2}}{e_{\rho-1}}{-v\!-\!\xi_{\rho-3}}}
\rput(0.5,2.5){\face {c_{\rho}}{c_{\rho-1}}~~{-u\!-\!\xi_{\rho-3}}}
\rput(3.5,3.5){\face {d_{4}}~~{e_{1}}{-v\!-\!\xi_1}}
\rput(3.5,2.5){\face {c_{4}}~~~{-u\!-\!\xi_1}}
\rput(4.5,3.5){\face {d_3}{d_2}{e_1}{e_2}{-v\!-\!\xi_0}}
\rput(4.5,2.5){\face {c_{3}}~~~{-u\!-\!\xi_0}}
\rput(7.5,3){\rtri ~~{e_1}~}
\end{pspicture}}\\
&=\quad\raisebox{-2.6cm}{\begin{pspicture}(-1.5,-0.5)(12.5,5.5)
\psset{unit=1.35}
\psline[linestyle=dashed](2,0)(9,0)
\psline[linestyle=dashed](1,1)(9,1)
\psline[linestyle=dashed](1,3)(9,3)
\psline[linestyle=dashed](1,2)(9,2)
\psline[linestyle=dashed](0,4)(9,4)
\rput(0,3){\DiagWt ~~{e_{\rho}}{d_{\rho+1}}{u-v}}
\rput(1,2){\DiagWt {b_{\rho}}~{d_{\rho}}{c_{\rho+1}}{\lambda-u-v}}
\pspolygon[fillstyle=solid,fillcolor=lightlightblue](4,1)(4,2)(6,2)(6,1)
\pspolygon[fillstyle=solid,fillcolor=lightlightblue](4,0)(4,1)(6,1)(6,0)
\rput(2.5,1.5){\face {b_{\rho}}{b_{\rho-1}}{c_{\rho-1}}{c_{\rho}}{-v\!-\!\xi_{\rho-1}}}
\rput(2.5,0.5){\face {a_{\rho}}~~~{v\!-\!\xi_{\rho-1}}}
\rput(3.5,1.5){\face {b_{\rho-1}}{b_{\rho-2}}~~{-v\!-\!\xi_{\rho-3}}}
\rput(3.5,0.5){\face {a_{\rho-1}}{a_{\rho-2}}~~{v\!-\!\xi_{\rho-2}}}
\rput(6.5,1.5){\face {b_{3}}{b_{2}}~~{-v\!-\!\xi_3}}
\rput(6.5,0.5){\face {a_{3}}~~~{v\!-\!\xi_2}}
\rput(7.5,1.5){\face {b_{2}}{b_1}{c_1}~{-v\!-\!\xi_0}}
\rput(7.5,0.5){\face {a_2}{a_1}~~{v\!-\!\xi_1}}
\rput(8.5,1){\rtri {a_1}~{c_1}~}
\pspolygon[fillstyle=solid,fillcolor=lightlightblue](4,3)(4,4)(6,4)(6,3)
\pspolygon[fillstyle=solid,fillcolor=lightlightblue](4,2)(4,3)(6,3)(6,2)
\rput(2.5,3.5){\face {d_{\rho}}{d_{\rho-1}}~{e_{\rho}}{-u\!-\!\xi_{\rho-2}}}
\rput(2.5,2.5){\face ~~~~{u\!-\!\xi_{\rho-1}}}
\rput(3.5,3.5){\face {d_{\rho-1}}{d_{\rho-2}}{e_{\rho-2}}{e_{\rho-1}}{-u\!-\!\xi_{\rho-3}}}
\rput(3.5,2.5){\face {c_{\rho-1}}{c_{\rho-2}}~~{u\!-\!\xi_{\rho-2}}}
\rput(6.5,3.5){\face {d_{3}}{d_{2}}~{e_{3}}{-u\!-\!\xi_1}}
\rput(6.5,2.5){\face {c_{3}}~~~{u\!-\!\xi_2}}
\rput(7.5,3.5){\face {d_{2}}{d_1}{e_1}{e_2}{-u\!-\!\xi_0}}
\rput(7.5,2.5){\face {c_2}~~~{u\!-\!\xi_1}}
\rput(8.5,3){\rtri ~~{e_1}~}
\end{pspicture}}=\;\mbox{RHS}
\end{align}
As we will see, the convenient normalization by $\eta^{(\rho)}(u,\xi)$, defined in (\ref{etarho}), ensures that $K^{(\rho)}(0,\xi)=I$.

The boundary diagram (\ref{Bxi}) can be opened up by rotating the lower row of faces anticlockwise by 45 degrees and the upper row of faces clockwise by 45 degrees. The orientation of each upper-row face is changed using the crossing symmetry (\ref{u}). This yields
\psset{unit=1.1cm}
\setlength{\unitlength}{1.1cm}
\bea
 K^{(\rho)}(u,\xi)=\frac{1}{\eta^{(\rho)}(u,\xi)}
 \quad
 \begin{pspicture}[shift=-3.6](0,-3.7)(4,3.8)
\psline[linewidth=1.pt,linecolor=blue](0.35,-3.5)(0.35,3.5)
\psline[linewidth=1.pt,linecolor=blue](1.05,-3.5)(1.05,-3)
\psline[linewidth=1.pt,linecolor=blue](1.05,3)(1.05,3.5)
\psline[linewidth=1.pt,linecolor=blue](1.75,-3.5)(1.75,3.5)
\psline[linewidth=1.pt,linecolor=blue](2.45,-3.5)(2.45,3.5)
\psline[linewidth=1.pt,linecolor=blue](3.15,-3.5)(3.15,3.5)
\pspolygon[linewidth=.25pt,fillstyle=solid,fillcolor=lightlightblue](0,2.8)(2.8,0)(3.5,0.7)(0.7,3.5)(0,2.8)
\pspolygon[linewidth=.25pt,fillstyle=solid,fillcolor=lightlightblue](0,-2.8)(2.8,0)(3.5,-0.7)(0.7,-3.5)(0,-2.8)
\psline[linewidth=.25pt](0.7,2.1)(1.4,2.8)
\psline[linewidth=.25pt](1.4,1.4)(2.1,2.1)
\psline[linewidth=.25pt](2.1,0.7)(2.8,1.4)
\psline[linewidth=.25pt](0.7,-2.1)(1.4,-2.8)
\psline[linewidth=.25pt](1.4,-1.4)(2.1,-2.1)
\psline[linewidth=.25pt](2.1,-0.7)(2.8,-1.4)
\psarc[linewidth=.25pt](0.7,2.1){.15}{45}{135}
\psarc[linewidth=.25pt](2.1,0.7){.15}{45}{135}
\psarc[linewidth=.25pt](2.8,0){.15}{45}{135}
\psarc[linewidth=.25pt](0.7,-3.5){.15}{45}{135}
\psarc[linewidth=.25pt](2.1,-2.1){.15}{45}{135}
\psarc[linewidth=.25pt](2.8,-1.4){.15}{45}{135}
\rput(0.65,0.05){{\color{blue} {\tiny $\bullet$}}}
\rput(1.05,0.05){{\color{blue} {\tiny $\bullet$}}}
\rput(1.45,0.05){{\color{blue} {\tiny $\bullet$}}}
\rput(1.4,-2.12){$\vvdots$}
\rput(1.4,2.23){$\dddots$}
\rput(.7,2.8){\fws $u\!+\!\xi_{\rho\!-\!1}$}
\rput(2.1,1.4){\fws $u\!+\!\xi_2$}
\rput(2.8,0.7){\fws $u\!+\!\xi_1$}
\rput(.7,-2.8){\fws $u\!-\!\xi_{\rho\!-\!1}$}
\rput(2.1,-1.4){\fws $u\!-\!\xi_2$}
\rput(2.8,-0.7){\fws $u\!-\!\xi_1$}
\end{pspicture}
\eea
The boundary operator can therefore be written in the linear (ordinary) TL algebra, generated by the identity $I$ and TL generators $e_j$ acting from the bottom to the top, as
\be
\eta^{(\rho)}(u,\xi) K_0^{(\rho)}(u,\xi)\!=\!X_0(u-\xi_{\rho\!-\!1})X_1(u-\xi_{\rho\!-\!2})\!\cdots\!X_{\rho\!-\!2}(u-\xi_1)
 X_{\rho\!-\!2}(u+\xi_1)X_{\rho\!-\!3}(u+\xi_2)\ldots X_0(u+\xi_{\rho\!-\!1})
\label{K0}
\ee
where the face operators are
\bea
X_j(u)=s(\lambda-u)I+s(u)e_j 
\eea
The index $0$ in $K_0^{(\rho)}(u,\xi)$ indicates that the leftmost strand is in position $j=0$. Note that a closed half-arc is allowed between the nodes in position $0$ and $1$.

The link states are restricted such that closed half-arcs between the rightmost $\rho-1$ nodes are disallowed. 
The set of link states is thus restricted in exactly the same way as in the case of $s$-type boundary conditions,
with $s$ replaced by $\rho$, and therefore give rise to a representation of the TL algebra.
Since this restriction applies to all the $r$-type link states it implies that, in expanding out the planar diagram, we should ignore
connectivity diagrams containing half-arcs between the $\rho-1$ nodes on the lower
edge of (\ref{Bxi}). We thus have the diagram expansion
\psset{unit=0.9cm}
\setlength{\unitlength}{0.9cm}
\bea
&&\hspace{1.3cm} K^{(\rho)}(u,\xi)\, =
\, \al_0^{(\rho)}
\quad
 \begin{pspicture}(0,.9)(4,2.1)
\facegrid{(0,0)}{(4,2)}
\psline[linewidth=1.5pt,linecolor=blue](-0.3,0.5)(0,0.5)
\psline[linewidth=1.5pt,linecolor=blue](-0.3,1.5)(0,1.5)
\psline[linewidth=1.5pt,linecolor=blue](0.5,0)(0.5,2)
\psline[linewidth=1.5pt,linecolor=blue](1.5,0)(1.5,2)
\psline[linewidth=1.5pt,linecolor=blue](2.5,0)(2.5,2)
\psline[linewidth=1.5pt,linecolor=blue](3.5,0)(3.5,2)
\psarc[linewidth=1.5pt,linecolor=blue](-.33,1){.61}{-60}{60}
\end{pspicture}
\ +\al_1^{(\rho)}
\quad
 \begin{pspicture}(0,.9)(4,2.3)
\facegrid{(0,0)}{(4,2)}
\psline[linewidth=1.5pt,linecolor=blue](-0.3,0.5)(0,0.5)
\psline[linewidth=1.5pt,linecolor=blue](-0.3,1.5)(0,1.5)
\psarc[linewidth=1.5pt,linecolor=blue](0,2){.5}{-90}{0}
\psarc[linewidth=1.5pt,linecolor=blue](0,0){.5}{0}{90}
\psline[linewidth=1.5pt,linecolor=blue](1.5,0)(1.5,2)
\psline[linewidth=1.5pt,linecolor=blue](2.5,0)(2.5,2)
\psline[linewidth=1.5pt,linecolor=blue](3.5,0)(3.5,2)
\end{pspicture}
\nonumber
\\
\label{Bexp}
\\[.4cm]
&&\hspace{-.65cm}
+\ \al_2^{(\rho)}
\quad
 \begin{pspicture}(0,.9)(4,2.3)
\facegrid{(0,0)}{(4,2)}
\psline[linewidth=1.5pt,linecolor=blue](-0.3,0.5)(0,0.5)
\psline[linewidth=1.5pt,linecolor=blue](-0.3,1.5)(0,1.5)
\psline[linewidth=1.5pt,linecolor=blue](2.5,0)(2.5,2)
\psline[linewidth=1.5pt,linecolor=blue](3.5,0)(3.5,2)
\rput(0,0){\loopb}
\rput(0,1){\loopb}
\psarc[linewidth=1.5pt,linecolor=blue](1,0){.5}{0}{90}
\psarc[linewidth=1.5pt,linecolor=blue](1,2){.5}{-90}{0}
\end{pspicture}
\ +\al_3^{(\rho)}
\quad
 \begin{pspicture}(0,.9)(4,2.3)
\facegrid{(0,0)}{(4,2)}
\psline[linewidth=1.5pt,linecolor=blue](-0.3,0.5)(0,0.5)
\psline[linewidth=1.5pt,linecolor=blue](-0.3,1.5)(0,1.5)
\psline[linewidth=1.5pt,linecolor=blue](3.5,0)(3.5,2)
\rput(0,0){\loopb}
\rput(0,1){\loopb}
\rput(1,0){\loopb}
\rput(1,1){\loopb}
\psarc[linewidth=1.5pt,linecolor=blue](2,0){.5}{0}{90}
\psarc[linewidth=1.5pt,linecolor=blue](2,2){.5}{-90}{0}
\end{pspicture}
\ +\ldots +\al_{\rho-1}^{(\rho)}
\quad
 \begin{pspicture}(0,.9)(4,2.3)
\facegrid{(0,0)}{(4,2)}
\psline[linewidth=1.5pt,linecolor=blue](-0.3,0.5)(0,0.5)
\psline[linewidth=1.5pt,linecolor=blue](-0.3,1.5)(0,1.5)
\rput(0,0){\loopb}
\rput(0,1){\loopb}
\rput(1,0){\loopb}
\rput(1,1){\loopb}
\rput(2,0){\loopb}
\rput(2,1){\loopb}
\psarc[linewidth=1.5pt,linecolor=blue](3,0){.5}{0}{90}
\psarc[linewidth=1.5pt,linecolor=blue](3,2){.5}{-90}{0}
\end{pspicture}
\nonumber
\\[.3cm]
\nonumber
\eea
The first diagram acts as the identity on the $\rho-1$ vertical strands. Opening up the other diagrams, 
we see that they coincide with the TL words $e_0e_1\ldots e_k$ for $k=0,1,\ldots,\rho-2$. 
All other TL words are killed by the restriction on link states. 
The coefficients $\al_i^{(\rho)}=\al_i^{(\rho)}(u)$ are straightforward to determine using
this {\em linear} TL algebra realization of the boundary diagrams.
Simplifying the boundary operator 
by implementing the condition on the link states yields the following proposition.\\[10pt]
{\bf Proposition:}\quad 
{\em Implementing the restriction on link states,  the boundary operator (\ref{K0}) simplifies to:}
\be
 K_0^{(\rho)}(u,\xi)= I+\frac{s(2u)}{s(u+\xi)s(u-\xi_{\rho})}\,P_0^{(\rho)}
\label{K0I}
\ee
{\em The generalized Temperley-Lieb projectors are}
\be
 P_0^{(\rho)}=\sum_{k=0}^{\rho-2}(-1)^k s\big((\rho-k-1)\lambda\big) e_0^{(k)},\qquad
 e_0^{(k)}= e_0e_1\ldots e_k=\prod_{j=0}^k e_j
\ee
{\em where $s\big((\rho+1)\lambda\big)=U_\rho(\beta/2)$ are Chebyshev polynomials of the second kind of order $\rho$ with $\beta=2\cos\lambda$.}\\[10pt]
{\bf Proof:}\quad For one column, that is $\rho=2$, the result follows immediately from the identity
\be
 X_j(u-v)X_j(u+v)=s(u-v-\lambda)s(u+v-\lambda)I+s(2u)e_j
\ee
with $v=\xi_1$. For $\rho>2$, we proceed by induction in $\rho$ using
\be
 K_0^{(\rho+1)}(u,\xi)=\frac{\eta^{(\rho)}(u,\xi)}{\eta^{(\rho+1)}(u,\xi)}\,
   X_0(u-\xi_{\rho})K_1^{(\rho)}(u,\xi)X_0(u+\xi_{\rho})
\ee
with
\bea
{\eta^{(\rho+1)}(u,\xi)\over \eta^{(\rho)}(u,\xi)}=s(u+\xi_{\rho-1})s(u-\xi_{\rho+1})
\eea
\be
 K_1^{(\rho)}(u,\xi)= I+\frac{s(2u)}{s(u+\xi)s(u-\xi_{\rho})}\,P_1^{(\rho)},\qquad
 P_1^{(\rho)}=\sum_{k=1}^{\rho-1}(-1)^{k-1}s\big((\rho-k)\lambda\big)
  e_1\ldots e_k 
\ee
Explicitly,
\bea
&\!\!\!\!\!\!\!&s(u\!+\!\xi)s(u\!-\!\xi_\rho)\,{\eta^{(\rho+1)}(u,\xi)\over \eta^{(\rho)}(u,\xi)}\,K_0^{(\rho+1)}(u,\xi)
\,=\,s(u\!+\!\xi)s(u\!-\!\xi_\rho) X_0(u\!-\!\xi_\rho)K_1^{(\rho)}(u,\xi)X_0(u\!+\!\xi_\rho)\nonumber\\
&\!\!=\!\!&s(u\!+\!\xi)s(u\!-\!\xi_\rho)X_0(u\!-\!\xi_\rho)X_0(u\!+\!\xi_\rho)
+s(2u)X_0(u\!-\!\xi_\rho)P_1^{(\rho)}X_0(u\!+\!\xi_\rho)\qquad\nonumber\\[4pt]
&\!\!=\!\!&s(u\!+\!\xi)s(u\!-\!\xi_\rho)\big[s(u\!+\!\xi_{\rho-1})s(u\!-\!\xi_{\rho+1})I+s(2u)e_0\big]\\
&&\mbox{}+s(2u)s(u\!-\!\xi_\rho)s(u\!+\!\xi_{\rho-1})
[P_0^{(\rho+1)}\!-\!U_{\rho-1}(\beta/2)e_0]
+s(2u)s(u\!-\!\xi_\rho)s(u\!+\!\xi_\rho)U_{\rho-2}(\beta/2)e_0\nonumber\\
&\!\!=\!\!&s(u+\xi_{\rho-1})s(u-\xi_\rho) [s(u\!+\!\xi)s(u\!-\!\xi_{\rho+1})I\!+\!s(2u) P_0^{(\rho+1)}]\nonumber
\eea
where all terms involving $e_0$ cancel out. The term proportional to $P_1^{(\rho)}$ is discarded since it is killed by the restriction on link states. To simplify some of the products involving the generalized TL projectors, we used (\ref{e00ID}) and (\ref{e0ID}) below.  $\Box$
\medskip

The first few generalized TL projectors are
\bea
P_j^{(2)}=e_j,\qquad P_j^{(3)}=\beta e_j\!-\!e_je_{j+1},\qquad P_j^{(4)}=(\beta^2\!-\!1)e_j\!-\!\beta e_je_{j+1}\!+\!e_je_{j+1}e_{j+2} 
\label{genTLproj}
\eea
These generalized projectors satisfy the properties
\bea
&P_j^{(\rho')}\,P_j^{(\rho)}=U_{\rho-1}(\beta/2)\,P_j^{(\rho)}&\label{e00ID}\\[4pt]
&P_1^{(\rho)}e_0=0,\qquad e_0 P_1^{(\rho)}=U_{\rho-1}(\beta/2) e_0-P_0^{(\rho+1)},\qquad e_0\,P_1^{(\rho)}e_0=U_{\rho-2}(\beta/2)e_0&
\label{e0ID}
\eea
Although we call the $P_j^{(\rho)}$ projectors, strictly speaking they can only be properly normalized in the case $U_{\rho-1}(\beta/2)\ne 0$, that is, $\rho$ not a multiple of $p'$. \\[10pt]
{\bf Corollary:}\quad 
{\em The expansion coefficients $\al_i^{(\rho)}$ appearing in (\ref{Bexp}) are given by}
\be
 \al^{(\rho)}_0=1,\qquad
 \al^{(\rho)}_i=\frac{(-1)^{i-1}s\big((\rho-i)\lambda\big)s(2u)}{s(u+\xi)s(u-\xi_{\rho})}
  =\frac{(-1)^{i-1}U_{\rho-i-1}(\beta/2)s(2u)}{s(u+\xi)s(u-\xi_{\rho})},\qquad i=1,\ldots,\rho-1
\ee
This follows since opening up the planar diagrams on the right side of (\ref{Bexp}) for $i=k+1=1,\ldots,\rho-1$ gives the ordered TL products $e_0^{(k)}$. This also establishes that $K_0^{(\rho)}(0,\xi)=I$.

\section{Proof of the Inversion Identity}
\label{InvIdProof}

In this appendix, we prove the inversion identity (\ref{DDI}) in the planar algebra. Diagrammatically, the left side of (\ref{DDI}) corresponds to
\psset{unit=0.9cm}
\setlength{\unitlength}{0.9cm}
\be
\vec D(u)\vec D(u+\lambda)
\;=
\qquad 
\begin{pspicture}[shift=-2.](0,-.1)(5.5,4.1)
\facegrid{(0,0)}{(5,4)}
\put(3,0){\color{lightpurple}\rule{2\unitlength}{4\unitlength}}
\leftarc{(0,1)}
\leftarc{(0,3)}
\rightarc{(5,1)}
\rightarc{(5,3)}
\rput(1.5,.5){\scriptsize row 1}
\rput(1.5,1.5){\scriptsize row 2}
\rput(1.5,2.5){\scriptsize row 3}
\rput(1.5,3.5){\scriptsize row 4}
\psline[linewidth=1.5pt](3,0)(3,4)
\rput(3,0){\loopa}
\rput(3,1){\loopb}
\rput(3,2){\loopa}
\rput(3,3){\loopb}
\rput(4,0){\loopa}
\rput(4,1){\loopb}
\rput(4,2){\loopa}
\rput(4,3){\loopb}
\rput(1.5,-.3){$\underbrace{\rule{3\unitlength}{0pt}}_{N+\rho-1}$}
\rput(4,-.3){$\underbrace{\rule{2\unitlength}{0pt}}_{s-1}$}
\end{pspicture}
\label{DDD1}
\ee
where $\lambda=\tfrac{\pi}{2}$ and
\psset{unit=1.07cm}
\setlength{\unitlength}{1.07cm}
\be
\begin{pspicture}(0,1.9)(4,4)
\facegrid{(0,0)}{(3,4)}
\leftarc{(0,1)}
\leftarc{(0,3)}
\rightarc{(3,1)}
\rightarc{(3,3)}
\rput(1.5,.5){\scriptsize row 1}
\rput(1.5,1.5){\scriptsize row 2}
\rput(1.5,2.5){\scriptsize row 3}
\rput(1.5,3.5){\scriptsize row 4}
\rput(1.5,-.3){$\underbrace{\rule{3\unitlength}{0pt}}_{N+\rho-1}$}
\end{pspicture}
=
\qquad\ \,
\begin{pspicture}(0,1.95)(6.5,4)
\facegrid{(0,0)}{(6,4)}
\leftarc{(0,1)}
\leftarc{(0,3)}
\rightarc{(6,1)}
\rightarc{(6,3)}
\psline[linewidth=1.5pt](3,0)(3,4)
\rput(0.5,.5){\fws $u$}
\rput(0.5,1.5){\fws $\lambda\!-\!u$}
\rput(0.5,2.5){\fws $\lambda\!+\!u$}
\rput(0.5,3.5){\fws $-u$}
\rput(1.55,.5){\fws $\cdots$}
\rput(1.55,1.5){\fws $\cdots$}
\rput(1.55,2.5){\fws $\cdots$}
\rput(1.55,3.5){\fws $\cdots$}
\rput(2.5,.5){\fws $u$}
\rput(2.5,1.5){\fws $\lambda\!-\!u$}
\rput(2.5,2.5){\fws $\lambda\!+\!u$}
\rput(2.5,3.5){\fws $-u$}
\rput(3.5,.5){\scriptsize $u\!-\!\xi_{\rho\!-\!1}$}
\rput(3.5,1.5){\scriptsize $-\!u\!\!-\!\!\xi_{\rho\!-\!2}$}
\rput(3.5,2.5){\scriptsize $u\!-\!\xi_{\rho\!-\!2}$}
\rput(3.5,3.5){\scriptsize $-\!u\!\!-\!\!\xi_{\rho\!-\!1}$}
\rput(4.55,.5){\fws $\cdots$}
\rput(4.55,1.5){\fws $\cdots$}
\rput(4.55,2.5){\fws $\cdots$}
\rput(4.55,3.5){\fws $\cdots$}
\rput(5.5,.5){\footnotesize $u\!-\!\xi_1$}
\rput(5.5,1.5){\footnotesize $-\!u\!\!-\!\!\xi_0$}
\rput(5.5,2.5){\footnotesize $u\!-\!\xi_0$}
\rput(5.5,3.5){\footnotesize $-\!u\!\!-\!\!\xi_1$}
\rput(1.5,-.3){$\underbrace{\rule{3\unitlength}{0pt}}_{N}$}
\rput(4.5,-.3){$\underbrace{\rule{3\unitlength}{0pt}}_{\rho-1}$}
\end{pspicture}
\label{DDD2}
\\[2.6cm]
\ee
Here and in the following, all unmarked face operators are meant to have their orientations marked in the bottom-left corner as in (\ref{D}). The total number of columns in (\ref{DDD2}) is $N+\rho-1$ and remains unchanged throughout the diagrammatic manipulations below. There are $N+\rho+s-2$ columns in (\ref{DDD1}), but the additional $s-1$ columns are trivial and will be ignored in the following.
It is also noted that, according to the discussion following the push-through property (\ref{pushthrough}), we do not need to worry about the restriction rule along the horizontal interface between $\vec D(u)$ and $\vec D(u+\lambda)$.

Now, following the proof of the inversion identity in \cite{PR0610},
we use the local inversion relation (\ref{Inv}) with $v=2u$ 
to insert a pair of faces with spectral parameters $\pm 2u$
\psset{unit=0.9cm}
\setlength{\unitlength}{0.9cm}
\bea
\begin{pspicture}(0,1.9)(4,4)
\facegrid{(0,0)}{(3,4)}
\leftarc{(0,1)}
\leftarc{(0,3)}
\rightarc{(3,1)}
\rightarc{(3,3)}
\rput(1.5,.5){\scriptsize row 1}
\rput(1.5,1.5){\scriptsize row 2}
\rput(1.5,2.5){\scriptsize row 3}
\rput(1.5,3.5){\scriptsize row 4}
\end{pspicture}
\!\!\!\!\!&\!\!\!=&\!\!\frac{1}{\cos^2 2u}\quad\ \
\begin{pspicture}(0,1.9)(10,4)
\facegrid{(0,0)}{(3,4)}
\leftarc{(0,1)}
\leftarc{(0,3)}
\rput(1.5,.5){\scriptsize row 1}
\rput(1.5,1.5){\scriptsize row 2}
\rput(1.5,2.5){\scriptsize row 3}
\rput(1.5,3.5){\scriptsize row 4}
\pspolygon[fillstyle=solid,fillcolor=lightlightblue](3,2)(4,1)(5,2)(4,3)(3,2)
\pspolygon[fillstyle=solid,fillcolor=lightlightblue](5,2)(6,1)(7,2)(6,3)(5,2)
\psline[linecolor=blue,linewidth=1.5pt](3,3.5)(7,3.5)
\psline[linecolor=blue,linewidth=1.5pt](3,0.5)(7,0.5)
\psline[linecolor=blue,linewidth=1.5pt](3,2.5)(3.5,2.5)
\psline[linecolor=blue,linewidth=1.5pt](3,1.5)(3.5,1.5)
\psline[linecolor=blue,linewidth=1.5pt](6.5,2.5)(7,2.5)
\psline[linecolor=blue,linewidth=1.5pt](6.5,1.5)(7,1.5)
\psarc[linecolor=blue,linewidth=1.5pt](5,2){0.71}{225}{315}
\psarc[linecolor=blue,linewidth=1.5pt](5,2){0.71}{45}{135}
\rput(4,2){\fws $2u$}
\rput(6,2){\fws $-2u$}
\psarc[linewidth=.25pt](3,2){.2}{-45}{45}
\psarc[linewidth=.25pt](5,2){.2}{-45}{45}
\facegrid{(7,0)}{(10,4)}
\rput(8.5,.5){\scriptsize row 1}
\rput(8.5,1.5){\scriptsize row 2}
\rput(8.5,2.5){\scriptsize row 3}
\rput(8.5,3.5){\scriptsize row 4}
\rightarc{(10,1)}
\rightarc{(10,3)}
\end{pspicture}
\nonumber
\\[2cm]
&\!\!\!=&\!\!\frac{1}{\cos^2 2u}\quad\ \
\begin{pspicture}(0.3,1.9)(7,4)
\facegrid{(2,0)}{(5,4)}
\leftarc{(0.5,1)}
\leftarc{(0.5,3)}
\pspolygon[fillstyle=solid,fillcolor=lightlightblue](0,2)(1,1)(2,2)(1,3)(0,2)
\rput(1,2){\fws $2u$}
\psarc[linewidth=.25pt](0,2){.2}{-45}{45}
\psline[linecolor=blue,linewidth=1.5pt](0.5,3.5)(2,3.5)
\psline[linecolor=blue,linewidth=1.5pt](0.5,0.5)(2,0.5)
\psline[linecolor=blue,linewidth=1.5pt](1.5,2.5)(2,2.5)
\psline[linecolor=blue,linewidth=1.5pt](1.5,1.5)(2,1.5)
\rput(3.5,.5){\scriptsize row 1}
\rput(3.5,1.5){\scriptsize row 3}
\rput(3.5,2.5){\scriptsize row 2}
\rput(3.5,3.5){\scriptsize row 4}
\pspolygon[fillstyle=solid,fillcolor=lightlightblue](5,2)(6,1)(7,2)(6,3)(5,2)
\psline[linecolor=blue,linewidth=1.5pt](5,2.5)(5.5,2.5)
\psline[linecolor=blue,linewidth=1.5pt](5,1.5)(5.5,1.5)
\psline[linecolor=blue,linewidth=1.5pt](5,3.5)(6.5,3.5)
\psline[linecolor=blue,linewidth=1.5pt](5,0.5)(6.5,0.5)
\rput(6,2){\fws $-2u$}
\psarc[linewidth=.25pt](5,2){.2}{-45}{45}
\rightarc{(6.5,1)}
\rightarc{(6.5,3)}
\end{pspicture}
\label{DDcosDD}
\\[1.1cm]
\nonumber
\eea
In the last rewriting, we have applied the YBE (\ref{YB}) repeatedly to bring the faces with spectral parameters $\pm2u$ to the left and right edges, in the process interchanging the two middle rows. 
Expanding these faces using (\ref{u}) then transforms the planar diagram in (\ref{DDcosDD}) into
\psset{unit=0.9cm}
\setlength{\unitlength}{0.9cm}
\bea
&-&\!\!\displaystyle{\frac{\sin^2 2u}{\cos^2 2u}}
\begin{pspicture}(-.5,1.9)(5.5,4.1)
\psellipse[linecolor=blue,linewidth=1.5pt](1,2)(1.2,1.53)
\psellipse[linecolor=blue,linewidth=1.5pt](4,2)(1.2,1.53)
\leftarc{(1,2)}
\rightarc{(4,2)}
\facegrid{(1,0)}{(4,4)}
\rput(2.5,.5){\scriptsize row 1}
\rput(2.5,1.5){\scriptsize row 3}
\rput(2.5,2.5){\scriptsize row 2}
\rput(2.5,3.5){\scriptsize row 4}
\end{pspicture}
+\,\displaystyle{\frac{\sin 2u\cos 2u}{\cos^2 2u}}
\begin{pspicture}(-.5,1.9)(5,4)
\psellipse[linecolor=blue,linewidth=1.5pt](1,2)(1.2,1.53)
\leftarc{(1,2)}
\rightarc{(4,1)}
\rightarc{(4,3)}
\facegrid{(1,0)}{(4,4)}
\rput(2.5,.5){\scriptsize row 1}
\rput(2.5,1.5){\scriptsize row 3}
\rput(2.5,2.5){\scriptsize row 2}
\rput(2.5,3.5){\scriptsize row 4}
\end{pspicture}
\nonumber
\\[2cm]
&-&\!\!\displaystyle{\frac{\cos 2u\sin 2u}{\cos^2 2u}}
\begin{pspicture}(0,1.9)(5.5,4.1)
\psellipse[linecolor=blue,linewidth=1.5pt](4,2)(1.2,1.53)
\leftarc{(1,1)}
\leftarc{(1,3)}
\rightarc{(4,2)}
\facegrid{(1,0)}{(4,4)}
\rput(2.5,.5){\scriptsize row 1}
\rput(2.5,1.5){\scriptsize row 3}
\rput(2.5,2.5){\scriptsize row 2}
\rput(2.5,3.5){\scriptsize row 4}
\end{pspicture}
+\,\displaystyle{\frac{\cos^2 2u}{\cos^2 2u}}
\begin{pspicture}(0,1.9)(4,4)
\leftarc{(1,1)}
\leftarc{(1,3)}
\rightarc{(4,1)}
\rightarc{(4,3)}
\facegrid{(1,0)}{(4,4)}
\rput(2.5,.5){\scriptsize row 1}
\rput(2.5,1.5){\scriptsize row 3}
\rput(2.5,2.5){\scriptsize row 2}
\rput(2.5,3.5){\scriptsize row 4}
\end{pspicture}
\label{DDDexp}
\\[1.1cm]
\nonumber
\eea
We now eliminate the last three diagrams by examining the consequences of having
a half-arc connecting the two left edges (or two right edges) of a two-column 
appearing in the two lower or two upper rows. Expanding in terms of connectivities, we thus
use that
\psset{unit=0.9cm}
\setlength{\unitlength}{0.9cm}
\be
\begin{pspicture}(0,.9)(1.2,2.2)
\leftarc{(0,1)}
\facegrid{(0,0)}{(1,2)}
\rput(.5,1.5){\small $\lambda\!+\!\eta$}
\rput(.5,.5){\small $\eta$}
\end{pspicture}
=-\sin^2 \!\eta
\begin{pspicture}(-0.8,.9)(1.2,2.2)
\leftarc{(0,1)}
\facegrid{(0,0)}{(1,2)}
\put(0,0){\loopb}
\put(0,1){\loopa}
\end{pspicture},\qquad\qquad
\begin{pspicture}(0,.9)(1.2,2.2)
\leftarc{(0,1)}
\facegrid{(0,0)}{(1,2)}
\rput(.5,1.5){\small $\eta$}
\rput(.5,.5){\small $\lambda\!+\!\eta$}
\end{pspicture}
=\cos^2 \!\eta
\begin{pspicture}(-0.8,.9)(1.2,2.2)
\leftarc{(0,1)}
\facegrid{(0,0)}{(1,2)}
\put(0,0){\loopb}
\put(0,1){\loopa}
\end{pspicture}
\\[.5cm]
\ee
and
\be
\begin{pspicture}(0,.9)(1.2,2.2)
\rightarc{(1,1)}
\facegrid{(0,0)}{(1,2)}
\rput(.5,1.5){\small $\lambda\!+\!\eta$}
\rput(.5,.5){\small $\eta$}
\end{pspicture}
\quad
=\cos^2 \!\eta
\begin{pspicture}(-0.3,.9)(1.2,2.2)
\rightarc{(1,1)}
\facegrid{(0,0)}{(1,2)}
\put(0,0){\loopa}
\put(0,1){\loopb}
\end{pspicture}\quad,\qquad\qquad
\begin{pspicture}(0,.9)(1.2,2.2)
\rightarc{(1,1)}
\facegrid{(0,0)}{(1,2)}
\rput(.5,1.5){\small $\eta$}
\rput(.5,.5){\small $\lambda\!+\!\eta$}
\end{pspicture}
\quad
=-\sin^2 \!\eta
\begin{pspicture}(-0.3,.9)(1.2,2.2)
\rightarc{(1,1)}
\facegrid{(0,0)}{(1,2)}
\put(0,0){\loopa}
\put(0,1){\loopb}
\end{pspicture}
\\[.8cm]
\ee
imply that such a half-arc will propagate. In all three diagrams under consideration, 
it is evident that this propagating property eventually will lead to a closed loop at the left or right edge, thereby yielding a vanishing contribution.
The surviving, top-left diagram in (\ref{DDDexp}) may be expanded as
\psset{unit=0.7cm}
\setlength{\unitlength}{0.7cm}
\bea
\begin{pspicture}(0.2,1.9)(5.5,4.1)
\psellipse[linecolor=blue,linewidth=1.5pt](1,2)(1.2,1.53)
\psellipse[linecolor=blue,linewidth=1.5pt](4,2)(1.2,1.53)
\leftarc{(1,2)}
\rightarc{(4,2)}
\facegrid{(1,0)}{(4,4)}
\rput(2.5,.5){\tiny row 1}
\rput(2.5,1.5){\tiny row 3}
\rput(2.5,2.5){\tiny row 2}
\rput(2.5,3.5){\tiny row 4}
\end{pspicture}
\!&\!\!\!=&\!\!-\sin^2\!u
\begin{pspicture}(-.5,1.9)(5.5,4.1)
\psellipse[linecolor=blue,linewidth=1.5pt](1,2)(1.2,1.53)
\psellipse[linecolor=blue,linewidth=1.5pt](4,2)(1.2,1.53)
\leftarc{(1,2)}
\rightarc{(4,2)}
\facegrid{(1,0)}{(4,4)}
\rput(1,2){\loopa}
\rput(1,3){\loopb}
\rput(2.5,.5){\tiny row 1}
\rput(2.5,1.5){\tiny row 3}
\rput(2.5,2.5){\tiny row 2}
\rput(2.5,3.5){\tiny row 4}
\end{pspicture}
+\cos^2\!u
\begin{pspicture}(-.5,1.9)(5.5,4.1)
\psellipse[linecolor=blue,linewidth=1.5pt](1,2)(1.2,1.53)
\psellipse[linecolor=blue,linewidth=1.5pt](4,2)(1.2,1.53)
\leftarc{(1,2)}
\rightarc{(4,2)}
\facegrid{(1,0)}{(4,4)}
\rput(1,2){\loopb}
\rput(1,3){\loopa}
\rput(2.5,.5){\tiny row 1}
\rput(2.5,1.5){\tiny row 3}
\rput(2.5,2.5){\tiny row 2}
\rput(2.5,3.5){\tiny row 4}
\end{pspicture}
\nonumber
\\[2cm]
\!&\!\!\!+&\!\! \g\left(\!\!
\begin{pspicture}(-.5,1.9)(5.5,4.1)
\psellipse[linecolor=blue,linewidth=1.5pt](1,2)(1.2,1.53)
\psellipse[linecolor=blue,linewidth=1.5pt](4,2)(1.2,1.53)
\leftarc{(1,2)}
\rightarc{(4,2)}
\facegrid{(1,0)}{(4,4)}
\rput(1,2){\loopa}
\rput(1,3){\loopa}
\rput(2.5,.5){\tiny row 1}
\rput(2.5,1.5){\tiny row 3}
\rput(2.5,2.5){\tiny row 2}
\rput(2.5,3.5){\tiny row 4}
\end{pspicture}
-
\begin{pspicture}(-.5,1.9)(5.5,4.1)
\psellipse[linecolor=blue,linewidth=1.5pt](1,2)(1.2,1.53)
\psellipse[linecolor=blue,linewidth=1.5pt](4,2)(1.2,1.53)
\leftarc{(1,2)}
\rightarc{(4,2)}
\facegrid{(1,0)}{(4,4)}
\rput(1,2){\loopb}
\rput(1,3){\loopb}
\rput(2.5,.5){\tiny row 1}
\rput(2.5,1.5){\tiny row 3}
\rput(2.5,2.5){\tiny row 2}
\rput(2.5,3.5){\tiny row 4}
\end{pspicture}
\!\!\right)
\label{DDDelephant}
\eea
where $\g=\cos u\sin u$.
Again due to the propagating property of half-arcs, the first two diagrams are readily seen to be proportional to the vertical
identity diagram $\Ib$:
\psset{unit=0.7cm}
\setlength{\unitlength}{0.7cm}
\bea
 -\sin^2\!u
\begin{pspicture}(-.5,1.9)(5.1,4.1)
\psellipse[linecolor=blue,linewidth=1.5pt](1,2)(1.2,1.53)
\psellipse[linecolor=blue,linewidth=1.5pt](4,2)(1.2,1.53)
\leftarc{(1,2)}
\rightarc{(4,2)}
\facegrid{(1,0)}{(4,4)}
\rput(1,2){\loopa}
\rput(1,3){\loopb}
\rput(2.5,.5){\tiny row 1}
\rput(2.5,1.5){\tiny row 3}
\rput(2.5,2.5){\tiny row 2}
\rput(2.5,3.5){\tiny row 4}
\end{pspicture}
&=&\!\!
\sin^{4N}\!u\;[\eta^{(\rho)}(u+\la,\xi)]^2\;
\begin{pspicture}(-1.3,1.9)(5.6,4)
\psellipse[linecolor=blue,linewidth=1.5pt](0,2)(1.2,1.53)
\psellipse[linecolor=blue,linewidth=1.5pt](3,2)(1.2,1.53)
\facegrid{(0,0)}{(3,4)}
\leftarc{(0,2)}
\rightarc{(3,2)}
\multiput(0,0)(1,0){3}{\loopb}
\multiput(0,1)(1,0){3}{\loopa}
\multiput(0,2)(1,0){3}{\loopa}
\multiput(0,3)(1,0){3}{\loopb}
\rput(1.5,-.4){$\underbrace{\rule{3\unitlength}{0pt}}_{N+\rho-1}$}
\end{pspicture}
\\[1.7cm]
\cos^2\!u
\begin{pspicture}(-.5,1.9)(5.1,4.1)
\psellipse[linecolor=blue,linewidth=1.5pt](1,2)(1.2,1.53)
\psellipse[linecolor=blue,linewidth=1.5pt](4,2)(1.2,1.53)
\leftarc{(1,2)}
\rightarc{(4,2)}
\facegrid{(1,0)}{(4,4)}
\rput(1,2){\loopb}
\rput(1,3){\loopa}
\rput(2.5,.5){\tiny row 1}
\rput(2.5,1.5){\tiny row 3}
\rput(2.5,2.5){\tiny row 2}
\rput(2.5,3.5){\tiny row 4}
\end{pspicture}
&=&\!\!
\cos^{4N}\! u\;[\eta^{(\rho)}(u,\xi)]^2\;
\begin{pspicture}(-1.3,1.9)(5.6,4)
\psellipse[linecolor=blue,linewidth=1.5pt](0,2)(1.2,1.53)
\psellipse[linecolor=blue,linewidth=1.5pt](3,2)(1.2,1.53)
\facegrid{(0,0)}{(3,4)}
\leftarc{(0,2)}
\rightarc{(3,2)}
\multiput(0,0)(1,0){3}{\loopa}
\multiput(0,1)(1,0){3}{\loopb}
\multiput(0,2)(1,0){3}{\loopb}
\multiput(0,3)(1,0){3}{\loopa}
\rput(1.5,-.4){$\underbrace{\rule{3\unitlength}{0pt}}_{N+\rho-1}$}
\end{pspicture}
\\[0.8cm]
\nonumber
\eea
where it is noted that the function $\eta^{(\rho)}(u,\xi)$ defined in (\ref{etarho}) satisfies
\be
 \eta^{(\rho)}(u+\la,\xi)=\prod_{j=1}^{\rho-1}\sin(u+\xi_j)\sin(u-\xi_j)
\ee
To analyse the contribution in brackets in (\ref{DDDelephant}), we introduce the 3-tangle
\psset{unit=0.7cm}
\setlength{\unitlength}{0.7cm}
\be
%
\begin{pspicture}(0,0.9)(1,2)
\pspolygon[fillstyle=solid,fillcolor=lightgray](0,0)(1,0)(1,2)(0,2)(0,0)
\psarc[linewidth=.25pt](0,0){.2}{0}{90}
\end{pspicture}
\ =\
\begin{pspicture}(0,0.9)(1,2)
\facegrid{(0,0)}{(1,2)}
\rput(0,0){\loopa}
\rput(0,1){\loopa}
\end{pspicture}
\ -\
\begin{pspicture}(0,0.9)(1,2)
\facegrid{(0,0)}{(1,2)}
\rput(0,0){\loopb}
\rput(0,1){\loopb}
\end{pspicture}
\ ,\qquad\qquad
\begin{pspicture}(-0.5,0.9)(1,2)
\pspolygon[fillstyle=solid,fillcolor=lightgray](0,0)(1,0)(1,2)(0,2)(0,0)
\psarc[linewidth=.25pt](0,0){.2}{0}{90}
\leftarc{(0,1)}
\end{pspicture}
\ =\
\begin{pspicture}(0,0.9)(1.5,2)
\pspolygon[fillstyle=solid,fillcolor=lightgray](0,0)(1,0)(1,2)(0,2)(0,0)
\psarc[linewidth=.25pt](0,0){.2}{0}{90}
\rightarc{(1,1)}
\end{pspicture}
\ =\, 0
\label{gray2}
\\[.4cm]
\ee
Due to the vanishing conditions in (\ref{gray2}) and the propagating property once again, we see that
\psset{unit=0.7cm}
\setlength{\unitlength}{0.7cm}
\bea
&&
\g\qquad\
\begin{pspicture}(0,1.9)(5,4)
\psellipse[linecolor=blue,linewidth=1.5pt](0,2)(1.2,1.53)
\psellipse[linecolor=blue,linewidth=1.5pt](4,2)(1.2,1.53)
\facegrid{(0,0)}{(4,4)}
\rput(2.5,.5){\tiny row 1}
\rput(2.5,1.5){\tiny row 3}
\rput(2.5,2.5){\tiny row 2}
\rput(2.5,3.5){\tiny row 4}
\pspolygon[fillstyle=solid,fillcolor=lightgray](0,2)(1,2)(1,4)(0,4)(0,2)
\psarc[linewidth=.25pt](0,2){.2}{0}{90}
\leftarc{(0,2)}
\rightarc{(4,2)}
\end{pspicture}
\ \ =\, \g^2\qquad\
\begin{pspicture}(0,1.9)(5,4)
\psellipse[linecolor=blue,linewidth=1.5pt](0,2)(1.2,1.53)
\psellipse[linecolor=blue,linewidth=1.5pt](4,2)(1.2,1.53)
\facegrid{(0,0)}{(4,4)}
\rput(2.5,.5){\tiny row 1}
\rput(2.5,1.5){\tiny row 3}
\rput(2.5,2.5){\tiny row 2}
\rput(2.5,3.5){\tiny row 4}
\pspolygon[fillstyle=solid,fillcolor=lightgray](0,2)(1,2)(1,4)(0,4)(0,2)
\psarc[linewidth=.25pt](0,2){.2}{0}{90}
\pspolygon[fillstyle=solid,fillcolor=lightgray](1,2)(2,2)(2,4)(1,4)(1,2)
\psarc[linewidth=.25pt](1,2){.2}{0}{90}
\leftarc{(0,2)}
\rightarc{(4,2)}
\end{pspicture}
\ \ =\, \ldots 
\nonumber\\[1.7cm]
&=&\!\g^N\prod_{j=1}^{\rho-1}\cos(u+\xi_j)\sin(u+\xi_j)\qquad\
\begin{pspicture}(0,1.9)(5,4)
\psellipse[linecolor=blue,linewidth=1.5pt](0,2)(1.2,1.53)
\psellipse[linecolor=blue,linewidth=1.5pt](4,2)(1.2,1.53)
\facegrid{(0,0)}{(4,2)}
\rput(2.5,.5){\tiny row 1}
\rput(2.5,1.5){\tiny row 3}
\pspolygon[fillstyle=solid,fillcolor=lightgray](0,2)(1,2)(1,4)(0,4)(0,2)
\psarc[linewidth=.25pt](0,2){.2}{0}{90}
\pspolygon[fillstyle=solid,fillcolor=lightgray](1,2)(2,2)(2,4)(1,4)(1,2)
\psarc[linewidth=.25pt](1,2){.2}{0}{90}
\pspolygon[fillstyle=solid,fillcolor=lightgray](2,2)(3,2)(3,4)(2,4)(2,2)
\psarc[linewidth=.25pt](2,2){.2}{0}{90}
\pspolygon[fillstyle=solid,fillcolor=lightgray](3,2)(4,2)(4,4)(3,4)(3,2)
\psarc[linewidth=.25pt](3,2){.2}{0}{90}
\leftarc{(0,2)}
\rightarc{(4,2)}
\end{pspicture}
\ \ =\, \ldots 
\\[1.1cm]
&=&\!(-1)^{N+\rho-1}\g^{2N}\prod_{j=1}^{\rho-1}\cos(u+\xi_j)\sin(u+\xi_j)\cos(u-\xi_j)\sin(u-\xi_j)\qquad\
\begin{pspicture}(0,1.9)(5,4)
\psellipse[linecolor=blue,linewidth=1.5pt](0,2)(1.2,1.53)
\psellipse[linecolor=blue,linewidth=1.5pt](4,2)(1.2,1.53)
\pspolygon[fillstyle=solid,fillcolor=lightgray](0,0)(1,0)(1,2)(0,2)(0,0)
\psarc[linewidth=.25pt](0,0){.2}{0}{90}
\pspolygon[fillstyle=solid,fillcolor=lightgray](1,0)(2,0)(2,2)(1,2)(1,0)
\psarc[linewidth=.25pt](1,0){.2}{0}{90}
\pspolygon[fillstyle=solid,fillcolor=lightgray](2,0)(3,0)(3,2)(2,2)(2,0)
\psarc[linewidth=.25pt](2,0){.2}{0}{90}
\pspolygon[fillstyle=solid,fillcolor=lightgray](3,0)(4,0)(4,2)(3,2)(3,0)
\psarc[linewidth=.25pt](3,0){.2}{0}{90}
\pspolygon[fillstyle=solid,fillcolor=lightgray](0,2)(1,2)(1,4)(0,4)(0,2)
\psarc[linewidth=.25pt](0,2){.2}{0}{90}
\pspolygon[fillstyle=solid,fillcolor=lightgray](1,2)(2,2)(2,4)(1,4)(1,2)
\psarc[linewidth=.25pt](1,2){.2}{0}{90}
\pspolygon[fillstyle=solid,fillcolor=lightgray](2,2)(3,2)(3,4)(2,4)(2,2)
\psarc[linewidth=.25pt](2,2){.2}{0}{90}
\pspolygon[fillstyle=solid,fillcolor=lightgray](3,2)(4,2)(4,4)(3,4)(3,2)
\psarc[linewidth=.25pt](3,2){.2}{0}{90}
\leftarc{(0,2)}
\rightarc{(4,2)}
\end{pspicture}
\nonumber
\\[0.6cm]
\nonumber
\eea
Since
\psset{unit=0.65cm}
\setlength{\unitlength}{0.65cm}
\be
\begin{pspicture}(0,1.9)(1,4)
\psellipse[linecolor=blue,linewidth=1.5pt](0,2)(1.2,1.53)
\pspolygon[fillstyle=solid,fillcolor=white,linecolor=white](0,0)(2,0)(2,4)(0,4)(0,0)
\facegrid{(0,0)}{(1,4)}
\leftarc{(0,2)}
\rput(0,0){\loopa}
\rput(0,1){\loopa}
\rput(0,2){\loopb}
\rput(0,3){\loopb}
\end{pspicture}
\ =\, 0,\qquad\qquad
\begin{pspicture}(0,1.9)(2,4)
\psellipse[linecolor=blue,linewidth=1.5pt](0,2)(1.2,1.53)
\facegrid{(0,0)}{(1,4)}
\leftarc{(0,2)}
\rput(0,0){\loopa}
\rput(0,1){\loopa}
\rput(0,2){\loopa}
\rput(0,3){\loopa}
\pspolygon[fillstyle=solid,fillcolor=lightgray](1,0)(2,0)(2,2)(1,2)(1,0)
\psarc[linewidth=.25pt](1,0){.2}{0}{90}
\pspolygon[fillstyle=solid,fillcolor=lightgray](1,2)(2,2)(2,4)(1,4)(1,2)
\psarc[linewidth=.25pt](1,2){.2}{0}{90}
\end{pspicture}
\ +\, \qquad
\begin{pspicture}(0,1.9)(2,4)
\psellipse[linecolor=blue,linewidth=1.5pt](0,2)(1.2,1.53)
\facegrid{(0,0)}{(1,4)}
\leftarc{(0,2)}
\rput(0,0){\loopb}
\rput(0,1){\loopb}
\rput(0,2){\loopb}
\rput(0,3){\loopb}
\pspolygon[fillstyle=solid,fillcolor=lightgray](1,0)(2,0)(2,2)(1,2)(1,0)
\psarc[linewidth=.25pt](1,0){.2}{0}{90}
\pspolygon[fillstyle=solid,fillcolor=lightgray](1,2)(2,2)(2,4)(1,4)(1,2)
\psarc[linewidth=.25pt](1,2){.2}{0}{90}
\end{pspicture}
\ =\, 0
\\[1.2cm]
\ee
it follows that
\psset{unit=0.65cm}
\setlength{\unitlength}{0.65cm}
\bea
&&
\begin{pspicture}(0.5,1.9)(5,4)
\psellipse[linecolor=blue,linewidth=1.5pt](0,2)(1.2,1.53)
\psellipse[linecolor=blue,linewidth=1.5pt](4,2)(1.2,1.53)
\pspolygon[fillstyle=solid,fillcolor=lightgray](0,0)(1,0)(1,2)(0,2)(0,0)
\psarc[linewidth=.25pt](0,0){.2}{0}{90}
\pspolygon[fillstyle=solid,fillcolor=lightgray](1,0)(2,0)(2,2)(1,2)(1,0)
\psarc[linewidth=.25pt](1,0){.2}{0}{90}
\pspolygon[fillstyle=solid,fillcolor=lightgray](2,0)(3,0)(3,2)(2,2)(2,0)
\psarc[linewidth=.25pt](2,0){.2}{0}{90}
\pspolygon[fillstyle=solid,fillcolor=lightgray](3,0)(4,0)(4,2)(3,2)(3,0)
\psarc[linewidth=.25pt](3,0){.2}{0}{90}
\pspolygon[fillstyle=solid,fillcolor=lightgray](0,2)(1,2)(1,4)(0,4)(0,2)
\psarc[linewidth=.25pt](0,2){.2}{0}{90}
\pspolygon[fillstyle=solid,fillcolor=lightgray](1,2)(2,2)(2,4)(1,4)(1,2)
\psarc[linewidth=.25pt](1,2){.2}{0}{90}
\pspolygon[fillstyle=solid,fillcolor=lightgray](2,2)(3,2)(3,4)(2,4)(2,2)
\psarc[linewidth=.25pt](2,2){.2}{0}{90}
\pspolygon[fillstyle=solid,fillcolor=lightgray](3,2)(4,2)(4,4)(3,4)(3,2)
\psarc[linewidth=.25pt](3,2){.2}{0}{90}
\leftarc{(0,2)}
\rightarc{(4,2)}
\end{pspicture}
\ \ =\ -\qquad\
\begin{pspicture}(0,1.9)(5,4)
\psellipse[linecolor=blue,linewidth=1.5pt](0,2)(1.2,1.53)
\psellipse[linecolor=blue,linewidth=1.5pt](4,2)(1.2,1.53)
\facegrid{(0,0)}{(1,4)}
\rput(0,0){\loopb}
\rput(0,1){\loopb}
\rput(0,2){\loopa}
\rput(0,3){\loopa}
\pspolygon[fillstyle=solid,fillcolor=lightgray](1,0)(2,0)(2,2)(1,2)(1,0)
\psarc[linewidth=.25pt](1,0){.2}{0}{90}
\pspolygon[fillstyle=solid,fillcolor=lightgray](2,0)(3,0)(3,2)(2,2)(2,0)
\psarc[linewidth=.25pt](2,0){.2}{0}{90}
\pspolygon[fillstyle=solid,fillcolor=lightgray](3,0)(4,0)(4,2)(3,2)(3,0)
\psarc[linewidth=.25pt](3,0){.2}{0}{90}
\pspolygon[fillstyle=solid,fillcolor=lightgray](1,2)(2,2)(2,4)(1,4)(1,2)
\psarc[linewidth=.25pt](1,2){.2}{0}{90}
\pspolygon[fillstyle=solid,fillcolor=lightgray](2,2)(3,2)(3,4)(2,4)(2,2)
\psarc[linewidth=.25pt](2,2){.2}{0}{90}
\pspolygon[fillstyle=solid,fillcolor=lightgray](3,2)(4,2)(4,4)(3,4)(3,2)
\psarc[linewidth=.25pt](3,2){.2}{0}{90}
\leftarc{(0,2)}
\rightarc{(4,2)}
\end{pspicture}
\ \ =\, \ldots 
\nonumber\\[1.7cm]
&=&\!(-1)^{N+\rho-2}\qquad\
\begin{pspicture}(0,1.9)(5,4)
\psellipse[linecolor=blue,linewidth=1.5pt](0,2)(1.2,1.53)
\psellipse[linecolor=blue,linewidth=1.5pt](4,2)(1.2,1.53)
\pspolygon[fillstyle=solid,fillcolor=white,linecolor=white](-0.5,1)(4.5,1)(4.5,3)(-0.5,3)(-0.5,1)
\facegrid{(0,0)}{(3,4)}
\rput(0,0){\loopb}
\rput(0,1){\loopb}
\rput(0,2){\loopa}
\rput(0,3){\loopa}
\rput(1,0){\loopb}
\rput(1,1){\loopb}
\rput(1,2){\loopa}
\rput(1,3){\loopa}
\rput(2,0){\loopb}
\rput(2,1){\loopb}
\rput(2,2){\loopa}
\rput(2,3){\loopa}
\pspolygon[fillstyle=solid,fillcolor=lightgray](3,0)(4,0)(4,2)(3,2)(3,0)
\psarc[linewidth=.25pt](3,0){.2}{0}{90}
\pspolygon[fillstyle=solid,fillcolor=lightgray](3,2)(4,2)(4,4)(3,4)(3,2)
\psarc[linewidth=.25pt](3,2){.2}{0}{90}
\leftarc{(0,2)}
\rightarc{(4,2)}
\end{pspicture}
\ \ =\, 2\,(-1)^{N+\rho-2}\Ib
\label{gelephant}
\\[0.5cm]
\nonumber
\eea
where we have used that
\psset{unit=0.65cm}
\setlength{\unitlength}{0.65cm}
\be
\ \ \ \
\begin{pspicture}(0,1.9)(2,4)
\psellipse[linecolor=blue,linewidth=1.5pt](0,2)(1.2,1.53)
\psellipse[linecolor=blue,linewidth=1.5pt](1,2)(1.2,1.53)
\pspolygon[fillstyle=solid,fillcolor=white,linecolor=white](-0.5,1)(1.5,1)(1.5,3)(-0.5,3)(-0.5,1)
\facegrid{(0,0)}{(1,4)}
\leftarc{(0,2)}
\rightarc{(1,2)}
\pspolygon[fillstyle=solid,fillcolor=lightgray](0,0)(1,0)(1,2)(0,2)(0,0)
\psarc[linewidth=.25pt](0,0){.2}{0}{90}
\pspolygon[fillstyle=solid,fillcolor=lightgray](0,2)(1,2)(1,4)(0,4)(0,2)
\psarc[linewidth=.25pt](0,2){.2}{0}{90}
\end{pspicture}
\ \; =\qquad \,
\begin{pspicture}(0,1.9)(2,4)
\psellipse[linecolor=blue,linewidth=1.5pt](0,2)(1.2,1.53)
\psellipse[linecolor=blue,linewidth=1.5pt](1,2)(1.2,1.53)
\pspolygon[fillstyle=solid,fillcolor=white,linecolor=white](-0.5,1)(1.5,1)(1.5,3)(-0.5,3)(-0.5,1)
\facegrid{(0,0)}{(1,4)}
\leftarc{(0,2)}
\rightarc{(1,2)}
\rput(0,0){\loopa}
\rput(0,1){\loopa}
\rput(0,2){\loopa}
\rput(0,3){\loopa}
\end{pspicture}
\ \; -\qquad \,
\begin{pspicture}(0,1.9)(2,4)
\psellipse[linecolor=blue,linewidth=1.5pt](0,2)(1.2,1.53)
\psellipse[linecolor=blue,linewidth=1.5pt](1,2)(1.2,1.53)
\pspolygon[fillstyle=solid,fillcolor=white,linecolor=white](-0.5,1)(1.5,1)(1.5,3)(-0.5,3)(-0.5,1)
\facegrid{(0,0)}{(1,4)}
\leftarc{(0,2)}
\rightarc{(1,2)}
\rput(0,0){\loopb}
\rput(0,1){\loopb}
\rput(0,2){\loopa}
\rput(0,3){\loopa}
\end{pspicture}
\ \; -\qquad \,
\begin{pspicture}(0,1.9)(2,4)
\psellipse[linecolor=blue,linewidth=1.5pt](0,2)(1.2,1.53)
\psellipse[linecolor=blue,linewidth=1.5pt](1,2)(1.2,1.53)
\pspolygon[fillstyle=solid,fillcolor=white,linecolor=white](-0.5,1)(1.5,1)(1.5,3)(-0.5,3)(-0.5,1)
\facegrid{(0,0)}{(1,4)}
\leftarc{(0,2)}
\rightarc{(1,2)}
\rput(0,0){\loopa}
\rput(0,1){\loopa}
\rput(0,2){\loopb}
\rput(0,3){\loopb}
\end{pspicture}
\ \; +\qquad \,
\begin{pspicture}(0,1.9)(2,4)
\psellipse[linecolor=blue,linewidth=1.5pt](0,2)(1.2,1.53)
\psellipse[linecolor=blue,linewidth=1.5pt](1,2)(1.2,1.53)
\pspolygon[fillstyle=solid,fillcolor=white,linecolor=white](-0.5,1)(1.5,1)(1.5,3)(-0.5,3)(-0.5,1)
\facegrid{(0,0)}{(1,4)}
\leftarc{(0,2)}
\rightarc{(1,2)}
\rput(0,0){\loopb}
\rput(0,1){\loopb}
\rput(0,2){\loopb}
\rput(0,3){\loopb}
\end{pspicture}
\ \; =\,2\
\begin{pspicture}(0,1.9)(1,4)
\pspolygon[fillstyle=solid,fillcolor=lightlightblue](0,0)(1,0)(1,4)(0,4)(0,0)
\psline[linecolor=blue,linewidth=1.5pt](0.5,0)(0.5,4)
\end{pspicture}
\\[1.2cm]
\ee
We thus conclude that $\vec D(u)\vec D(u+\lambda)$ is proportional to the vertical identity diagram.
Combining the weights following from the various steps above readily produces the right-hand side of
(\ref{DDI}):
\bea
 \Db(u)\Db(u+\lambda)&=&-\frac{\sin^2 2u}{\cos^2 2u}\Big(\!\sin^{4N}\!u\;[\eta^{(\rho)}(u+\la,\xi)]^2
  +\cos^{4N}\! u\;[\eta^{(\rho)}(u,\xi)]^2\nn
  &&\qquad\quad+\,(-1)^{N+\rho-1}\g^{2N}\prod_{j=1}^{\rho-1}\cos(u+\xi_j)\sin(u+\xi_j)\cos(u-\xi_j)\sin(u-\xi_j)\nn
  &&\qquad\quad\times\, 2\,(-1)^{N+\rho-2}\Big)
 \nn
 &=&-\tan^2 2u\,\Big(\!\cos^{2N}\!u\,\eta^{(\rho)}(u,\xi)-\sin^{2N}\!u\,\eta^{(\rho)}(u+\la,\xi)\!\Big)^{\!2}\Ib
\eea
$\Box$

\section{Double-Columns and $q$-Narayana Numbers}
\label{qNarayana}

\noindent
{\bf Definition:}\quad
{\em A single-column configuration of height $M$ consists of $M$ sites arranged as a column. The 
notion of occupants, signature, weight and corresponding monomial are defined as for 
double-column configurations in Section~\ref{qCatalan}.
\\[.2cm]
\noindent
{\bf Definition:}\quad 
The polynomial {\scriptsize $\sbinMm{M}{m}_{\!q}$} is defined 
as the sum of the monomials associated to the distinct single-column configurations of height $M$ 
with exactly $m$ occupants. If any of the inequalities $0\leq m\leq M$ is violated, we set
{\scriptsize $\sbinMm{M}{m}_{\!q}$} $=0$.}
\\[.2cm]
{\bf Proposition:}\quad
{\em The polynomial {\scriptsize $\sbinMm{M}{m}_{\!q}$} is given by}
\be
 \sbinMm{M}{m}_{\!q}=\sum_{1\leq j_1<j_2<\ldots<j_m\leq M}q^{j_1+\ldots+j_m}
   =q^{\frac{1}{2}m(m+1)}\gauss{M}{m}_{\!q}
\ee
{\bf Proof:}\quad
A single-column configuration of height $M$ with exactly $m$ occupants is characterized by its signature $(j_m,\ldots,j_1)$ where $1\leq j_1<j_2<\ldots<j_m\leq M$. Summing over all distinct configurations thus amounts to summing over all the associated signatures. This gives rise to the multiple sum. The unique column with only the $m$ lowest sites occupied has weight $\tfrac{1}{2}m(m+1)$, while the $q$-binomial {\scriptsize $\gauss{M}{m}_{\!q}$} by construction encapsulates the distribution of excitations of this configuration. This yields the second expression.
$\Box$
\\[.2cm]
\noindent
{\bf Definition:}\quad
{\em $D^M$ denotes the set of distinct double-column configurations of height $M$. 
Similarly, $A^M$ denotes the set of distinct admissible double-column configurations of height $M$. 
$D_{m,n}^M$ is the subset of $D^M$ consisting of the double-column configurations with exactly 
$m$ and $n$ occupants in the left and right columns, respectively. Similarly, $A_{m,n}^M$ is the 
subset of $A^M$ consisting of the (admissible) double-column configurations with exactly $m$ and 
$n$ occupants in the left and right columns, respectively.}
\\[.2cm]
\noindent
{\bf Lemma:}\quad
{\em The polynomial defined as the sum of the monomials associated to the elements of 
$D_{m,n}^M$ is given by}
\be
 \sbinMm{M}{m}_{\!q}\sbinMm{M}{n}_{\!q}=q^{\frac{1}{2}m(m+1)+\frac{1}{2}n(n+1)}
  \gauss{M}{m}_{\!q}\gauss{M}{n}_{\!q}
\label{MmMn}
\ee
{\bf Proof:}\quad
The distributions of occupancies in the two columns of an element of $D_{m,n}^M$ are independent. 
The polynomial is thus given by the product of the polynomials associated to the individual columns.
$\Box$
\\[.2cm]
\noindent
{\bf Theorem:}\quad
{\em For $0\leq m\leq n\leq M$, there is a weight-preserving bijection between 
$D_{m,n}^M\!\setminus\!A_{m,n}^M$ and $D_{m-1,n+1}^M$.}
\\[.2cm]
{\bf Proof:}\quad
For $m=0$, $D_{m,n}^M=A_{m,n}^M$ 
so $D_{m,n}^M\!\setminus\!A_{m,n}^M=\emptyset=D_{m-1,n+1}^M$ rendering the bijection trivial.
A non-admissible configuration with $m>0$ occupants in the left column violates
the inequality $L_j\leq R_j$ for at least one $j$ where $1\leq j\leq m$. Let $j_0$ denote the 
minimum such $j$. We now introduce the map
\be
 \varphi_{m,n}^M:\ (L,R)\ \mapsto\ ((R_1,\ldots,R_{j_0-1},L_{j_0+1},\ldots,L_m),
   (L_1,\ldots,L_{j_0},R_{j_0},\ldots,R_n))
\label{bij}
\ee
defined for every signature $(L,R)$ corresponding to an element of 
$D_{m,n}^M\!\setminus\!A_{m,n}^M$.
Since $R_{j_0-1}\geq L_{j_0-1}>L_{j_0+1}$ and $L_{j_0}>R_{j_0}$, the image of $(L,R)$ 
is the signature of an element of $D_{m-1,n+1}^M$, so
\be
 \varphi_{m,n}^M:\ D_{m,n}^M\!\setminus\!A_{m,n}^M \to\ D_{m-1,n+1}^M
\ee
This map is clearly injective. To show that it is surjective, we consider the signature 
$(\mathcal{L},\mathcal{R})$ of a given but arbitrary element of $D_{m-1,n+1}^M$. 
Since $m\leq n$, there exists $i_0$, $1\leq i_0\leq m$, such that $\mathcal{L}_i\geq\mathcal{R}_i$ 
for $i<i_0$ while $\mathcal{L}_{i_0}<\mathcal{R}_{i_0}$. We now introduce the map
\be
 \phi_{m-1,n+1}^M:\ (\mathcal{L},\mathcal{R})\ \mapsto\ ((\mathcal{R}_1,\ldots,\mathcal{R}_{i_0},
    \mathcal{L}_{i_0},\ldots,\mathcal{L}_{m-1}),
   (\mathcal{L}_1,\ldots,\mathcal{L}_{i_0-1},\mathcal{R}_{i_0+1},\ldots,\mathcal{R}_{n+1}))
\label{phii}
\ee
defined for every signature corresponding to an element of $D_{m-1,n+1}^M$.
As above, it follows that the image corresponds to an element of $D_{m,n}^M$. To see that it is 
non-admissible, we note that the image in (\ref{phii}), here denoted by $(L',R')$, satisfies 
$L'_i\leq R_i'$ for $i<i_0$, while $L'_{i_0}>R'_{i_0}$. The map $\phi_{m-1,n+1}^M$ is 
thus the inverse of the map $\varphi_{m,n}^M$ and the bijection has been established.
The bijection (\ref{bij}) is weight preserving since the sum of the signature entries of the image of 
$(L,R)$ equals the sum of the signature entries of $(L,R)$ itself.
$\Box$
\\[.2cm]
\noindent
{\bf Remark:}\quad
The constraint $m\leq n$ is important for the existence of the bijection. For $m>n$, 
$A_{m,n}^M=\emptyset$ and $D_{m,n}^M\!\setminus\!A_{m,n}^M=D_{m,n}^M$ which is generally 
not isomorphic to $D_{m-1,n+1}^M$. The proof above applies for $m\leq n$, but breaks down for
$m>n$ since the existence of $i_0$ is no longer ensured.
\\[.2cm]
\noindent
{\bf Remark:}\quad
There is a very simple configurational realization of this bijection. Viewed from above, one identifies 
the location, $j_0$, of the first violation of admissibility in the non-admissible double-column 
configuration with signature $(L,R)$. 
One then divides the configuration into two with a horizontal cut just below the height 
where the violation occurred, i.e.$\!\,$ just below the height $L_{j_0}$ (which for $L_{j_0}=1$ would be 
just below the entire double-column configuration). One finally interchanges the parts of the two 
columns appearing above the cut while the parts below the cut stay put.
\\[.2cm]
\noindent
{\bf Theorem:}\quad
{\em The (generalized) $q$-Narayana numbers defined as the sum of the monomials associated to the elements of 
$A_{m,n}^M$ is given by}
\be
  \sbinlr{M}{m,n}_{\!q}=q^{\hf m(m+1)+\hf n(n+1)}
   \bigg(\gauss{M}{m}_{\!q}\gauss{M}{n}_{\!q}-q^{n-m+1}\gauss{M}{m-1}_{\!q}
    \gauss{M}{n+1}_{\!q}\bigg)
\label{AdmMin}
\ee
{\bf Proof:}\quad
Since $D_{m,n}^M$ is the disjoint union of $A_{m,n}^M$ and $D_{m,n}^M\!\setminus\! A_{m,n}^M$, 
the polynomial {\scriptsize $\sbinlr{M}{m,n}_{\!q}$} can be written as the sum of the monomials 
associated to the elements of $D_{m,n}^M$ minus the sum of the monomials associated to 
the elements of $D_{m,n}^M\!\setminus\! A_{m,n}^M$. The expression (\ref{AdmMin})
now follows from the previous Theorem and (\ref{MmMn}).
$\Box$

\section{Proofs of $q$-Catalan Combinatorial Identities}
\label{CatIdProofs}

The finitized partition functions, for general $(r,s)$ boundary conditions,  in (\ref{catId1}) and (\ref{catId2}) are decompositions into irreducible blocks given by $q$-Catalan polynomials. In this appendix, we prove combinatorial identities to simplify these decompositions to the explicit finitized characters given by (\ref{finitizedZ}). 
This is achieved by separately expanding out the left and right sides of the identities as sums of a single $q$-binomial. 
We note that the $q$-binomials satisfy the elementary properties
\bea
\gauss{a}{b}_{\!q} = \frac{(1-q^{a-b+1})}{(1-q^{b})} \gauss{a}{b-1}_{\!q} = \frac{(1-q^{a})}{(1-q^{a-b})} \gauss{a-1}{b}_{\!q}=
 \frac{(1-q^{a})}{(1-q^{b})} \gauss{a-1}{b-1}_{\!q}
\label{qbinprop}
\eea
and use the notation
\bea
\tilde Z^{(N)}_{r,s}(q)=q^{c\over 24} Z^{(N)}_{(1,1)|(r,s)}(q),\qquad t=
\begin{cases}
\frac{1}{2} (N-2r+s), &\mbox{$\rho=2r$}\\[2pt]
\frac{1}{2} (N-2r+s+1), &\mbox{$\rho=2r-1$}
\end{cases}
\label{tee}
\eea
We consider four cases depending on the parities of $\rho$ and $s$.

\subsection{$\rho$ even, $s$ odd}
We telescope the partition function (\ref{finitizedZ}) as follows
\bea
\tilde Z_{r,s}^{(N)} (q) &=& q^{-\frac{1}{8}}\bigg(\! q^{\frac{1}{8} (2r-s)^2}\!\! \gauss{N}{t}_{\!q}\!\!\! -\! q^{\frac{1}{8} (2r-s+2)^2}\!\! \gauss{N}{\!t\!-\!1\!}_{\!q}\!\!\!
+ q^{\frac{1}{8} (2r-s+2)^2} \!\!\gauss{N}{\!t\!-\!1\!}_{\!q} -\!\!\! \nonumber\\
&&\qquad \cdots +\! q^{\frac{1}{8} (2r+s-2)^2}\!\! \gauss{N}{\!t\!-\!s\!+\!1\!}_{\!q}\! -\! q^{\frac{1}{8} (2r+s)^2} \!\!\gauss{N}{\!t\!-\!s\!}_{\!q}\!\bigg)\nonumber\\
&=& q^{-\frac{1}{8}} \sum_{k=1}^{s} \bigg( q^{\frac{1}{8} (2r-s-2+2k)^2} \gauss{N}{t+1-k}_{\!q} -q^{\frac{1}{8} (2r-s+2k)^2} \gauss{N}{t-k}_{\!q} \bigg)
\eea
where $t=\frac{1}{2} (N-2r+s)$. 
Using the first part of (\ref{qbinprop}) on the first $q$-binomial, the two terms combine
\bea
\tilde{Z}_{r,s}^{(N)} (q)&\!\!=\!\!& q^{-\frac{1}{8}}\sum_{k=1}^{s} \frac{q^{\frac{1}{8} (2r-s+2k)^2}}{(1-q^{t+1-k})}\gauss{N}{t-k}_{\!q} \Big(q^{\frac{1}{2}(-2r+s+1-2k)} \big( 1 -q^{N-t+k} \big) - \big( 1 - q^{t+1-k} \big) \Big)\qquad\nonumber\\
&\!\!=\!\!& q^{-\frac{1}{8}}\sum_{k=1}^{s} \frac{q^{\frac{1}{8} (2r-s+2k)^2}}{(1-q^{t+1-k})}\gauss{N}{t-k}_{\!q} (q^{\frac{1}{2}(-2r+s+1-2k)}-1)(1+q^{\frac{N+1}{2}})
\label{Zevodd}
\eea
Using the definition of the the $q$-Catalan polynomials in (\ref{qCat}), the decomposition (\ref{catId1}) of the finitized partition function with $\rho=2r$ can be expanded as
\bea
\sum_{k=1}^{s} \qqcat{\frac{N-1}{2}}{\frac{2r-s-1+2k}{2}}{q}
&\!\!=\!\!&\sum_{k=1}^{s} q^{\frac{1}{8}(2r-s-1+2k)(2r-s-3+2k)} \frac{(1-q^{\frac{1}{2}(2r-s-1+2k)})}{(1-q^{\frac{N+1}{2}})} \gauss{N+1}{t+1-k}_{\!q}\\
&\!\!=\!\!& q^{-\frac{1}{8}} \sum_{k=1}^{s} q^{\frac{1}{8} (2r-s+2k)^2} q^{\frac{1}{2}(-2r+s+1-2k)} \frac{(1-q^{\frac{1}{2} (2r-s-1+2k)})}{(1-q^{\frac{N+1}{2}})} \gauss{N+1}{t+1-k}_{\!q}\qquad\nonumber
\eea
Applying the third part of (\ref{qbinprop}), we can simplify this expression to obtain the desired result in agreement with (\ref{Zevodd})
\bea
\sum_{k=1}^{s} \qqcat{\frac{N-1}{2}}{\frac{2r-s-1+2k}{2}}{q}
&\!\!=\!\!& q^{-\frac{1}{8}}\sum_{k=1}^{s} \frac{q^{\frac{1}{8} (2r-s+2k)^2}}{(1-q^{t+1-k})} \gauss{N}{t-k}_{\!q}  \frac{(q^{\frac{1}{2}(-2r+s+1-2k)}-1)(1-q^{N+1})}{(1-q^{\frac{N+1}{2}})}\qquad
\eea

\subsection{$\rho$ odd, $s$ odd}
We telescope the partition function (\ref{finitizedZ}) as follows
\bea
\tilde Z_{r,s}^{(N)} (q) &=& q^{-\frac{1}{8}}\bigg(\! q^{\frac{1}{8} (2r-s)^2}\!\! \gauss{N}{t}_{\!q}\!\!\! -\! q^{\frac{1}{8} (2r-s+2)^2}\!\! \gauss{N}{\!t\!-\!1\!}_{\!q}\!\!\!
+ q^{\frac{1}{8} (2r-s+2)^2} \!\!\gauss{N}{\!t\!-\!1\!}_{\!q} - \!\!\! \nonumber\\
&&\qquad \cdots +\! q^{\frac{1}{8} (2r+s-2)^2}\!\! \gauss{N}{\!t\!-\!s\!+\!1\!}_{\!q}\! -\! q^{\frac{1}{8} (2r+s)^2} \!\!\gauss{N}{\!t\!-\!s\!}_{\!q}\!\bigg)\nonumber\\
&=& q^{-\frac{1}{8}} \sum_{k=1}^{s} \bigg( q^{\frac{1}{8} (2r-s-2+2k)^2} \gauss{N}{t+1-k}_{\!q} -q^{\frac{1}{8} (2r-s+2k)^2} \gauss{N}{t-k}_{\!q} \bigg)
\eea
where $t=\frac{1}{2} (N-2r+s+1)$.
Using the first part of (\ref{qbinprop}) on the first $q$-binomial, the two terms combine
\bea
\tilde{Z}_{r,s}^{(N)} (q)&\!\!=\!\!& q^{-\frac{1}{8}}\sum_{k=1}^{s} \frac{q^{\frac{1}{8} (2r-s+2k)^2}}{(1-q^{t+1-k})}\gauss{N}{t-k}_{\!q} \Big(q^{\frac{1}{2}(-2r+s+1-2k)} \big( 1 -q^{N-t+k} \big) - \big( 1 - q^{t+1-k} \big) \Big)\qquad\nonumber\\
&\!\!=\!\!& q^{-\frac{1}{8}}\sum_{k=1}^{s} \frac{q^{\frac{1}{8} (2r-s+2k)^2}}{(1-q^{t+1-k})}\gauss{N}{t-k}_{\!q} \Big(q^{\frac{1}{2}(-2r+s+1-2k)}-1-q^{\frac{N}{2}}+q^{t+1-k}\Big)
\label{Zoddodd}
\eea
Using the definition of the the $q$-Catalan polynomials in (\ref{qCat}), the decomposition (\ref{catId1}) of the finitized partition function with $\rho=2r-1$ can be expanded as
\bea
&&\sum_{k=1}^{s} \big[\qqcat{\frac{N-2}{2}}{\frac{2r-s-1+2k}{2}}{q}+q^{\frac{N}{2}} \qqcat{\frac{N-2}{2}}{\frac{2r-s-3+2k}{2}}{q} \big] \nonumber\\
&=&\sum_{k=1}^{s} \bigg(q^{\frac{1}{8}(2r-s-1+2k)(2r-s-3+2k)} \frac{(1-q^{\frac{1}{2}(2r-s-1+2k)})}{(1-q^{N/2})} \gauss{N}{t-k}_{\!q}\nonumber \\
&&\qquad + q^{\frac{N}{2}} q^{\frac{1}{8}(2r-s-3+2k)(2r-s-5+2k)} \frac{(1-q^{\frac{1}{2}(2r-s-3+2k)})}{(1-q^{N/2})}\gauss{N}{t+1-k}_{\!q} \bigg) \nonumber\\
&=&q^{-\frac{1}{8}} \sum_{k=1}^{s} \bigg(q^{\frac{1}{8} (2r-s+2k)^2} q^{\frac{1}{2}(-2r+s+1-2k)} \frac{(1-q^{\frac{1}{2}(2r-s-1+2k)})}{(1-q^{N/2})} \gauss{N}{t-k}_{\!q}\nonumber \\
&&\qquad + q^{\frac{N}{2}} q^{\frac{1}{8} (2r-s+2k)^2} q^{-2r+s+2-2k} \frac{(1-q^{\frac{1}{2}(2r-s-3+2k)})}{(1-q^{N/2})}\gauss{N}{t+1-k}_{\!q} \bigg)
\eea
Applying the first part of (\ref{qbinprop}) to the second $q$-binomial, we can combine the two terms to obtain the desired result in agreement with (\ref{Zoddodd})
\bea
&&\sum_{k=1}^{s} \big[\qqcat{\frac{N-2}{2}}{\frac{2r-s-1+2k}{2}}{q}+q^{\frac{N}{2}} \qqcat{\frac{N-2}{2}}{\frac{2r-s-3+2k}{2}}{q} \big] \nonumber \\
&=&\!\!\!q^{-\frac{1}{8}}\!\sum_{k=1}^{s}\! \frac{q^{\frac{1}{8}(2r-s+2k)^2}}{(1-q^{N/2})(1-q^{t+1-k})} \!\!\gauss{N}{t-k}_{\!q}\!\!\! \Big(q^{\frac{1}{2}(-2r+s+1-2k)}(1\!\!-\!\!q^{\frac{1}{2}(2r-s-1+2k)})(1\!\!-\!\!q^{t+1-k}) \nonumber \\
&&\qquad + q^{\frac{N}{2}} q^{-2r+s+2-2k}(1-q^{\frac{1}{2}(2r-s-3+2k)})(1-q^{N-t+k})\Big)\nonumber\\
&=& q^{-\frac{1}{8}}\sum_{k=1}^{s} \frac{q^{\frac{1}{8} (2r-s+2k)^2}}{(1-q^{t+1-k})}\gauss{N}{t-k}_{\!q} \Big(q^{\frac{1}{2}(-2r+s+1-2k)}  - 1 -q^{\frac{N}{2}} + q^{t+1-k} \Big)
\eea

\subsection{$\rho$ even, $s$ even}
We telescope the partition function (\ref{finitizedZ}) as follows
\bea
\tilde Z_{r,s}^{(N)} (q) &=& q^{-\frac{1}{8}}\bigg(\! q^{\frac{1}{8} (2r-s)^2}\!\! \gauss{N}{t}_{\!q}\!\!\! -\! q^{\frac{1}{8} (2r-s+4)^2}\!\! \gauss{N}{\!t\!-\!2\!}_{\!q}\!\!\!
+ q^{\frac{1}{8} (2r-s+4)^2} \!\!\gauss{N}{\!t\!-\!2\!}_{\!q} -\!\!\! \nonumber \\
&&\qquad\cdots+\! q^{\frac{1}{8} (2r+s-4)^2}\!\! \gauss{N}{\!t\!-\!s\!+\!2\!}_{\!q}\! -\! q^{\frac{1}{8} (2r+s)^2} \!\!\gauss{N}{\!t\!-\!s\!}_{\!q}\!\bigg) \nonumber\\
&=& q^{-\frac{1}{8}} \sum_{k=1}^{s/2} \bigg( q^{\frac{1}{8} (2r-s-4+4k)^2} \gauss{N}{t+2-2k}_{\!q} -q^{\frac{1}{8} (2r-s+4k)^2} \gauss{N}{t-2k}_{\!q} \bigg)
\eea
where $t=\frac{1}{2} (N-2r+s)$. Using the first part of (\ref{qbinprop}) twice on the first $q$-binomial, the two terms combine
\bea
\tilde Z_{r,s}^{(N)} (q) \!&=&\! q^{-\frac{1}{8}}\sum_{k=1}^{s/2} \frac{q^{\frac{1}{8} (2r-s+4k)^2}}{(1\!-\!q^{t+2-2k})(1\!-\!q^{t+1-2k})}\!\gauss{N}{\!t\!-\!2k\!}_{\!q}\!\!\! \Big(q^{-2r+s+2-4k} - 1+ q^{N+1}-q^{N-2r+s+3-4k}  \Big)\nonumber\\
&=& q^{-\frac{1}{8}} \sum_{k=1}^{s/2} \frac{q^{\frac{1}{8}(2r-s+4k)^2}}{(1-q^{t+2-2k})(1-q^{t+1-2k})}\gauss{N}{t-2k}_{\!q}(q^{-2r+s+2-4k}-1)(1-q^{N+1})
\label{Zevev}
\eea
Using the definition of the the $q$-Catalan polynomials in (\ref{qCat}), the decomposition (\ref{catId2}) of the finitized partition function with $\rho=2r$ can be expanded as
\bea
q^{-\frac{1}{8}} \sum_{k=1}^{s/2} \qqcatt{\frac{N}{2}}{\frac{2r-s-2+4k}{2}}{q}&=& q^{-\frac{1}{8}} \sum_{k=1}^{s/2} q^{\frac{1}{8} (2r-s-4+4k)^2} \frac{(1-q^{2r-s-2+4k})}{(1-q^{N-t+2k})} \gauss{N+1}{t+2-2k}_{\!q}
\eea
Applying the third and first parts of (\ref{qbinprop}), we can simplify this expression to obtain the desired result in agreement with (\ref{Zevev})
\bea
q^{-\frac{1}{8}}\sum_{k=1}^{s/2}\! \qqcatt{\frac{N}{2}}{\frac{2r-s-2+4k}{2}}{q}
\!\!\!\!&=&\!\!\! q^{-\frac{1}{8}}\sum_{k=1}^{s/2} q^{\frac{1}{8} (2r-s+4k)^2} \!q^{-2r+s+2-4k}
\nonumber\\
&&\qquad\mbox{}\times \frac{(1-q^{2r-s-2+4k})(1-q^{N+1})(1-q^{N-t+2k})}{(1-q^{N-t+2k})(1-q^{t+2-2k})(1-q^{t+1-2k})}\!\! \gauss{N}{t-2k}_{\!q}\\
\!\!\!\!&=&\!\!\!q^{-\frac{1}{8}} \sum_{k=1}^{s/2}\! \frac{q^{\frac{1}{8}(2r-s+4k)^2}}{(1-q^{t+2-2k})(1-q^{t+1-2k})}\!\!\gauss{N}{t-2k}_{\!q}\!\!(q^{-2r+s+2-4k}\!-\!1)(1-q^{N+1})\nonumber
\eea

\subsection{$\rho$ odd, $s$ even}
We telescope the partition function (\ref{finitizedZ}) as follows
\bea
\tilde Z_{r,s}^{(N)} (q) &=& q^{-\frac{1}{8}}\bigg(\! q^{\frac{1}{8} (2r-s)^2}\!\! \gauss{N}{t}_{\!q}\!\!\! -\! q^{\frac{1}{8} (2r-s+4)^2}\!\! \gauss{N}{\!t\!-\!2\!}_{\!q}\!\!\!
+ q^{\frac{1}{8} (2r-s+4)^2} \!\!\gauss{N}{\!t\!-\!2\!}_{\!q} \!\!\! -\nonumber \\
&&\qquad \cdots +\! q^{\frac{1}{8} (2r+s-4)^2}\!\! \gauss{N}{\!t\!-\!s\!+\!2\!}_{\!q}\! -\! q^{\frac{1}{8} (2r+s)^2} \!\!\gauss{N}{\!t\!-\!s\!}_{\!q}\!\bigg)\nonumber\\
&=& q^{-\frac{1}{8}} \sum_{k=1}^{s/2} \bigg( q^{\frac{1}{8} (2r-s-4+4k)^2} \gauss{N}{t+2-2k}_{\!q} -q^{\frac{1}{8} (2r-s+4k)^2} \gauss{N}{t-2k}_{\!q} \bigg)
\eea
where $t=\frac{1}{2} (N-2r+s+1)$.
Using the first part of (\ref{qbinprop}) twice on the first $q$-binomial, the two terms combine
\bea
\tilde{Z}_{r,s}^{(N)} (q) &=& q^{-\frac{1}{8}}\sum_{k=1}^{s/2} \frac{q^{\frac{1}{8} (2r-s+4k)^2}}{(1-q^{t+2-2k})(1-q^{t+1-2k})}\gauss{N}{t-2k}_{\!q} \nonumber \\ &&\qquad \times \Big(q^{-2r+s+2-4k} (1-q^{N-t-1+2k})(1-q^{N-t+2k}) - (1-q^{t+2-2k})(1-q^{t+1-2k})\Big) \nonumber\\
&=& q^{-\frac{1}{8}}\sum_{k=1}^{s/2} \frac{q^{\frac{1}{8} (2r-s+4k)^2}}{(1-q^{t+2-2k})(1-q^{t+1-2k})}\gauss{N}{t-2k}_{\!q}  \nonumber \\   &&\qquad \times \Big(q^{-2r+s+2-4k} - 1-q^{t-2k} + q^{N}+q^{t+2-2k}-q^{N-2r+s+4-4k}  \Big)
\label{Zoddev}
\eea
Using the definition of the the $q$-Catalan polynomials in (\ref{qCat}), the decomposition (\ref{catId2}) of the finitized partition function with $\rho=2r-1$ can be expanded as
\bea
&& q^{-\frac{1}{8}} \sum_{k=1}^{s/2} \big[\qqcatt{\frac{N-1}{2}}{\frac{2r-s-2+4k}{2}}{q}+q^{\frac{N}{2}} \qqcatt{\frac{N-1}{2}}{\frac{2r-s-4+4k}{2}}{q} \big]\nonumber \\
&=& q^{-\frac{1}{8}}\sum_{k=1}^{s/2}\bigg( q^{\frac{1}{8} (2r-s-4+4k)^2}\frac{(1-q^{2r-s-2+4k})}{(1-q^{N-t+2k})}\gauss{N}{t+1-2k}_{\!q}\nonumber \\ 
&&\qquad \mbox{}+ q^{\frac{N}{2}}q^{\frac{1}{8} (2r-s-6+4k)^2} \frac{(1-q^{2r-s-4+4k})}{(1-q^{N-t-1+2k})}\gauss{N}{t+2-2k}_{\!q}\bigg) \nonumber \\
&=& q^{-\frac{1}{8}}\sum_{k=1}^{s/2}\bigg( q^{\frac{1}{8} (2r-s+4k)^2}q^{-2r+s+2-4k}\,\frac{(1-q^{2r-s-2+4k})}{(1-q^{N-t+2k})}\gauss{N}{t+1-2k}_{\!q}\nonumber \\ 
&&\qquad \mbox{}+ q^{\frac{1}{8} (2r-s+4k)^2}q^{\frac{N}{2}-3r+\frac{3s}{2}+\frac{36}{8}-6k}\,\frac{(1-q^{2r-s-4+4k})}{(1-q^{N-t-1+2k})}\gauss{N}{t+2-2k}_{\!q}\bigg) 
\eea
Applying the first part of (\ref{qbinprop}) to the first $q$-binomial, combining the two terms and then applying the identity again, we obtain the desired result in agreement with (\ref{Zoddev})
\bea
&& q^{-\frac{1}{8}} \sum_{k=1}^{s/2} \big[\qqcatt{\frac{N-1}{2}}{\frac{2r-s-2+4k}{2}}{q}+q^{\frac{N}{2}} \qqcatt{\frac{N-1}{2}}{\frac{2r-s-4+4k}{2}}{q} \big]\nonumber \\
&=&q^{-\frac{1}{8}} \sum_{k=1}^{s/2} q^{\frac{1}{8} (2r-s+4k)^2}\gauss{N}{t+1-2k}_{\!q} \bigg(q^{-2r+s+2-4k}\,\frac{(1-q^{2r-s-2+4k})}{(1-q^{N-t+2k})}\nonumber \\
&&\qquad \mbox{}+ q^{\frac{N}{2}-3r+\frac{3s}{2}+\frac{36}{8}-6k}\,\frac{(1-q^{2r-s-4+4k})}{(1-q^{t+2-2k})} \bigg)\nonumber\\
&=&q^{-\frac{1}{8}} \sum_{k=1}^{s/2} \frac{q^{\frac{1}{8} (2r-s+4k)^2}}{(1-q^{t+2-2k})(1-q^{N-t+2k})}\gauss{N}{t+1-2k}_{\!q} \bigg( q^{-2r+s+2-4k} \big(1-q^{2r-s-2+4k}\big)\big(1-q^{t+2-2k}\big) \nonumber \\
&&\qquad\mbox{} + q^{\frac{N}{2}-3r+\frac{3s}{2}+\frac{36}{8}-6k} \big(1-q^{2r-s-4+4k}\big)\big(1-q^{N-t+2k}\big) \bigg) \nonumber\\
&=& q^{-\frac{1}{8}}\sum_{k=1}^{s/2} \frac{q^{\frac{1}{8} (2r-s+4k)^2}}{(1-q^{t+2-2k})(1-q^{t+1-2k})}\gauss{N}{t-2k}_{\!q} \nonumber \\
&&\qquad \times \Big(q^{-2r+s+2-4k} - 1-q^{t-2k} + q^{N}+q^{t+2-2k}-q^{N-2r+s+4-4k} \Big)
\eea


\end{document}